%% file: paper.tex
\documentclass[onefignum,onetabnum]{siamonline250211}

\input{shared}

\ifpdf
\hypersetup{
  pdftitle={The Gaussian Latent Machine: Efficient Prior and Posterior Sampling for Inverse Problems},
  pdfauthor={M. Kuric, M. Zach, A. Habring, M. Unser, and T. Pock}
}
\fi

\begin{document}

\maketitle

% REQUIRED
\begin{abstract}
We consider the problem of sampling from a product-of-experts-type model that encompasses many standard prior and posterior distributions commonly found in Bayesian imaging. We show that this model can be easily lifted into a novel latent variable model,  which we refer to as a Gaussian latent machine. This leads to a general sampling approach that unifies and generalizes many existing sampling algorithms from the literature. Most notably,  it yields a highly efficient and effective two-block Gibbs sampling approach in the general case, while also specializing to direct sampling algorithms in particular cases. Finally, we present detailed numerical experiments that demonstrate the efficiency and effectiveness of our proposed sampling approach across a wide range of prior and posterior sampling problems from Bayesian imaging.
\end{abstract}

% REQUIRED

% REQUIRED

% Introduction
\input{sections/introduction.tex}

% Preliminaries
\input{sections/preliminaries.tex}

% Proposed Approach
\input{sections/approach.tex}

% Experimental Results
\input{sections/results.tex}

% Conclusions
\input{sections/conclusion.tex}

% Appendix
\newpage
\appendix
\input{sections/appendix.tex}

% Acknowledgements
\section*{Acknowledgments}
\input{sections/acknowledgements.tex}

% Bibliography
\bibliographystyle{siamplain}
\bibliography{references}
\end{document}

\maketitle
\section{Example of Different Measure in \eqref{def_fmp}}%
\label{sec:sm_measure}
Assume that a factor $\phi: \R \to \Rpp$ is a univariate \gls{gmm} of the form
\begin{align*}
	\phi\br{t} \ceq \sum_{i = 1}^d w_i \cdot \nd*{t; \mu_i, \sigma^2_i},	
\end{align*}
where $w \in \Delta_d$, $\mu_1, \ldots, \mu_d \in \R$, and $\sigma^2_1, \ldots, \sigma^2_d \in \Rpp$. Then $\mathcal{Z} = \set{1, \ldots, d}$, $f\br{z} = w_z$ for $z \in \set{1, \ldots, d}$, $\mu\br*{\zeta}= \mu_{\zeta}$, and $\sigma^2\br*{\zeta} = \sigma^2_{\zeta}$ for $\zeta \in \set{1, \ldots, d}$ and the measure in \eqref{def_fmp} is to be interpreted as counting measure and thus \eqref{def_fmp} yields
\begin{align*}
	\phi\br*{t} = \int_{\mathcal{Z}}  g\br*{t, z} \f{}{z}\,\mathrm{d} z = \sum_{z = 1}^d g(t, z) f(z) = \sum_{i = 1}^d w_i \cdot \nd*{t; \mu_i, \sigma^2_i}.
\end{align*}

\section{Example of \texorpdfstring{$\f_i \neq \f_{Z_i}$}{f\_i != Z\_i} in GLMs}%
\label{sec:sm_fi_fzi}
Consider the overcomplete GLM
\begin{align}
	\fxz{x, z} \propto& \prod_{i = 1}^2 g_i\br*{x, z_i} \cdot \f{i}{z_i},
	\label{glm_example}
\end{align}
where $K \ceq I_2$, $X \in \R$, $Z_1, Z_2 \in \set{0, 1}$, 
\begin{align*}
	g_1\br{x, z_1} \ceq& \begin{cases}
		\nd*{x; 0, \frac{1}{4 \pi}}, &\text{if}\ z_1 = 0 \\
		\nd*{x; \sqrt{\log 2 / \pi}, \frac{1}{4 \pi}}, &\text{otherwise}, 
	\end{cases}
	\\
	g_2\br{x, z_2} \ceq& \begin{cases}
		\nd*{x; -\sqrt{\log 2 / \pi}, \frac{1}{4 \pi}}, &\text{if } z_2 = 0 \\
		\nd*{x; 0, \frac{1}{4 \pi}}, &\text{otherwise},
	\end{cases}
\end{align*}
\begin{align*}
	\textstyle
	f_1\br{z_1} \ceq \begin{cases}
		3/10, &\text{if } z_1 = 0 \\
		7/10, &\text{otherwise}, 
	\end{cases} \quad \text{and} \quad f_2\br{z_2} \ceq \begin{cases}
		4/5, &\text{if } z_2 = 0 \\
		1/5, &\text{otherwise}.
	\end{cases}
\end{align*}

Denoting \( \gamma \coloneqq \sqrt{\log 2 / \pi} \), by plugging in the definitions of $g_1$, $g_2$, $f_1$, and $f_2$ into \eqref{glm_example}, it follows that
\begin{align}
		\fxz{x, z} \propto {} & \prod_{i = 1}^2 g_i\br*{x, z_i} \cdot \f{i}{z_i} \nonumber \\
		= {} & \br*{\tfrac{3}{10} \cdot \ib{z_1 = 0} \cdot \nd*{x; 0, \tfrac{1}{4 \pi}} + \tfrac{7}{10}\cdot \ib{z_1 = 1} \cdot \nd*{x; \gamma, \tfrac{1}{4 \pi}}} \nonumber \\
			&\cdot\br*{\tfrac{4}{5} \cdot \ib{z_2 = 0} \cdot \nd*{x; -\gamma, \tfrac{1}{4 \pi}} + \tfrac{1}{5} \cdot \ib{z_2 = 1} \cdot \nd*{x; 0, \tfrac{1}{4 \pi}}} \nonumber \\
		= {} & \tfrac{12}{50} \cdot \ib{z_1 = 0} \cdot \ib{z_2 = 0} \cdot \nd*{x; 0, \tfrac{1}{4 \pi}} \cdot \nd*{x; -\gamma, \tfrac{1}{4 \pi}} \nonumber \\
		&+ \tfrac{3}{50} \cdot \ib{z_1 = 0} \cdot \ib{z_2 = 1} \cdot \nd*{x; 0, \tfrac{1}{4 \pi}} \cdot \nd*{x; 0, \tfrac{1}{4 \pi}} \nonumber \\
		&+ \tfrac{28}{50} \cdot \ib{z_1 = 1} \cdot \ib{z_2 = 0} \cdot \nd*{x; \gamma, \tfrac{1}{4 \pi}} \cdot \nd*{x; -\gamma, \tfrac{1}{4 \pi}} \nonumber  \\
		&+ \tfrac{7}{50} \cdot \ib{z_1 = 1} \cdot \ib{z_2 = 1} \cdot \nd*{x; \gamma, \tfrac{1}{4 \pi}} \cdot \nd*{x; 0, \tfrac{1}{4 \pi}}.
	\label{algebra}
\end{align}
Via the following technical result regarding the integral of the product of two univariate normal distributions\footnote{The closed-form expression for a product of two univariate Gaussian distributions is derived in \cite[Section 1]{Bromiley:2014}. Hence, this technical result can be easily obtained by integrating out over this closed-form expression.}
\begin{align*}
	\int_{\R} \nd{x; \mu_1, \sigma_1^2} \cdot \nd{x; \mu_2, \sigma_2^2} \;dx = \nd{\mu_1; \mu_2, \sigma^2_1 + \sigma^2_2},
\end{align*}
where $\mu_1, \mu_2 \in \R$ are the means and $\sigma^2_1, \sigma^2_2 \in \Rpp$ are the variances, the following four results follow:
\begin{align*}
		&\int_{\R} \nd*{x; 0, \tfrac{1}{4 \pi}} \cdot \nd*{x; -\gamma, \tfrac{1}{4 \pi}} \;dx = \nd*{0; -\gamma, \tfrac{1}{2 \pi}} = \tfrac{1}{2}, \\
		&\int_{\R} \nd*{x; 0, \tfrac{1}{4 \pi}} \cdot \nd*{x; 0, \tfrac{1}{4 \pi}} \;dx = \nd*{0; 0, \tfrac{1}{2 \pi}} = 1,	\\
		&\int_{\R} \nd*{x; \gamma, \tfrac{1}{4 \pi}} \cdot \nd*{x; -\gamma, \tfrac{1}{4 \pi}} \;dx = \nd*{\gamma; -\gamma, \tfrac{1}{2 \pi}} = \tfrac{1}{16}, \\	
		&\int_{\R} \nd*{x; \gamma, \tfrac{1}{4 \pi}} \cdot \nd*{x; 0, \tfrac{1}{4 \pi}} \;dx = \nd*{\gamma; 0, \tfrac{1}{2 \pi}} = \tfrac{1}{2}.
\end{align*}
Based on these intermediate results, it follows from \eqref{algebra} that
\begin{equation}
	\begin{aligned}
		\fz{z} \propto {} & \int_{\R} \fxz{x, z} \;dx \\
		= {} & \tfrac{12}{50} \cdot \tfrac{1}{2} \cdot \ib{z_1 = 0} \cdot \ib{z_2 = 0} + \tfrac{3}{50} \cdot 1 \cdot \ib{z_1 = 0} \cdot \ib{z_2 = 1} \\
		&+ \tfrac{28}{50} \cdot \tfrac{1}{16} \cdot \ib{z_1 = 1} \cdot \ib{z_2 = 0} + \tfrac{7}{50} \cdot \tfrac{1}{2} \cdot \ib{z_1 = 1} \cdot \ib{z_2 = 1} \\
		= {} & \tfrac{96}{800} \cdot \ib{z_1 = 0} \cdot \ib{z_2 = 0} + \tfrac{48}{800} \cdot \ib{z_1 = 0} \cdot \ib{z_2 = 1} \\
		&+ \tfrac{28}{800} \cdot \ib{z_1 = 1} \cdot \ib{z_2 = 0} + \tfrac{56}{800} \cdot \ib{z_1 = 1} \cdot \ib{z_2 = 1} \\
		\propto {} & 96 \cdot \ib{z_1 = 0} \cdot \ib{z_2 = 0} + 48 \cdot \ib{z_1 = 0} \cdot \ib{z_2 = 1} \\
		&+ 28 \cdot \ib{z_1 = 1} \cdot \ib{z_2 = 0} + 56 \cdot \ib{z_1 = 1} \cdot \ib{z_2 = 1}.
		\label{def_latents_unnormalized}
	\end{aligned}
\end{equation}
Hence, it follows that
\begin{equation*}
	f_{Z_1}\br{z_1} \propto \sum_{z_2 \in \set{0, 1}} \fz{z_1, z_2} = 144 \cdot \ib{z_1 = 0} + 84 \cdot \ib{z_1 = 1},
\end{equation*}
and
\begin{equation*}
	f_{Z_2}\br{z_2} \propto \sum_{z_1 \in \set{0, 1}} \fz{z_1, z_2} = 124 \cdot \ib{z_2 = 0} + 104 \cdot \ib{z_2 = 1},
\end{equation*}
and, consequently, that
\begin{equation}
	\textstyle
	f_{Z_1}\br{z_1} = \begin{cases}
		12/19, &\text{if } z_1 = 0 \\
		7/19, &\text{otherwise}, 
	\end{cases} \quad \text{and} \quad f_{Z_2}\br{z_2} = \begin{cases}
		31/57, &\text{if } z_2 = 0 \\
		26/57, &\text{otherwise}.
	\end{cases}
    \label{def_latent_marginals}
\end{equation}
This shows that $f_1 \neq f_{Z_1}$ and $f_2 \neq f_{Z_2}$, as desired.

\section{Example of Dependent Latents in GLMs}%
\label{sec:dependent_latents}
Consider again the same example as in \Cref{sec:sm_fi_fzi}. One one hand, by renormalizing~\eqref{def_latents_unnormalized} and simplifying the resulting fractions, the latent distribution is given by
\begin{equation*}
	\begin{aligned}
		\fz{z} = {} & \frac{24}{57} \cdot \ib{z_1 = 0} \cdot \ib{z_2 = 0} + \frac{12}{57} \cdot \ib{z_1 = 0} \cdot \ib{z_2 = 1} \\
					& + \frac{7}{57} \cdot \ib{z_1 = 1} \cdot \ib{z_2 = 0} + \frac{14}{57} \cdot \ib{z_1 = 1} \cdot \ib{z_2 = 1}.
	\end{aligned}
\end{equation*}
On the other hand, after simplifying the resulting fractions from~\eqref{def_latent_marginals}, the product of the individual latent distributions is given by
\begin{equation*}
	\begin{aligned}
		f_{Z_1}\br{z_1} \cdot f_{Z_2}\br{z_2} = {} & \frac{124}{361} \cdot \ib{z_1 = 0} \cdot \ib{z_2 = 0} + \frac{104}{361} \cdot \ib{z_1 = 0} \cdot \ib{z_2 = 1} \\
		& +\frac{217}{1083} \cdot \ib{z_1 = 1} \cdot \ib{z_2 = 0} + \frac{182}{1083} \cdot \ib{z_1 = 1} \cdot \ib{z_2 = 1}.
	\end{aligned}
\end{equation*}
Since $\fz \neq f_{Z_1} \cdot f_{Z_2}$, we can conclude that the components $Z_1$ and $Z_2$ of the latent variable $Z$ are mutually dependent. 

\section{PDF Derivation for the Symmetrized Gamma Distribution} 
\label{sec:sgamma}
Suppose that $X \da \SymGammaDist\br{\alpha, \beta}$ with parameters $\alpha, \beta \in \Rpp$. Therefore, $X$ can be expressed as a Gaussian scale mixture via the conditional Gaussian distribution $f_{X \mid V}\br{x \mid v} = \nd{x; 0, v}$ and the latent distribution $V \da \GammaDist\br{\alpha, \beta}$. Consequently, its density is given by
\begin{align}
	\fx{x} =& \int^{\infty}_{0} f_{X \mid V}\br{x \mid v} f_V\br{v} \;dv \nonumber \\
		=& \int^{\infty}_{0} \frac{1}{\sqrt{2 \pi v}} \cdot \exp\br*{-\frac{x^2}{2 v}} \cdot \frac{\beta^\alpha}{\Gamma\br*{\alpha}} \cdot v^{\alpha - 1} \cdot \exp\br*{-\beta v} \;dv \nonumber \\
		=& \frac{\beta^\alpha}{\sqrt{2 \pi} \cdot \Gamma\br*{\alpha}} \cdot \int^{\infty}_{0} v^{\alpha - \frac{1}{2} - 1} \cdot \exp\br*{-\frac{x^2}{2 v} -\beta v} \;dv.
		\label{eq_pdf_sgamma}
\end{align}
Based on the following technical result \cite[Equation 3.471.9]{Gradshteyn:2007}
\begin{align*}
	\int^{\infty}_{0} z^{a - 1} \cdot \exp\br*{-\frac{b}{z} -c z} \;dz = 2 \cdot \br*{\frac{b}{c}}^{\frac{a}{2}} \cdot K_a\br*{2 \cdot \sqrt{bc}}
\end{align*}
that holds for any $b, c \in \Rpp$ and $K_a$ denotes the modified Bessel function of the second kind with parameter $a \in \R$, it follows by the relabeling $a \ceq \alpha - \frac{1}{2}$, $b \ceq \frac{x^2}{2}$ and $c \ceq \beta$ that
\begin{align*}
	\int^{\infty}_{0} v^{\alpha - \frac{1}{2} - 1} \cdot \exp\br*{-\frac{x^2}{2 v} -\beta v} \;dv =& 2 \cdot \br*{\frac{x^2}{2 \beta}}^{\frac{\alpha - \frac{1}{2}}{2}} \cdot K_{\alpha - \frac{1}{2}}\br*{2 \cdot \sqrt{\frac{\beta x^2}{2}}} \\
	=& 2 \cdot \br*{\frac{\abs{x}}{\sqrt{2 \beta}}}^{\alpha - \frac{1}{2}} \cdot K_{\alpha - \frac{1}{2}}\br*{\sqrt{2 \beta} \cdot \abs{x}},
\end{align*}
and consequently from \eqref{eq_pdf_sgamma} that
\begin{align*}
	\fx{x} = \frac{\sqrt{2} \cdot \beta^{\alpha}}{\sqrt{\pi} \cdot \Gamma\br{\alpha}} \cdot \br*{\frac{\abs{x}}{\sqrt{2 \beta}}}^{\alpha - \frac{1}{2}} \cdot K_{\alpha - \frac{1}{2}}\br*{\sqrt{2 \beta} \cdot \abs{x}},
\end{align*}
as desired.

\section{Derivations of the Conditional Latent Distributions} 
\label{sec:cond_latent_derivation}
The derivations for each factor type are given in the following subsections.
\subsection{Laplace} Observe that
\begin{align*}
	\f_{Z_i \mid X}\br*{z_i \mid x} \propto& g_i\br*{\br{Kx}_i,  z_i} \cdot \f{i}{z_i}	 \\
	=& \frac{1}{\sqrt{2 \pi z_i}} \cdot \exp\br*{-\frac{\br{Kx}^2_i}{2 z_i}} \cdot \frac{1}{2 b^2} \cdot \exp\br*{\frac{z_i}{2 b^2}} \\
	\propto& z^{-\frac{1}{2}}_i \cdot \exp\br*{-\frac{\frac{z_i}{b^2} + \frac{\br{Kx}^2_i}{z_i}}{2}},
\end{align*}
which immediately implies that $\f_{Z_i \mid X = x} = \GIGDist\br*{\frac{1}{b^2}, \br{Kx}^2_i, \frac{1}{2}}$,  as desired.

\subsection{Student-t}
Observe that
\begin{align*}
	\f_{Z_i \mid X}\br*{z_i \mid x} \propto& g_i\br*{\br{Kx}_i,  z_i} \cdot \f{i}{z_i} \\
	=& \frac{\sqrt{z_i}}{\sqrt{2 \pi}} \cdot \exp\br*{-\frac{z_i}{2} \br{Kx}^2_i} \cdot \frac{\br{\frac{\nu}{2}}^{\frac{\nu}{2}}}{\Gamma\br{\frac{\nu}{2}}} \cdot z^{\frac{\nu}{2} - 1}_i \cdot \exp\br*{-\frac{\nu}{2} z_i} \\
	\propto& z^{\frac{\nu + 1}{2} - 1}_i \cdot \exp\br*{-\frac{\nu + \br{Kx}^2_i}{2} \cdot z_i},
\end{align*}
which immediately implies that $\f_{Z_i \mid X = x} = \GammaDist\br*{\frac{\nu + 1}{2}, \frac{\nu + \br{Kx}^2_i}{2}}$,  as desired.

\subsection{Symmetrized Gamma} Observe that
\begin{align*}
	\f_{Z_i \mid X}\br*{z_i \mid x} \propto& g_i\br*{\br{Kx}_i,  z_i} \cdot \f{i}{z_i}	\\
		=& \frac{1}{\sqrt{2 \pi z_i}} \cdot \exp\br*{-\frac{\br{Kx}^2_i}{2 z_i}} \cdot \frac{\beta^{\alpha}}{\Gamma\br{\alpha}} \cdot z^{\alpha - 1}_i \cdot \exp\br{-\beta z_i} \\
		\propto& \cdot z^{\alpha -\frac{1}{2} - 1}_i \cdot \exp\br*{-\frac{2 \beta z_i + \frac{\br{Kx}^2_i}{z_i}}{2}},
\end{align*}
which immediately implies that $\f_{Z_i \mid X = x} = \GIGDist\br*{2 \beta, \br{Kx}^2_i, \alpha - \frac{1}{2}}$, as desired.

\subsection{Gaussian Mixture Model} Observe that
\begin{align*}
	\f_{Z_i \mid X}\br*{z_i \mid x} \propto& g_i\br*{\br{Kx}_i,  z_i} \cdot \f{i}{z_i} \\
		=& \sum_{j = 1}^d w_j \cdot \nd{\br{Kx}_i; \mu_j, \sigma^2_j} \cdot \ib{z_i = j},
\end{align*}
which immediately implies that $\f_{Z_i \mid X = x} = \CatDist\br*{\bar{w}}$ where $\bar{w}_j \ceq \frac{w_j \cdot \nd{\br{Kx}^2_i; \mu_j, \sigma^2_j}}{\sum_{k = 1}^d w_k \cdot \nd{\br{Kx}^2_i; \mu_k, \sigma^2_k}}$ for $j = 1, \ldots, d$, as desired.

\section{Derivations of the Ground Truth Marginals in the Baseline Experiments}
\label{sec:baseline_derivations}
For the sake of simplicity, we assume symmetric factors in our derivations. Furthermore, we rely on the following technical result
\begin{align*}
	\int_{-\infty}^{\infty} \f{x - y} \g{y - z} dy = \br*{f * g}\br*{x - z},
\end{align*}
which follows trivially from the change of variables $u = x - y$, where $f$ and $g$ are univariate functions and $*$ denotes the convolution operator. 

\subsection{Loop Topology} By symmetry of the factors and the loop topology, it follows that the marginal distributions $X_2 - X_1$, $X_4 - X_3$, $X_3 - X_1$ and $X_4 - X_2$ are the same. Therefore, we will only derive the marginal distribution of $X_2 - X_1$. This factor graph defines an improper distribution on $\R^4$, and therefore we can, as per \Cref{prop:smart_tie_breaking_works_too}, add a factor acting on the first component of $X$ to obtain a proper density without modifying the marginals on the original improper distribution. This leads to proper distributions of the form
\begin{align*}
	\fx{x} \propto \phi\br{x_1} \cdot \phi\br{x_2 - x_1} \cdot \phi\br{x_4 - x_2} \cdot \phi\br{x_4 - x_3} \cdot \phi\br{x_3 - x_1},
\end{align*}
which after marginalizing out over $X_3$ and $X_4$ yields
\begin{align*}
	& \int_{-\infty}^{\infty} \int_{-\infty}^{\infty} \fx{x} dx_3 dx_4 \\
	=& \phi\br{x_1} \cdot \phi\br{x_2 - x_1} \cdot \int_{-\infty}^{\infty} \phi\br{x_4 - x_2} \underbrace{\int_{-\infty}^{\infty} \phi\br{x_4 - x_3} \phi\br{x_3 - x_1} dx_3}_{= \br{\phi * \phi}\br{x_4 - x_1}} dx_4 \\
	=& \phi\br{x_1} \cdot \phi\br{x_2 - x_1} \cdot \int_{-\infty}^{\infty} \underbrace{\phi\br{x_4 - x_2}}_{= {\phi}\br{x_2 - x_4}} \br{\phi * \phi}\br{x_4 - x_1} dx_4 \\
	=& \phi\br{x_1} \cdot \phi\br{x_2 - x_1} \cdot \underbrace{\int_{-\infty}^{\infty} {\phi}\br{x_2 - x_4} \br{\phi * \phi}\br{x_4 - x_1} dx_4}_{\br{{\phi} * \phi * \phi}\br{x_2 - x_1}} \\
	=& \phi\br{x_1} \cdot \br*{\br{{\phi} * \phi * \phi} \cdot \phi}\br{x_2 - x_1}.
\end{align*}
By introducing the change of variables $U_1 = X_1$ and $U_2 = X_2 - X_1$ and marginalizing out over $U_1$, it trivially follows that
\begin{align*}
	\f_{X_2 - X_1} \propto \br*{\phi * \phi * \phi} \cdot \phi,
\end{align*}
as desired.

\subsection{Grid Topology} Similarly, by symmetry of the factors and the grid topology, it follows that the marginal distributions $X_2 - X_1$, $X_3 - X_2$, $X_5 - X_4$, $X_6 - X_5$, $X_4 - X_1$ and $X_6 - X_3$ are the same. Therefore, we will derive the inner marginal distribution $X_5 - X_2$ and the outer marginal distribution $X_2 - X_1$. This factor graph defines an improper distribution on $\R^4$, and therefore we can, as per \Cref{prop:smart_tie_breaking_works_too}, add a factor acting on the second component of $X$ to obtain a proper density without modifying the marginals on the original improper distribution. This leads to proper distributions of the form
\begin{align*}
	\fx{x} \propto \phi\br{x_2} \cdot \phi\br{x_3 - x_2} \cdot \phi\br{x_5 - x_4} \cdot \phi\br{x_6 - x_5} \cdot \phi\br{x_4 - x_1} \cdot \phi\br{x_5 - x_2}.
\end{align*}

\subsubsection{Deriving the Marginal Distribution of \texorpdfstring{$X_5 - X_2$}{X\_5 - X\_2}} Marginalizing out $X_1$, $X_3$, $X_4$, and $X_6$ from $\fx$ yields
\begin{align*}
	& \int_{-\infty}^{\infty} \int_{-\infty}^{\infty} \int_{-\infty}^{\infty} \int_{-\infty}^{\infty} \fx{x} dx_1 dx_3 dx_4 dx_6 \\
=& \phi\br{x_2} \cdot \phi\br{x_5 - x_2} \int_{-\infty}^{\infty} \phi\br{x_6 - x_5} \underbrace{\int_{-\infty}^{\infty} \phi\br{x_6 - x_3} \phi\br{x_3 - x_2} dx_3}_{= \br{\phi * \phi}\br{x_6 - x_2}} dx_6 \\
	&\int_{-\infty}^{\infty} \phi\br{x_2 - x_1} \underbrace{\int_{-\infty}^{\infty} \phi\br{x_5 - x_4} \phi\br{x_4 - x_1} dx_4}_{= \br{\phi * \phi}\br{x_5 - x_1}} dx_1 \\
	=& \phi\br{x_2} \cdot \phi\br{x_5 - x_2} \int_{-\infty}^{\infty} \underbrace{\phi\br{x_6 - x_5}}_{= {\phi}\br{x_5 - x_6}} \br{\phi * \phi}\br{x_6 - x_2} dx_6 \int_{-\infty}^{\infty} \underbrace{\phi\br{x_2 - x_1}}_{= {\phi}\br{x_1 - x_2}} \br{\phi * \phi}\br{x_5 - x_1} dx_1
\end{align*}
\begin{align*}
	=& \phi\br{x_2} \cdot \phi\br{x_5 - x_2} \cdot \underbrace{\int_{-\infty}^{\infty} {\phi}\br{x_5 - x_6} \br{\phi * \phi}\br{x_6 - x_2} dx_6}_{= \br{{\phi} * \phi * \phi}\br{x_5 - x_2}} \underbrace{\int_{-\infty}^{\infty} \br{\phi * \phi}\br{x_5 - x_1} {\phi}\br{x_1 - x_2}  dx_1}_{= \br{\phi * \phi * {\phi}}\br{x_5 - x_2}} \\
	=& \phi\br{x_2} \cdot \br*{\br{\phi * \phi * {\phi}} \cdot \phi \cdot \br{\phi * \phi * {\phi}}}\br{x_5 - x_2}.
\end{align*}
By introducing the change of variables $U_1 = X_2$ and $U_2 = X_5 - X_2$ and marginalizing out over $U_1$, it trivially follows that
\begin{align*}
	\f_{X_5 - X_2} \propto \br*{\phi * \phi * \phi} \cdot \phi \cdot \br*{\phi * \phi * \phi},
\end{align*}
as desired.

\subsubsection{Deriving the Marginal Distribution of \texorpdfstring{$X_2 - X_1$}{X\_2 - X\_1}} Similarly, marginalizing out $X_3$, $X_4$, $X_5$, and $X_6$ from $\fx$ yields
\begin{align*}
		& \phi\br{x_2} \cdot \int_{-\infty}^{\infty} \int_{-\infty}^{\infty} \int_{-\infty}^{\infty} \int_{-\infty}^{\infty} \fx{x} dx_3 dx_4 dx_5 dx_6 \\
	&= \phi\br{x_2} \cdot \phi\br{x_2 - x_1} \cdot \int_{-\infty}^{\infty} \int_{-\infty}^{\infty} \br*{\int_{-\infty}^{\infty} \phi\br{x_6 - x_5} \underbrace{\int_{-\infty}^{\infty} \phi\br{x_6 - x_3} \phi\br{x_3 - x_2} dx_3}_{= \br{\phi * \phi}\br{x_6 - x_2}} dx_5} \\
	& \phi\br{x_5 - x_2} \phi\br{x_5 - x_4} \phi\br{x_4 - x_1} dx_5 dx_4 \\
	&= \phi\br{x_2} \cdot \phi\br{x_2 - x_1} \cdot \int_{-\infty}^{\infty} \int_{-\infty}^{\infty} \br*{\int_{-\infty}^{\infty} \underbrace{\phi\br{x_6 - x_5}}_{= {\phi}\br{x_5 - x_6}} \br{\phi * \phi}\br{x_6 - x_2} dx_5} \\
	& \phi\br{x_5 - x_2} \phi\br{x_5 - x_4} \phi\br{x_4 - x_1} dx_5 dx_4 \\
	&= \phi\br{x_2} \cdot \phi\br{x_2 - x_1} \cdot \int_{-\infty}^{\infty} \int_{-\infty}^{\infty} \underbrace{\br*{\int_{-\infty}^{\infty} {\phi}\br{x_5 - x_6} \br{\phi * \phi}\br{x_6 - x_2} dx_5}}_{= \br{{\phi} * \phi * \phi}\br{x_5 - x_2}} \\
	& \phi\br{x_5 - x_2} \phi\br{x_5 - x_4} \phi\br{x_4 - x_1} dx_5 dx_4 \\
	&= \phi\br{x_2} \cdot \phi\br{x_2 - x_1} \cdot \int_{-\infty}^{\infty} \int_{-\infty}^{\infty} \underbrace{\br{{\phi} * \phi * \phi}\br{x_5 - x_2} \phi\br{x_5 - x_2}}_{= \br{\br{{\phi} * \phi * \phi} \cdot \phi}\br{x_5 - x_2}} \phi\br{x_5 - x_4} \phi\br{x_4 - x_1} dx_5 dx_4 \\
	&= \phi\br{x_2} \cdot \phi\br{x_2 - x_1} \cdot \int_{-\infty}^{\infty} \int_{-\infty}^{\infty} \br{\br{{\phi} * \phi * \phi} \cdot \phi}\br{x_5 - x_2} \phi\br{x_5 - x_4} \phi\br{x_4 - x_1} dx_5 dx_4 \\
	&= \phi\br{x_2} \cdot \phi\br{x_2 - x_1} \cdot \int_{-\infty}^{\infty} \phi\br{x_4 - x_1} \int_{-\infty}^{\infty} \underbrace{\phi\br{x_5 - x_4}}_{={\phi}\br{x_4 - x_5}} \br{\br{{\phi} * \phi * \phi} \cdot \phi} \br{x_5 - x_2} dx_5 dx_4 \\
	&= \phi\br{x_2} \cdot \phi\br{x_2 - x_1} \cdot \int_{-\infty}^{\infty} \phi\br{x_4 - x_1} \underbrace{\int_{-\infty}^{\infty} {\phi}\br{x_4 - x_5} \br{\br{{\phi} * \phi * \phi} \cdot \phi} \br{x_5 - x_2} dx_5}_{= \br{\br{\br{{\phi} * \phi * \phi} \cdot \phi} * \phi}\br{x_4 - x_2}} dx_4
\end{align*}

\begin{align*}
	&= \phi\br{x_2} \cdot \phi\br{x_2 - x_1} \cdot \int_{-\infty}^{\infty} \underbrace{\phi\br{x_4 - x_1}}_{= \phi\br{x_1 - x_4}} \br{\br{\br{{\phi} * \phi * \phi} \cdot \phi} * \phi}\br{x_4 - x_2} dx_4 \\
	&= \phi\br{x_2} \cdot \phi\br{x_2 - x_1} \cdot \underbrace{\int_{-\infty}^{\infty} \phi\br{x_1 - x_4} \br{\br{\br{{\phi} * \phi * \phi} \cdot \phi} * \phi}\br{x_4 - x_2} dx_4}_{= \br{\br{\br{{\phi} * \phi * \phi} \cdot \phi} * \phi * \phi}\br{x_1 - x_2}} \\
	&= \phi\br{x_2} \cdot \phi\br{x_2 - x_1} \cdot \underbrace{\br{\br{\br{{\phi} * \phi * \phi} \cdot \phi} * \phi * \phi}\br{x_1 - x_2}}_{=\br{\br{\br{{\phi} * \phi * \phi} \cdot \phi} * \phi * \phi}\br{x_2 - x_1}} \\
	&= \phi\br{x_2} \cdot \br{\br{\br{\br{{\phi} * \phi * \phi} \cdot \phi} * \phi * \phi} \cdot \phi}\br{x_2 - x_1},
\end{align*}
where in the penultimate equality we have used the fact that products and convolutions of symmetric functions preserve symmetry. Finally, by introducing the change of variables $U_1 = X_2$ and $U_2 = X_2 - X_1$ and marginalizing out over $U_1$, it trivially follows that
\begin{align*}
	\f_{X_2 - X_1} \propto \br*{\br*{\br*{\phi * \phi * \phi} \cdot \phi} * \phi * \phi} \cdot \phi,
\end{align*}
as desired.

\bibliographystyle{siamplain}
\bibliography{references}

%% file: shared.tex
\usepackage{glossaries-extra}
\glsdisablehyper%
\newabbreviation{glm}{GLM}{Gaussian latent machine}
\newabbreviation{gmm}{GMM}{Gaussian mixture model}
\newabbreviation{gsm}{GSM}{Gaussian scale mixture}
\newabbreviation{mcmc}{MCMC}{Markov chain Monte Carlo}
\newabbreviation{map}{MAP}{maximum a-posteriori}
\newabbreviation{sde}{SDE}{stochastic differential equation}
\newabbreviation{ula}{ULA}{unadjusted Langevin algorithm}
\newabbreviation{hmc}{HMC}{Hamiltonian Monte Carlo}
\newabbreviation{mala}{MALA}{Metropolis-adjusted Langevin algorithm}
\newabbreviation{psnr}{PSNR}{peak signal-to-noise ratio}
\newabbreviation{dct}{DCT}{discrete cosine transform}
\usepackage{amsfonts}
\usepackage{graphicx}
\usepackage{epstopdf}
\usepackage{bbm}
\ifpdf
  \DeclareGraphicsExtensions{.eps,.pdf,.png,.jpg}
\else
  \DeclareGraphicsExtensions{.eps}
\fi

\usepackage{enumitem}
\setlist[enumerate]{leftmargin=.5in}
\setlist[itemize]{leftmargin=.5in}

\usepackage{adjustbox}
\usepackage{subcaption}
\usepackage{csquotes}
\usepackage{booktabs}
\usepackage{stmaryrd}
\usepackage{siunitx}
\usepackage{microtype}
\usepackage{amssymb}
\usepackage{threeparttable, adjustbox}
\usepackage{xcolor}
\definecolor{RedViolet}{rgb}{0.77,0.08,0.52}
\usepackage{tikz}
\usetikzlibrary{backgrounds, positioning, plotmarks}
\usepackage{pgfplots}
\usepgfplotslibrary{groupplots, statistics, fillbetween}
\pgfplotsset{compat=1.17}
\usepgfplotslibrary{external}
\usepackage{algpseudocode}
\usepackage{enumitem}
\usepackage{placeins}

\renewcommand{\P}{\mathbb{P}}
\newcommand{\1}{\mathbb{1}} % Changed to make it compile

\newsiamremark{remark}{Remark}
\newsiamremark{hypothesis}{Hypothesis}
\crefname{hypothesis}{Hypothesis}{Hypotheses}
\newsiamthm{claim}{Claim}

\headers{The Gaussian Latent Machine}{M. Kuric, M. Zach, A. Habring, M. Unser, and T. Pock}

\title{The Gaussian Latent Machine:{\\}Efficient Prior and Posterior Sampling for Inverse Problems\thanks{Submitted to the editors May 20, 2025.
  \funding{This work has received funding from the European Union’s EIC Pathfinder Challenges 2022 programme under grant agreement No~101115317 (NEO).}
}}

\author{Muhamed Kuric\thanks{Institute of Visual Computing, Graz University of Technology, 8010 Graz, Austria  
  (\email{muhamed.kuric@tugraz.at, andreas.habring@tugraz.at, thomas.pock@tugraz.at}).}
\and Martin Zach\thanks{Biomedical Imaging Group and Center for Biomedical Imaging, École Polytechnique Fédérale de Lausanne, 1015 Lausanne, Switzerland 
  (\email{martin.zach@epfl.ch, michael.unser@epfl.ch}).}
\and Andreas Habring\footnotemark[2]
\and Michael Unser\footnotemark[3]
\and Thomas Pock\footnotemark[2]
}

\usepackage{amsopn}

\input defs.tex

\Crefname{ALC@unique}{Line}{Lines}

\newtheorem{assumption}{Assumption}

\definecolor{maincolor}{HTML}{4C72B0}
\definecolor{secondarycolor}{HTML}{55A868}
\definecolor{tertiarycolor}{HTML}{DD8452}
\definecolor{quaternarycolor}{HTML}{DA8BC3}
\definecolor{quinarycolor}{HTML}{8172B3}
\tikzset{line join=bevel}

\pgfplotsset{%
	/tikz/line join=bevel,
	compat/labels=pre 1.3,
	every boxplot/.style={mark=o,every mark/.append style={mark size=1.5pt}},
	marginal group plot/.style={
		group/group size=4 by 4,
		group/x descriptions at=edge bottom,
		group/vertical sep=3mm,
		enlargelimits=false,
		enlarge y limits={value=0.0125,lower},
		width=2.7cm,
		height=2.16cm,
		scale only axis,
		ylabel style={text depth=0pt},
		title style={text depth=0pt},
		ticklabel style={font=\tiny},
		xticklabel={
    			\ifdim \tick pt < 0pt
      		\pgfmathparse{abs(\tick)}%
      		\llap{$-{}$}\pgfmathprintnumber{\pgfmathresult}
   			\else
      		\pgfmathprintnumber{\tick}
   			\fi
		},
	},
	asmarginal group plot/.style={
		/tikz/line join=bevel,
		group/group size=4 by 4,
		group/x descriptions at=edge bottom,
		group/y descriptions at=edge left,
		group/vertical sep=3mm,
		group/horizontal sep=3mm,
		enlargelimits=false,
		enlarge y limits={value=0.0125,lower},
		width=2.7cm,
		height=2.16cm,
		scale only axis,
		ylabel style={text depth=0pt},
		title style={text depth=0pt},
		ticklabel style={font=\tiny},
		xticklabel={
    			\ifdim \tick pt < 0pt
      		\pgfmathparse{abs(\tick)}%
      		\llap{$-{}$}\pgfmathprintnumber{\pgfmathresult}
   			\else
      		\pgfmathprintnumber{\tick}
   			\fi
		},
	},
	convplots group plot/.style={
		/tikz/line join=bevel,
		group/group size=2 by 3,
		group/vertical sep=6mm,
		group/horizontal sep=8mm,
		group/x descriptions at=all,
		enlargelimits=false,
		enlarge y limits={value=0.0125,lower},
		width=2.7cm,
		height=2.16cm,
		scale only axis,
		ylabel style={text depth=0pt},
		title style={text depth=0pt},
		ticklabel style={font=\tiny},
		xticklabel={
    			\ifdim \tick pt < 0pt
      		\pgfmathparse{abs(\tick)}%
      		\llap{$-{}$}\pgfmathprintnumber{\pgfmathresult}
   			\else
      		\pgfmathprintnumber{\tick}
   			\fi
		},
	},
	wasserstein group plot/.style={
		/tikz/line join=bevel,
		group/group size=4 by 4,
		group/vertical sep=6mm,
		group/x descriptions at=all,
		enlargelimits=false,
		enlarge y limits={value=0.05,lower},
		width=2.7cm,
		height=2.16cm,
		scale only axis,
		ylabel style={text depth=0pt},
		title style={text depth=0pt},
		ticklabel style={font=\tiny},
		xticklabel={
    			\ifdim \tick pt < 0pt
      		\pgfmathparse{abs(\tick)}%
      		\llap{$-{}$}\pgfmathprintnumber{\pgfmathresult}
   			\else
      		\pgfmathprintnumber{\tick}
   			\fi
		},
	}
}

\newcommand{\drawcolorbar}{
	\pgfplotscolorbardrawstandalone[
		scale=0.315, colormap={blackwhite}{gray(0cm)=(0); gray(.6cm)=(1)}, colorbar style={ticks=none},
	]%
}
\newcommand{\drawcolorbarh}[1]{%
	\pgfplotscolorbardrawstandalone[
		scale=0.315, colorbar horizontal, colormap={blackwhite}{gray(0cm)=(0); gray(.6cm)=(1)}, colorbar style={ticks=none, width=#1},
	]%
}

%% file: defs.tex
\usepackage{xparse}

\usepackage{mathtools}
\usepackage{bbm}

\usepackage{suffix}

\newcommand{\x}{\times}
\renewcommand{\1}{\mathbbm{1}}

\DeclareMathOperator{\ExpDist}{Exp}
\DeclareMathOperator{\LaplaceDist}{Laplace}
\DeclareMathOperator{\tDist}{St}
\DeclareMathOperator{\GammaDist}{Gamma}
\DeclareMathOperator{\SymGammaDist}{SymGamma}
\DeclareMathOperator{\GIGDist}{GIG}
\DeclareMathOperator{\CatDist}{Cat}
\DeclareMathOperator{\GMMDist}{GMM}
\DeclareMathOperator{\PSNR}{PSNR}

\newcommand{\ceq}{\coloneq}						% Assignment operator
\newcommand{\eqc}{\eqqcolon}					% Reversed assignment operator
\newcommand{\deq}{\triangleq}					% Equal by definition
\DeclarePairedDelimiter{\br}{(}{)}					% brackets ()
\DeclarePairedDelimiter{\sbr}{[}{]}					% square brackets []
\DeclarePairedDelimiter{\cbr}{\{}{\}}				% curly brackets {}
\DeclarePairedDelimiter{\ocbr}{(}{]}					% open closed brackets
\DeclarePairedDelimiter{\cobr}{[}{)}					% closed open brackets
\DeclarePairedDelimiter{\ipbr}{\langle}{\rangle}		% inner product brackets
\DeclarePairedDelimiter\vertb{\lvert}{\rvert}			% vertical "brackets"
\DeclarePairedDelimiter\Vertb{\lVert}{\rVert}			% double vertical "brackets"
\DeclarePairedDelimiter{\ib}{\llbracket}{\rrbracket}	% Iverson brackets

\newcommand{\R}{\mathbb{R}} 				% set of real numbers
\newcommand{\Rp}{\mathbb{R}_{+}} 		% set of nonnegative real numbers
\newcommand{\Rpp}{\mathbb{R}_{++}} 		% set of positive real numbers
\newcommand{\symmpp}{\mathbb{S}_{++}} 	% set of symmetric positive definite matrices
\newcommand{\oi}[3][]{\br[#1]{#2 , #3}} 			% open interval
\newcommand{\ci}[3][]{\sbr[#1]{#2 , #3}}	 		% closed interval
\newcommand{\oci}[3][]{\ocbr[#1]{#2 , #3}} 		% open-closed interval
\newcommand{\coi}[3][]{\cobr[#1]{#2 , #3}} 		% closed-open interval

\WithSuffix\newcommand\oi*[2]{\br*{#1 , #2}}	
\WithSuffix\newcommand\ci*[2]{\sbr*{#1 , #2}}
\WithSuffix\newcommand\oci*[2]{\ocbr*{#1 , #2}}
\WithSuffix\newcommand\coi*[2]{\cobr*{#1 , #2}}

\DeclarePairedDelimiter{\tuple}{(}{)}			% multiple arguments can be separated with a comma

\NewDocumentCommand{\set}{ O{} s g g} {
	\ensuremath{
		\IfBooleanTF{#2}			
			{
				\IfNoValueTF{#3}{\cbr*}{\IfNoValueTF{#4}{\cbr*{#3}}{\cbr*{#3 \mid #4}}}		
			}
			{
				\IfNoValueTF{#3}{\cbr[#1]}{\IfNoValueTF{#4}{\cbr[#1]{#3}}{\cbr[#1]{#3 \mid #4}}}
			}
	}
}

\NewDocumentCommand{\ip}{ O{} s G{\cdot} G{\cdot} } {
	\ensuremath{
		\IfBooleanTF{#2}			
		{
			\ipbr*{#3, #4}		
		}
		{
			\ipbr[#1]{#3, #4}	
		}
	}
}

\NewDocumentCommand{\ipn}{ O{} s G{\cdot} G{\cdot} } {
	\ensuremath{ 
		\IfBooleanTF{#2}			
		{
			\ipbr*{#3, #4}_{\R^n}		
		}
		{
			\ipbr[#1]{#3, #4}_{\R^n}
		}
	}
}

\NewDocumentCommand{\ipm}{ O{} s G{\cdot} G{\cdot} } {
	\ensuremath{ 
		\IfBooleanTF{#2}			
		{
			\ipbr*{#3, #4}_{\R^m}		
		}
		{
			\ipbr[#1]{#3, #4}_{\R^m}
		}
	}
}

\NewDocumentCommand{\ipmn}{ O{} s G{\cdot} G{\cdot} } {
	\ensuremath{ 
		\IfBooleanTF{#2}			
		{
			\ipbr*{#3, #4}_{\R^{m \x n}}		
		}
		{
			\ipbr[#1]{#3, #4}_{\R^{m \x n}}
		}
	}
}

\NewDocumentCommand{\ipf}{ O{} s G{\cdot} G{\cdot} } {
	\ensuremath{ 
		\IfBooleanTF{#2}			
		{
			\ipbr*{#3, #4}_{F}
		}
		{
			\ipbr[#1]{#3, #4}_{F}
		}
	}
}

\NewDocumentCommand{\abs}{ O{} s G{\cdot} } {
	\ensuremath{
		\IfBooleanTF{#2}			
		{
			\vertb*{#3}		
		}
		{
			\vertb[#1]{#3}
		}
	}
}

\NewDocumentCommand{\huber}{ O{} s G{\cdot} } {
	\ensuremath{
		\IfBooleanTF{#2}			
		{
			\vertb*{#3}_{\varepsilon}	
		}
		{
			\vertb[#1]{#3}_{\varepsilon}
		}
	}
}

\NewDocumentCommand{\norm}{ O{} s G{\cdot} } {
	\ensuremath{
		\IfBooleanTF{#2}			
		{
			\Vertb*{#3}
		}
		{
			\Vertb[#1]{#3}
		}
	}
}

\NewDocumentCommand{\dnorm}{ O{} s G{\cdot} } {
	\ensuremath{
		\IfBooleanTF{#2}			
		{
			\Vertb*{#3}_{\star}
		}
		{
			\Vertb[#1]{#3}_{\star}
		}
	}
}

\NewDocumentCommand{\normz}{ O{} s G{\cdot} } {
	\ensuremath{
		\IfBooleanTF{#2}			
		{
			\Vertb*{#3}_{0}
		}
		{
			\Vertb[#1]{#3}_{0}
		}
	}
}

\NewDocumentCommand{\normo}{ O{} s G{\cdot} } {
	\ensuremath{
		\IfBooleanTF{#2}			
		{
			\Vertb*{#3}_{1}
		}
		{
			\Vertb[#1]{#3}_{1}
		}
	}
}

\NewDocumentCommand{\normh}{ O{} s G{\cdot} } {
	\ensuremath{
		\IfBooleanTF{#2}			
		{
			\Vertb*{#3}_{1, \varepsilon}
		}
		{
			\Vertb[#1]{#3}_{1, \varepsilon}
		}
	}
}

\NewDocumentCommand{\normi}{ O{} s G{\cdot} } {
	\ensuremath{
		\IfBooleanTF{#2}			
		{
			\Vertb*{#3}_{\infty}
		}
		{
			\Vertb[#1]{#3}_{\infty}
		}
	}
}

\NewDocumentCommand{\normt}{ O{} s G{\cdot} } {
	\ensuremath{
		\IfBooleanTF{#2}			
		{
			\Vertb*{#3}_{2}
		}
		{
			\Vertb[#1]{#3}_{2}
		}
	}
}

\NewDocumentCommand{\dist}{ O{} s g g g } {
	\ensuremath{
		\IfBooleanTF{#2}			
		{
			\IfNoValueTF{#3}{d}{\IfNoValueTF{#4}{d_{#3}}{\IfNoValueTF{#5}{d \br*{#3 , #4} }{d_{#3} \br*{#4 , #5} }}}
		}
		{
			\IfNoValueTF{#3}{d}{\IfNoValueTF{#4}{d_{#3}}{\IfNoValueTF{#5}{d \br[#1]{#3 , #4} }{d_{#3} \br[#1]{#4 , #5} }}}
		}
	}
}

\NewDocumentCommand{\D}{ O{} s g g g } {
	\ensuremath{
		\IfBooleanTF{#2}			
		{
			\IfNoValueTF{#3}{D}{\IfNoValueTF{#4}{D_{#3}}{\IfNoValueTF{#5}{D \br*{#3 \,\Vert\, #4} }{D_{#3} \br*{#4 \,\Vert\, #5} }}}
		}
		{
			\IfNoValueTF{#3}{D}{\IfNoValueTF{#4}{D_{#3}}{\IfNoValueTF{#5}{D \br[#1]{#3 \,\Vert\, #4} }{D_{#3} \br[#1]{#4 \,\Vert\, #5} }}}
		}
	}
}

\NewDocumentCommand{\Dkl}{ O{} s g g } {
	\ensuremath{
		\IfBooleanTF{#2}			
		{
			\IfNoValueTF{#3}{D_{\text{KL}}}{\IfNoValueTF{#4}{D_{\text{KL}} \br*{#3 \,\Vert\, \cdot} }{D_{\text{KL}} \br*{#3 \,\Vert\, #4} }}
		}
		{
			\IfNoValueTF{#3}{D_{\text{KL}}}{\IfNoValueTF{#4}{D_{\text{KL}} \br[#1]{#3 \,\Vert\, \cdot} }{D_{\text{KL}} \br[#1]{#3 \,\Vert\, #4} }}
		}
	}
}

\NewDocumentCommand{\Df}{ O{} s g g } {
	\ensuremath{
		\IfBooleanTF{#2}			
		{
			\IfNoValueTF{#3}{D_f}{\IfNoValueTF{#4}{D_f \br*{#3 \,\Vert\, \cdot} }{D_f \br*{#3 \,\Vert\, #4} }}
		}
		{
			\IfNoValueTF{#3}{D_f}{\IfNoValueTF{#4}{D_f \br[#1]{#3 \,\Vert\, \cdot} }{D_f \br[#1]{#3 \,\Vert\, #4} }}
		}
	}
}

\NewDocumentCommand{\Da}{ O{} s g g } {
	\ensuremath{
		\IfBooleanTF{#2}			
		{
			\IfNoValueTF{#3}{D_{\alpha}}{\IfNoValueTF{#4}{D_{\alpha} \br*{#3 \,\Vert\, \cdot} }{D_{\alpha} \br*{#3 \,\Vert\, #4} }}
		}
		{
			\IfNoValueTF{#3}{D_{\alpha}}{\IfNoValueTF{#4}{D_{\alpha} \br[#1]{#3 \,\Vert\, \cdot} }{D_{\alpha} \br[#1]{#3 \,\Vert\, #4} }}
		}
	}
}

\NewDocumentCommand{\loh}{ O{} s g } {
	\ensuremath{
		\IfBooleanTF{#2}			
		{
			\IfNoValueTF{#3}{o}{o\br*{#3}}
		}
		{
			\IfNoValueTF{#3}{o}{o\br[#1]{#3}}
		}
	}
}

\NewDocumentCommand{\lOh}{ O{} s g } {
	\ensuremath{
		\IfBooleanTF{#2}			
		{
			\IfNoValueTF{#3}{\mathcal{O}}{\mathcal{O}\br*{#3}}
		}
		{
			\IfNoValueTF{#3}{\mathcal{O}}{\mathcal{O}\br[#1]{#3}}
		}
	}
}

\NewDocumentCommand{\lomega}{ O{} s g } {
	\ensuremath{
		\IfBooleanTF{#2}			
		{
			\IfNoValueTF{#3}{\omega}{\omega\br*{#3}}
		}
		{
			\IfNoValueTF{#3}{\omega}{\omega\br[#1]{#3}}
		}
	}
}

\NewDocumentCommand{\lOmega}{ O{} s g } {
	\ensuremath{
		\IfBooleanTF{#2}			
		{
			\IfNoValueTF{#3}{\Omega}{\Omega\br*{#3}}
		}
		{
			\IfNoValueTF{#3}{\Omega}{\Omega\br[#1]{#3}}
		}
	}
}

\NewDocumentCommand{\lTheta}{ O{} s g } {
	\ensuremath{
		\IfBooleanTF{#2}			
		{
			\IfNoValueTF{#3}{\Theta}{\Theta\br*{#3}}
		}
		{
			\IfNoValueTF{#3}{\Theta}{\Theta\br[#1]{#3}}
		}
	}
}

\NewDocumentCommand{\ltilde}{ g } {
	\ensuremath{ 
		\IfNoValueTF{#1}{\sim}{\sim #1}
	}
}

\DeclarePairedDelimiter{\seq}{(}{)}

\NewDocumentCommand{\ev}{ O{} s g g } {
	\ensuremath{
		\IfBooleanTF{#2}			
		{
			\IfNoValueTF{#3}{\lambda}{\IfNoValueTF{#4}{\lambda_{#3}}{\lambda_{#3} \br*{#4} }}
		}
		{
			\IfNoValueTF{#3}{\lambda}{\IfNoValueTF{#4}{\lambda_{#3}}{\lambda_{#3} \br[#1]{#4} }}
		}
	}
}

\NewDocumentCommand{\evmin}{ O{} s g } {
	\ensuremath{
		\IfBooleanTF{#2}			
		{
			\IfNoValueTF{#3}{\lambda_{\text{min}}}{\lambda_{\text{min}} \br*{#3} }
		}
		{
			\IfNoValueTF{#3}{\lambda_{\text{min}}}{\lambda_{\text{min}} \br[#1]{#3} }
		}
	}
}

\NewDocumentCommand{\evmax}{ O{} s g } {
	\ensuremath{
		\IfBooleanTF{#2}			
		{
			\IfNoValueTF{#3}{\lambda_{\text{max}}}{\lambda_{\text{max}} \br*{#3} }
		}
		{
			\IfNoValueTF{#3}{\lambda_{\text{max}}}{\lambda_{\text{max}} \br[#1]{#3} }
		}
	}
}

\NewDocumentCommand{\sv}{ O{} s g g } {
	\ensuremath{
		\IfBooleanTF{#2}			
		{
			\IfNoValueTF{#3}{\sigma}{\IfNoValueTF{#4}{\sigma_{#3}}{\sigma_{#3} \br*{#4}}}
		}
		{
			\IfNoValueTF{#3}{\sigma}{\IfNoValueTF{#4}{\sigma_{#3}}{\sigma_{#3} \br[#1]{#4}}}
		}
	}
}

\NewDocumentCommand{\svmin}{ O{} s g } {
	\ensuremath{
		\IfBooleanTF{#2}			
		{
			\IfNoValueTF{#3}{\sigma_{\text{min}}}{\sigma_{\text{min}} \br*{#3} }
		}
		{
			\IfNoValueTF{#3}{\sigma_{\text{min}}}{\sigma_{\text{min}} \br[#1]{#3} }
		}
	}
}

\NewDocumentCommand{\svmax}{ O{} s g } {
	\ensuremath{
		\IfBooleanTF{#2}			
		{
			\IfNoValueTF{#3}{\sigma_{\text{max}}}{\sigma_{\text{max}} \br*{#3} }
		}
		{
			\IfNoValueTF{#3}{\sigma_{\text{max}}}{\sigma_{\text{max}} \br[#1]{#3} }
		}
	}
}

\newcommand{\inv}[1]{#1^{-1}} 

\NewDocumentCommand{\pinv}{ g } {
	\ensuremath{ 
		\IfNoValueTF{#1}{\dagger}{#1^{\dagger}}
	}
}

\NewDocumentCommand{\adj}{ g } {
	\ensuremath{ 
		\IfNoValueTF{#1}{\star}{#1^{\star}}
	}
}

\DeclareMathOperator{\diagop}{diag}
\NewDocumentCommand{\diag}{ O{} s g } {
	\ensuremath{
		\IfBooleanTF{#2}			
		{
			\IfNoValueTF{#3}{\diagop}{\diagop \br*{#3} }
		}
		{
			\IfNoValueTF{#3}{\diagop}{\diagop \br[#1]{#3} }
		}
	}
}

\DeclareMathOperator{\rankop}{rank}
\NewDocumentCommand{\rank}{ O{} s g } {
	\ensuremath{
		\IfBooleanTF{#2}			
		{
			\IfNoValueTF{#3}{\rankop}{\rankop \br*{#3} }
		}
		{
			\IfNoValueTF{#3}{\rankop}{\rankop \br[#1]{#3} }
		}
	}
}

\DeclareMathOperator{\rangeop}{range}
\NewDocumentCommand{\range}{ O{} s g } {
	\ensuremath{
		\IfBooleanTF{#2}			
		{
			\IfNoValueTF{#3}{\rangeop}{\rangeop \br*{#3} }
		}
		{
			\IfNoValueTF{#3}{\rangeop}{\rangeop \br[#1]{#3} }
		}
	}
}

\DeclareMathOperator{\spanop}{span}
\NewDocumentCommand{\spanv}{ O{} s g } {
	\ensuremath{
		\IfBooleanTF{#2}			
		{
			\IfNoValueTF{#3}{\spanop}{\spanop \br*{#3} }
		}
		{
			\IfNoValueTF{#3}{\spanop}{\spanop \br[#1]{#3} }
		}
	}
}

\DeclareMathOperator{\nullspop}{null}
\NewDocumentCommand{\nullsp}{ O{} s g } {
	\ensuremath{ 
		\IfNoValueTF{#3}{\nullspop}{\nullspop \br{#3} }
	}
}

\NewDocumentCommand{\Det}{ O{} s g } {
	\ensuremath{
		\IfBooleanTF{#2}			
		{
			\IfNoValueTF{#3}{\det}{\det \br*{#3} }
		}
		{
			\IfNoValueTF{#3}{\det}{\det \br[#1]{#3} }
		}
	}
}

\DeclareMathOperator{\trop}{tr}
\NewDocumentCommand{\tr}{ O{} s g } {
	\ensuremath{
		\IfBooleanTF{#2}			
		{
			\IfNoValueTF{#3}{\trop}{\trop \br*{#3} }
		}
		{
			\IfNoValueTF{#3}{\trop}{\trop \br[#1]{#3} }
		}
	}
}

\NewDocumentCommand{\cond}{ O{} s g } {
	\ensuremath{
		\IfBooleanTF{#2}			
		{
			\IfNoValueTF{#3}{\kappa}{\kappa \br*{#3} }
		}
		{
			\IfNoValueTF{#3}{\kappa}{\kappa \br[#1]{#3} }
		}
	}
}

\NewDocumentCommand{\opt}{ g } {
	\ensuremath{ 
		\IfNoValueTF{#1}{\star}{#1^{\star}}
	}
}

\NewDocumentCommand{\lag}{ O{} s g } {
	\ensuremath{
		\IfBooleanTF{#2}			
		{
			\IfNoValueTF{#3}{\mathcal{L}}{\mathcal{L} \br*{#3} }
		}
		{
			\IfNoValueTF{#3}{\mathcal{L}}{\mathcal{L} \br[#1]{#3} }
		}
	}
}

\DeclareMathOperator{\projop}{proj}
\NewDocumentCommand{\proj}{ O{} s g g } {
	\ensuremath{
		\IfBooleanTF{#2}			
		{
			\IfNoValueTF{#3}{\projop}{\IfNoValueTF{#4}{\projop \br*{#3}}{\projop_{#3} \br*{#4}}}
		}
		{
			\IfNoValueTF{#3}{\projop}{\IfNoValueTF{#4}{\projop \br[#1]{#3}}{\projop_{#3} \br[#1]{#4}}}
		}
	}
}

\DeclareMathOperator{\proxop}{prox}
\NewDocumentCommand{\prox}{ O{} s g g } {
	\ensuremath{
		\IfBooleanTF{#2}			
		{
			\IfNoValueTF{#3}{\proxop}{\IfNoValueTF{#4}{\proxop_{#3}}{\proxop_{#3} \br*{#4}}}
		}
		{
			\IfNoValueTF{#3}{\proxop}{\IfNoValueTF{#4}{\proxop_{#3}}{\proxop_{#3} \br[#1]{#4}}}
		}
	}
}

\NewDocumentCommand{\f}{ O{} s g g g } {
	\ensuremath{
		\IfBooleanTF{#2}			
		{
			\IfNoValueTF{#3}{f}{\IfNoValueTF{#4}{f \br*{#3}}{\IfNoValueTF{#5}{f_{#3} \br*{#4}  }{f^{#4}_{#3} \br*{#5} }}}
		}
		{
			\IfNoValueTF{#3}{f}{\IfNoValueTF{#4}{f \br[#1]{#3}}{\IfNoValueTF{#5}{f_{#3} \br[#1]{#4}  }{f^{#4}_{#3} \br[#1]{#5} }}}
		}
	}
}

\NewDocumentCommand{\g}{ O{} s g g g } {
	\ensuremath{
		\IfBooleanTF{#2}			
		{
			\IfNoValueTF{#3}{g}{\IfNoValueTF{#4}{g \br*{#3}}{\IfNoValueTF{#5}{g_{#3} \br*{#4}  }{g^{#4}_{#3} \br*{#5} }}}
		}
		{
			\IfNoValueTF{#3}{g}{\IfNoValueTF{#4}{g \br[#1]{#3}}{\IfNoValueTF{#5}{g_{#3} \br[#1]{#4}  }{g^{#4}_{#3} \br[#1]{#5} }}}
		}
	}
}

\NewDocumentCommand{\h}{ O{} s g g g } {
	\ensuremath{
		\IfBooleanTF{#2}			
		{
			\IfNoValueTF{#3}{h}{\IfNoValueTF{#4}{h \br*{#3}}{\IfNoValueTF{#5}{h_{#3} \br*{#4}  }{h^{#4}_{#3} \br*{#5} }}}
		}
		{
			\IfNoValueTF{#3}{h}{\IfNoValueTF{#4}{h \br[#1]{#3}}{\IfNoValueTF{#5}{h_{#3} \br[#1]{#4}  }{h^{#4}_{#3} \br[#1]{#5} }}}
		}
	}
}

\NewDocumentCommand{\vphi}{ O{} s g g g } {
	\ensuremath{
		\IfBooleanTF{#2}			
		{
			\IfNoValueTF{#3}{\varphi}{\IfNoValueTF{#4}{\varphi \br*{#3}}{\IfNoValueTF{#5}{\varphi_{#3} \br*{#4}  }{h^{#4}_{#3} \br*{#5} }}}
		}
		{
			\IfNoValueTF{#3}{\varphi}{\IfNoValueTF{#4}{\varphi \br[#1]{#3}}{\IfNoValueTF{#5}{\varphi_{#3} \br[#1]{#4}  }{\varphi^{#4}_{#3} \br[#1]{#5} }}}
		}
	}
}

\newcommand{\da}{\sim}					% distributed as
		
\NewDocumentCommand{\nd}{ O{} s g } {
	\ensuremath{
		\IfBooleanTF{#2}			
		{
			\IfNoValueTF{#3}{\mathcal{N}}{\mathcal{N} \br*{#3} }
		}
		{
			\IfNoValueTF{#3}{\mathcal{N}}{\mathcal{N} \br[#1]{#3} }
		}
	}
}

\DeclareMathOperator{\Eop}{\mathbb{E}}
\NewDocumentCommand{\E}{ O{} s g g } {
	\ensuremath{
		\IfBooleanTF{#2}			
		{
			\IfNoValueTF{#3}{\Eop}{\IfNoValueTF{#4}{\Eop \sbr*{#3}}{\Eop_{#3} \sbr*{#4}}}
		}
		{
			\IfNoValueTF{#3}{\Eop}{\IfNoValueTF{#4}{\Eop \sbr[#1]{#3}}{\Eop_{#3} \sbr[#1]{#4}}}
		}
	}
}

\DeclareMathOperator{\varop}{var}
\NewDocumentCommand{\var}{ O{} s g } {
	\ensuremath{
		\IfBooleanTF{#2}			
		{
			\IfNoValueTF{#3}{\varop}{\varop \br*{#3} }
		}
		{
			\IfNoValueTF{#3}{\varop}{\varop \br[#1]{#3} }
		}
	}
}

\DeclareMathOperator{\covop}{cov}
\NewDocumentCommand{\cov}{ O{} s g } {
	\ensuremath{
		\IfBooleanTF{#2}			
		{
			\IfNoValueTF{#3}{\covop}{\covop \br*{#3} }
		}
		{
			\IfNoValueTF{#3}{\covop}{\covop \br[#1]{#3} }
		}
	}
}

\DeclareMathOperator{\entop}{H}
\NewDocumentCommand{\ent}{ O{} s g } {
	\ensuremath{
		\IfBooleanTF{#2}			
		{
			\IfNoValueTF{#3}{\entop}{\entop \sbr*{#3} }
		}
		{
			\IfNoValueTF{#3}{\entop}{\entop \sbr[#1]{#3} }
		}
	}
}

\newcommand{\XGZ}{X \mid Z}	

\newcommand{\xgy}{x \mid y}	
\newcommand{\ygx}{y \mid x}

\newcommand{\xgz}{x \mid z}	
\newcommand{\zgx}{z \mid x}

\NewDocumentCommand{\p}{ O{} s g g} {
	\ensuremath{
		\IfBooleanTF{#2}			
			{
				\IfNoValueTF{#3}{p}{\IfNoValueTF{#4}{p \br*{#3}}{p_{#3} \br*{#4}}}		
			}
			{
				\IfNoValueTF{#3}{p}{\IfNoValueTF{#4}{p \br[#1]{#3}}{p_{#3} \br[#1]{#4}}}
			}
	}
}

\NewDocumentCommand{\q}{ O{} s g g} {
	\ensuremath{
		\IfBooleanTF{#2}			
			{
				\IfNoValueTF{#3}{q}{\IfNoValueTF{#4}{q \br*{#3}}{q_{#3} \br*{#4}}}		
			}
			{
				\IfNoValueTF{#3}{q}{\IfNoValueTF{#4}{q \br[#1]{#3}}{q_{#3} \br[#1]{#4}}}
			}
	}
}

\NewDocumentCommand{\px}{ O{} s g } {
	\ensuremath{
		\IfBooleanTF{#2}			
		{
			\IfNoValueTF{#3}{p_X}{ p_X \br*{#3}}
		}
		{
			\IfNoValueTF{#3}{p_X}{p_X \br[#1]{#3}}
		}
	}
}

\NewDocumentCommand{\py}{ O{} s g } {
	\ensuremath{
		\IfBooleanTF{#2}			
		{
			\IfNoValueTF{#3}{p_Y}{ p_Y \br*{#3}}
		}
		{
			\IfNoValueTF{#3}{p_Y}{p_Y \br[#1]{#3}}
		}
	}
}

\NewDocumentCommand{\pz}{ O{} s g } {
	\ensuremath{
		\IfBooleanTF{#2}			
		{
			\IfNoValueTF{#3}{p_Z}{ p_Z \br*{#3}}
		}
		{
			\IfNoValueTF{#3}{p_Z}{p_Z \br[#1]{#3}}
		}
	}
}

\NewDocumentCommand{\pxgy}{ O{} s g } {
	\ensuremath{
		\IfBooleanTF{#2}			
		{
			\IfNoValueTF{#3}{p_{X \mid Y}}{ p_{X \mid Y} \br*{#3}}
		}
		{
			\IfNoValueTF{#3}{p_{X \mid Y}}{p_{X \mid Y} \br[#1]{#3}}
		}
	}
}

\NewDocumentCommand{\pygx}{ O{} s g } {
	\ensuremath{
		\IfBooleanTF{#2}			
		{
			\IfNoValueTF{#3}{p_{Y \mid X}}{ p_{Y \mid X} \br*{#3}}
		}
		{
			\IfNoValueTF{#3}{p_{Y \mid X}}{p_{Y \mid X} \br[#1]{#3}}
		}
	}
}

\NewDocumentCommand{\pzgy}{ O{} s g } {
	\ensuremath{
		\IfBooleanTF{#2}			
		{
			\IfNoValueTF{#3}{p_{Z \mid Y}}{ p_{Z \mid Y} \br*{#3}}
		}
		{
			\IfNoValueTF{#3}{p_{Z \mid Y}}{p_{Z \mid Y} \br[#1]{#3}}
		}
	}
}

\NewDocumentCommand{\pygz}{ O{} s g } {
	\ensuremath{
		\IfBooleanTF{#2}			
		{
			\IfNoValueTF{#3}{p_{Y \mid Z}}{ p_{Y \mid Z} \br*{#3}}
		}
		{
			\IfNoValueTF{#3}{p_{Y \mid Z}}{p_{Y \mid Z} \br[#1]{#3}}
		}
	}
}

\NewDocumentCommand{\fx}{ O{} s g } {
	\ensuremath{
		\IfBooleanTF{#2}			
		{
			\IfNoValueTF{#3}{f_X}{f_X \br*{#3}}
		}
		{
			\IfNoValueTF{#3}{f_X}{f_X \br[#1]{#3}}
		}
	}
}

\NewDocumentCommand{\Fx}{ O{} s g } {
	\ensuremath{
		\IfBooleanTF{#2}			
		{
			\IfNoValueTF{#3}{F_X}{F_X \br*{#3}}
		}
		{
			\IfNoValueTF{#3}{F_X}{F_X \br[#1]{#3}}
		}
	}
}

\NewDocumentCommand{\fy}{ O{} s g } {
	\ensuremath{
		\IfBooleanTF{#2}			
		{
			\IfNoValueTF{#3}{f_Y}{f_Y \br*{#3}}
		}
		{
			\IfNoValueTF{#3}{f_Y}{f_Y \br[#1]{#3}}
		}
	}
}

\NewDocumentCommand{\fz}{ O{} s g } {
	\ensuremath{
		\IfBooleanTF{#2}			
		{
			\IfNoValueTF{#3}{f_Z}{f_Z \br*{#3}}
		}
		{
			\IfNoValueTF{#3}{f_Z}{f_Z \br[#1]{#3}}
		}
	}
}

\NewDocumentCommand{\fzi}{ O{} s g } {
	\ensuremath{
		\IfBooleanTF{#2}			
		{
			\IfNoValueTF{#3}{f_{Z_i}}{f_{Z_i} \br*{#3}}
		}
		{
			\IfNoValueTF{#3}{f_{Z_i}}{f_{Z_i} \br[#1]{#3}}
		}
	}
}

\NewDocumentCommand{\fxz}{ O{} s g } {
	\ensuremath{
		\IfBooleanTF{#2}			
		{
			\IfNoValueTF{#3}{f_{X, Z}}{f_{X, Z} \br*{#3}}
		}
		{
			\IfNoValueTF{#3}{f_{X, Z}}{f_{X, Z} \br[#1]{#3}}
		}
	}
}

\NewDocumentCommand{\fxgy}{ O{} s g } {
	\ensuremath{
		\IfBooleanTF{#2}			
		{
			\IfNoValueTF{#3}{f_{X \mid Y}}{ f_{X \mid Y} \br*{#3}}
		}
		{
			\IfNoValueTF{#3}{f_{X \mid Y}}{f_{X \mid Y} \br[#1]{#3}}
		}
	}
}

\NewDocumentCommand{\fygx}{ O{} s g } {
	\ensuremath{
		\IfBooleanTF{#2}			
		{
			\IfNoValueTF{#3}{f_{Y \mid X}}{ f_{Y \mid X} \br*{#3}}
		}
		{
			\IfNoValueTF{#3}{f_{Y \mid X}}{f_{Y \mid X} \br[#1]{#3}}
		}
	}
}

\NewDocumentCommand{\fzgy}{ O{} s g } {
	\ensuremath{
		\IfBooleanTF{#2}			
		{
			\IfNoValueTF{#3}{f_{Z \mid Y}}{ f_{Z \mid Y} \br*{#3}}
		}
		{
			\IfNoValueTF{#3}{f_{Z \mid Y}}{f_{Z \mid Y} \br[#1]{#3}}
		}
	}
}

\NewDocumentCommand{\fygz}{ O{} s g } {
	\ensuremath{
		\IfBooleanTF{#2}			
		{
			\IfNoValueTF{#3}{f_{Y \mid Z}}{ f_{Y \mid Z} \br*{#3}}
		}
		{
			\IfNoValueTF{#3}{f_{Y \mid Z}}{f_{Y \mid Z} \br[#1]{#3}}
		}
	}
}

\NewDocumentCommand{\fzgx}{ O{} s g } {
	\ensuremath{
		\IfBooleanTF{#2}			
		{
			\IfNoValueTF{#3}{f_{Z \mid X}}{ f_{Z \mid X} \br*{#3}}
		}
		{
			\IfNoValueTF{#3}{f_{Z \mid X}}{f_{Z \mid X} \br[#1]{#3}}
		}
	}
}

\NewDocumentCommand{\fxgz}{ O{} s g } {
	\ensuremath{
		\IfBooleanTF{#2}			
		{
			\IfNoValueTF{#3}{f_{X \mid Z}}{ f_{X \mid Z} \br*{#3}}
		}
		{
			\IfNoValueTF{#3}{f_{X \mid Z}}{f_{X \mid Z} \br[#1]{#3}}
		}
	}
}

\newcommand{\node}{\mathcal{V}}

\newcommand{\edge}{\mathcal{E}}			% E is used already for expectation

\NewDocumentCommand{\graph}{ O{} s g g } {
	\ensuremath{
		\IfBooleanTF{#2}			
		{
			\IfNoValueTF{#3}{\br*{\node , \edge}}{\IfNoValueTF{#4}{\br*{#3 , \edge}}{\br*{#3 , #4}}}
		}
		{
			\IfNoValueTF{#3}{\br[#1]{\node , \edge}}{\IfNoValueTF{#4}{\br[#1]{#3 , \edge}}{\br[#1]{#3 , #4}}}
		}
	}
}

\usepackage{xspace}
\makeatletter
\DeclareRobustCommand\onedot{\futurelet\@let@token\@onedot}
\def\@onedot{\ifx\@let@token.\else.\null\fi\xspace}

\def\eg{\emph{e.g}\onedot}
\def\ie{\emph{i.e}\onedot}

%% file: sections/introduction.tex
\section{Introduction}%
\label{sec:intro}
We address the problem of sampling from probability distributions that admit a density $\fx$ of the form
\begin{align}
	\fx{x} \propto \prod_{i = 1}^{m} \phi_i\bigl(\br*{K x}_i\bigr),
	\label{def_poe} \tag{PoE}
\end{align}
where $\phi_1, \ldots, \phi_m: \R \to \Rpp$ are integrable univariate functions and $K\in \R^{m\times n}$ is a linear operator. 
Depending on the kernel of $K$, the integral $\int_{\R^n} \fx{x}\;\mathrm{d}x$ might be finite or infinite. In the former case, we call $\fx$ a \emph{proper} density on $\R^n$. In the latter case, $\fx$ is only proper on a linear subspace of $\R^n$, and we call it \emph{improper}. 
We consider both cases in this work and, to simplify our presentation, we assume without loss of generality that $\phi_1, \ldots, \phi_m$ are univariate densities. 

The density defined by~\eqref{def_poe} is a variant of the well-known products-of-experts model~\cite{Hinton:1999, Hinton:2002} introduced by Hinton and corresponds to a two-layer neural network with activation functions $\log \phi_i$ that computes $\log \fx{x} + \text{const}$. 
Interestingly, many distributions that emerge from the Bayesian treatment of inverse problems with standard imaging models are of this form.
To see this connection, let $\mathcal{F}:\R^n\rightarrow \R^d$ denote a forward operator and consider the problem of recovering an unknown image $X\in \R^n$ from its noisy observation
\begin{align}
	Y = \mathcal{F}(X) + \eta,
	\label{eq_mm}
\end{align}
where $\eta\in\R^d$ is measurement noise.
Although the realization of the measurement noise $\eta$ is usually unknown, it is typically assumed that its distribution is known.
In the Bayesian framework, it is assumed that the unknown image is a realization of a random variable with distribution \( f_X \) and, consequently, inverse problems are addressed by analyzing the posterior distribution $\fxgy$, which via Bayes' theorem can be expressed as
\[
	\fxgy{\xgy} \propto \fygx{\ygx} \cdot \fx{x}.
\]
This shifts the focus from modeling the posterior distribution $\fxgy$ to modeling the prior distribution $\fx$, since the likelihood $\fygx$ can typically be derived from the measurement model~\eqref{eq_mm}.
The prior captures information about the unknown image $X$ that we want to recover, and many standard image priors (\eg, anisotropic total variation~\cite{Rudin:1992}, fields-of-experts~\cite{Roth:2005, Roth:2009}, models based on sparse representations~\cite{Aharon:2006}, anisotropic total generalized variation~\cite{Bredies:2010}, cosparse analysis models~\cite{Rubinstein:2013}, and some more recent neural network priors like the one in~\cite{Goujon:2023}) are Gibbs distributions of the form
\begin{align*}
	\fx{x} \propto \exp\bigl(-\mathcal{R}\br{x}\bigr),
\end{align*}
where the so-called \emph{regularizer} $\mathcal{R} : \R^n \to \R$ can be expressed as
\begin{align*}
	\mathcal{R}\br{x} \coloneqq \sum_{i = 1}^r \rho^{\mathrm{reg}}_i\bigl(\br{K_{\mathrm{reg}} x}_i\bigr),
\end{align*}
where $K_{\mathrm{reg}} \in \R^{r \times n}$ is a linear operator and $\rho^{\mathrm{reg}}_1, \ldots, \rho^{\mathrm{reg}}_r: \R \to \R$ are univariate functions.
Such priors can be directly mapped to~\eqref{def_poe} by choosing $K \coloneqq K_{\mathrm{reg}}$ and $\phi_i\br{t} \propto \exp\bigl( -\rho^{\mathrm{reg}}_i\br{t} \bigr)$ for $i = 1, \ldots, r$.
Similarly, many standard likelihood terms $\fygx$ in imaging are Gibbs distributions of the form
\begin{align*}
	\fygx{\ygx} \propto \exp\br{-\mathcal{D}\br{\mathcal{F}(x),  y}}
\end{align*}
where $\mathcal{D} : \R^d \times \R^d \to \R$ is a \enquote{distance} (\eg, the mean squared error) that can be expressed as
\begin{align*}
	\mathcal{D}\br{z,  y} \coloneqq \sum_{i = 1}^d \rho^{\mathrm{data}}_i\br{z_i - y_i},
\end{align*}
where $\rho^{\mathrm{data}}_1, \ldots, \rho^{\mathrm{data}}_d: \R \to \R$ are univariate functions.
If the forward operator $\mathcal{F} $ is linear with matrix representation \( K_{\mathrm{data}} \in \R^{d \times n} \)---which is the case in denoising, deblurring, computed tomography, and many other relevant applications---then the likelihood is again of the form~\eqref{def_poe} and, consequently, the entire posterior distribution
\begin{align*}
	\fxgy{\xgy} \propto \exp\br{-\mathcal{D}\br{K_{\mathrm{data}}x,  y} - \mathcal{R}\br{x}}
\end{align*}
can be expressed in the form~\eqref{def_poe} by choosing 
\begin{align*}
	K \coloneqq \begin{bmatrix}
		K_{\mathrm{data}} \\
		K_{\mathrm{reg}}
	\end{bmatrix}
\end{align*}
and defining $\phi_i\br{t} \propto \exp\bigl(-\rho^{\mathrm{data}}_i\br{t - y_i}\bigr)$ for $i = 1, \ldots, d$ and $\phi_{d + i}\br{t} \propto \exp\bigl(-\rho^{\mathrm{reg}}_i\br{t}\bigr)$ for $i = 1, \ldots, r$.

A major benefit of the Bayesian framework is that it naturally provides uncertainty quantification. 
This is critical in high-stakes applications such as medical imaging, where confidence levels in predictions directly influence decision making. 
Beyond solving inverse problems, efficient sampling methods can also be used to explore the statistical properties of priors~\cite{Schmidt:2010}, to learn priors directly from image data~\cite{Zhu:1997b, Roth:2009, Schmidt:2010, Zach:2023}, and to evaluate the ability of a prior to generate photo-realistic images~\cite{Du:2019, Song:2019}. 
This makes prior sampling particularly valuable for tasks like the validation of generative models, the assessment of the expressiveness of different image priors, and the development of principled methods for learning priors from data. 
This broad range of applications makes the development of efficient and general sampling approaches for models of the form~\eqref{def_poe} a highly appealing research direction.

\subsection{Proposed Sampling Approach}%
\label{intro:summary}
The fundamental idea of our approach is to sensibly lift the density $\fx$ defined by~\eqref{def_poe} to a latent variable model $\fxz$ that has favourable structure for sampling.
More precisely, we construct a joint distribution $\fxz$ that admits $\fx$ as its marginal distribution and, consequently, reduces the problem of sampling from $\fx$ to the problem of sampling from the corresponding latent variable model $\fxz$. We will show in \Cref{sec:approach} that a latent variable model $\fxz$ with particularly favourable structure for sampling can be obtained under the following simple assumption on the factors\footnote{%
	We often refer to the univariate densities $\phi_1, \ldots, \phi_m$ as factors, since they correspond to factors in a factor graph representation~\cite{Kschischang:2001} of the~\eqref{def_poe} model.%
} in the~\eqref{def_poe} model, and we will refer to such latent variable models as \emph{\glspl{glm}}.

\begin{assumption}%
\label{assumption_fmp}
	All factors $\phi_1, \ldots, \phi_m$ in the~\eqref{def_poe} model satisfy the following factor marginalization property.
\end{assumption}

\begin{definition}[Factor marginalization property]
	A univariate density $\phi : \R \to \Rpp$ satisfies the \emph{factor marginalization property} if it admits a representation of the form
	\begin{equation}
		\phi\br*{t} = \int_{\mathcal{Z}}  g\br*{t, z} \f{}{z}\,\mathrm{d} z \label{def_fmp} \tag{FMP},
	\end{equation}
	where $f: \mathcal{Z} \to \R$ ($\mathcal{Z} \subseteq \R$) is a univariate density and $g: \R \times \R \to \R$ is a univariate Gaussian factor of the form
	\begin{equation}
		g\br*{t, \zeta} \coloneqq \nd*{t; \mu\br*{\zeta}, \sigma^2\br*{\zeta}},
	\end{equation}
	where $\mu: \mathcal{Z} \to \R$ and $\sigma^2: \mathcal{Z} \to \Rpp$ are univariate functions.
\end{definition}

\remark For notational convenience, we use the Lebesgue measure in \eqref{def_fmp} and similar expressions throughout the remainder of this work. However, the underlying measure of $f$ may also be an arbitrary context-dependent measure. An example where the underlying measure is a counting measure can be found in Section SM1 of the supplementary materials, and in that case $f$ corresponds to the probability mass function of a disrete random variable. 

\Cref{assumption_fmp} is fulfilled if all factors $\phi_1, \ldots, \phi_m$ in the~\eqref{def_poe} model can be exactly expressed as univariate \glspl{gmm}, composed of finite, countably infinite, or uncountably infinite number of components.
Such an exact representation is possible in many cases of practical interest (see \Cref{sec:glm_sampling} and \cref{tab:latents} for more information).
Alternatively, it is relatively simple to approximate a factor with a \gls{gmm} to arbitrary accuracy~\cite{Plataniotis:2001} in cases when an exact representation is not possible or is too difficult to obtain.
This makes it easy to substitute any model of the form~\eqref{def_poe} with one that satisfies~\cref{assumption_fmp} and, consequently, makes the proposed sampling approach appealing for all such models.
Such approximations are particularly interesting in the context of diffusion models where \glspl{gmm} have recently been shown to have favourable properties~\cite{Zach:2024}.

The structure of the latent variable model $\fxz$ that emerges under~\Cref{assumption_fmp} allows for highly efficient two-block Gibbs sampling that uses $X$ and $Z$ as blocks.
The corresponding Gibbs subproblems reduce to drawing samples from the conditional distributions $\fxgz$ and $\fzgx$.
Specifically, sampling from $\fxgz$ reduces to sampling from a multivariate Gaussian distribution on $\R^n$, and sampling from $\fzgx$ reduces to independently sampling from $m$ univariate distributions.
Both of these can be performed efficiently even in the high-dimensional setting typically found in imaging.
Furthermore, we show that a minor modification of the approach can be used to handle some specific nonlinear extensions of the \eqref{def_poe} model.
This renders the proposed approach highly practical.
Moreover, our proposed sampling approach does not require knowledge of the normalization constant $\int_{\R^n} \prod_{i = 1}^{m} \phi_i\br*{\br*{K x}_i}\,\mathrm{d}x$ in~\eqref{def_poe}, which is hard to compute in the general case.

Finally, for some special cases of interest, it is possible to construct direct sampling approaches for models of the form~\eqref{def_poe}.
In particular, efficient prior sampling algorithms with linear time and memory complexities for pairwise priors defined over chain and tree topologies are discussed in~\Cref{sec:direct_sampling}.

\subsection{Related Work}
Sampling is one of the most popular approaches for inference in complex probabilistic models, and is therefore a highly active and well-developed area of research. We will briefly review the fundamentals and highlight the most popular approaches in the field of Bayesian imaging.  We refer readers to \cite{Gelman:2013,Green:2015,Pereyra:2016,Robert:2004} for more detailed introductions.

The most general sampling approaches belong to the class of \gls{mcmc} methods, where a Markov chain $\seq{X_k}$ is constructed such that its stationary distribution is a desired target distribution $\fx$.  A general framework to construct such Markov chains is the well-known Metropolis-Hastings algorithm.  It combines a proposal distribution with an accept/reject step that ensures that the resulting Markov chain has $\fx$ as its stationary distribution.
The proposal distribution on its own already defines a Markov chain, and therefore Metropolis-Hastings can be thought of as a mechanism to transform a Markov chain into another one with desired stationary distribution. Consequently, the efficiency of these approaches is usually dictated by the quality of the proposal distribution,  and most research efforts have focused on designing proposal distributions that induce stationary distributions close to $\fx$,  often by incorporating local information such as gradients and curvature. 

In the high-dimensional problems typically encountered in imaging, the most popular approaches are based on the overdamped Langevin diffusion \gls{sde}, which is defined for \( t \geq 0 \) as
\begin{align*}
	\mathrm{d} X_t = \nabla\log \fx{X_t} \mathrm{d} t + \sqrt{2} \mathrm{d} W_t,
\end{align*}
where $\seq{W_t}$ denotes Brownian motion.
Under relatively mild conditions (\textit{e.g.}, that $-\log\fx$ is differentiable with Lipschitz continuous gradient that satisfies certain growth conditions) the resulting Markov chain is ergodic and admits $\fx$ as its unique stationary distribution~\cite{Roberts:1996, Fruehwirth:2024}.
An Euler-Maruyama discretization of the overdamped Langevin \gls{sde} leads to the well-known \gls{ula}, which is defined for \( k = 0, 1, \ldots \) as
\begin{align*}
	X_{k+1} = X_k + \tau \nabla\log\fx{X_k} + \sqrt{2\tau} Z_k
\end{align*}
where $\tau>0$ is a step-size parameter and $\seq{Z_k}$ is an i.i.d.~sequence of standard normal distributed random vectors.
Though the discretization biases the stationary distribution, under some conditions the corresponding Markov chain is ergodic with stationary distribution $\fx^\tau$ that depends on the step-size $\tau$ and $\fx^\tau$ approaches $\fx$ as $\tau$ approaches $\num{0}$ in various \enquote{distances}, such as the Wasserstein-2 distance, the total variation distance, or the Kullback-Leibler divergence~\cite{Fruehwirth:2024, Durmus:2017, Dalalyan:2016}.
Consequently, it is common practice to use the \gls{ula} as proposal mechanism within Metropolis-Hastings.
This leads to the popular \gls{mala}.

Significant research efforts have been made to extend \gls{ula} to broader classes of target densities and to improve its efficiency.
For instance, in~\cite{Durmus:2022, Durmus:2018} the authors replace the nondifferentiable part of a convex potential $-\log\fx$ with its Moreau-Yosida envelope.
This enables the application of \gls{ula}, although the target distribution is altered due to the smoothing.
In~\cite{Durmus:2019, Habring:2024}, subgradient and proximal-gradient versions of \gls{ula} are directly analyzed without resorting to approximations via Moreau-Yosida envelopes.
The work~\cite{Cheng:2018} shows that Langevin sampling can be accelerated by discretizing the underdamped Langevin \gls{sde} when $-\log\fx$ is strongly convex,  twice continuously differentiable, and admits a Lipschitz continuous gradient.
Lastly, the recent work~\cite{Habring:2025} accelerates Langevin sampling by combining ideas from diffusion models~\cite{Song:2019} with Moreau-Yosida approximations.

Contrary to Langevin based methods, our sampling approach follows an alternative and less-known line of research based on Gibbs sampling in latent variable models. Hence, we will now briefly summarize the historical development of ideas that led to our proposed sampling approach.

A two-block Gibbs sampling approach adapted to image priors of the form~\eqref{def_poe} with Student-t factors was proposed more than 20 years ago in the context of image prior learning~\cite{Welling:2002}. Their learning approach requires prior samples at each iteration, and the authors realized that introducing a latent variable for each factor leads to a lifted model that allows for efficient two-block Gibbs sampling to generate those samples. They also briefly noted the similarities and differences between their lifted model and \gls{gsm} representations.

Similarly, a two-block Gibbs sampling approach adapted to fields-of-experts priors with factors that are represented as finite-component \glspl{gsm} was introduced in~\cite{Schmidt:2010}. The authors demonstrated the versatility of their approach through a variety of applications. For instance, they noted that the approach can be used for approximate sampling by approximating factors with finite-dimensional \glspl{gsm}. This allowed them to approximately investigate the statistical properties of existing image priors and to learn image priors that more accurately reflect the statistics of natural images. They also showed that their approach can be easily adapted to posterior sampling in image denoising problems.

Inspired by~\cite{Schmidt:2010} and the earlier approaches~\cite{Rue:2001, Rue:2005} from the statistics literature, the authors of~\cite{Papandreou:2010} investigated an efficient direct sampling approach for a particular form of a Gaussian Markov random field model. Their model is equivalent to the~\eqref{def_poe} model with Gaussian factors, while the sampling problem is equivalent to the first subproblem in our two-stage Gibbs sampling approach. In contrast to the aforementioned earlier works, the authors focused on iterative approaches that scale well to high-dimensional sampling problems, which are also one of the key enablers behind our approach. Following~\cite{Welling:2002} and~\cite{Schmidt:2010}, the authors also realized the utility of their approach in the context of two-block Gibbs sampling. In particular, they noted that the key assumption behind the two-block Gibbs sampling approach from~\cite{Welling:2002} was that the factors can be expressed as \glspl{gsm}, which is a particular instance of~\Cref{assumption_fmp}. Based on that, they derived a two-block Gibbs sampler for posterior sampling of one-dimensional\footnote{One-dimensional in this context refers to signals with a single index dimension like time series data.} denoising problems with total variation regularizers. Furthermore, they pointed out that this approach can be extended to posterior sampling for image denoising problems with isotropic TV regularizers.

In a recent line of research, the authors of~\cite{Bohra:2023} independently developed direct prior sampling algorithms and Gibbs sampling algorithms for posterior sampling problems in one-dimensional signals. They assume priors that are discretized versions of Lévy processes and from this viewpoint obtain direct prior sampling algorithms that amount to randomly sampling an initial condition and performing a random walk under the corresponding innovation distribution of the Lévy process. They then showed that this approach can be mapped to standard one-dimensional counterparts of image regularizers like total variation.  The discretization of their Lévy process however leads to a particular boundary condition that is typically not found in standard image priors. Their Gibbs sampling algorithms for posterior sampling were developed for linear Gaussian likelihood terms and heavily rely on \gls{gsm} representations of the corresponding innovation distributions. As a special case, their Gibbs sampling approach recovers the one-dimensional posterior sampling approach of~\cite{Papandreou:2010} for denoising problems with total variation regularizers, although modified by the aforementioned boundary condition in the prior.

Our work provides a simple unifying viewpoint of the sampling approaches proposed in~\cite{Welling:2002,  Schmidt:2010,  Papandreou:2010,  Bohra:2023} and substantially generalizes them in several ways. For instance,  our approach is agnostic to signal structure and can, therefore, be applied to one-dimensional signals,  images,  higher-dimensional signals like video data, or even more specialized structures like signals defined over tree topologies or more general factor graphs.  It supports a considerably wider range of prior models and likelihood terms beyond Gaussian or linear Gaussian.  Finally, it can be applied to both proper and improper probabilistic models.

Some of the related work involves settings that are not covered by the theory in this work. For instance,~\cite{Bohra:2023} covers Bernoulli-Laplace factors in one-dimensional signals. It is possible to extend our approach to more general discrete-continuous factors that contain Bernoulli-Laplace distributions as a special case. However, we leave such developments for future work since the mathematical framework required to develop such generalizations substantially deviates from the one used in this work.  The approach in~\cite{Bardsley:2014} covers~\eqref{def_poe} models with nonlinear operators $K$ and Gaussian factors. They partly linearize their nonlinear operator at each iteration, thereby obtaining a partly nonlinear Gaussian Markov random field similar to the one in~\cite{Papandreou:2010}, and use this distribution as a proposal for Metropolis-Hastings. Their follow-up work~\cite{Wang:2017} further generalizes this approach to probabilistic models with nonlinear Gaussian likelihood terms and priors with Laplace factors. They show that, under some mild technical conditions, a nonlinear transformation can be used to transform the Laplace factors into Gaussian factors and use the inverse transform to absorb this nonlinearity into the nonlinear operator in their likelihood term.

Finally, there is an interesting connection between our approach and lifting approaches commonly found in variational imaging, such as half-quadratic splitting~\cite{Chambolle:1997}. The basic idea behind such variational lifting approaches is to lift optimization problems of the form 
\begin{align*}
	\min_{x \in \R^n} E\br{x},
\end{align*}
where $E: \R^n \to \R$ is an energy function to optimization problems of the form 
\begin{align*}
		\min_{x \in \R^n,  z \in \mathcal{Z}} \overline{E}\br{x, z},
\end{align*}
where $z \in \mathcal{Z} \subseteq \R^m$ are auxiliary variables often obtained through some form of duality and $\overline{E}: \R^n \times \mathcal{Z} \to \R$ is a lifted energy function with property that
\begin{align*}
	E\br{x} = \min_{z \in \mathcal{Z}} \overline{E}\br{x, z}
\end{align*}
for any $x \in \R^n$.  If we consider the Gibbs distribution counterparts of those energies given by $\fx{x} \propto \exp\br{-E\br{x}}$ and $\f_{X, Z}\br{x, z} \propto \exp\br{-\overline{E}\br{x, z}}$, then this form of lifting is equivalent to lifting a distribution $\fx$ to a joint distribution $\f_{X, Z}$ such that the maximization property
\begin{align*}
	\fx{x} = \max_{z \in \mathcal{Z}} \f_{X, Z}\br{x, z}
\end{align*}
holds for any $x \in \R^n$. By contrast, our approach lifts $\fx$ to a latent variable model $\f_{X, Z}$ such that the marginalization property
\begin{align*}
	\fx{x} = \int_{\mathcal{Z}} \fxz{x, z}\,\mathrm{d}z
\end{align*}
holds for any $x \in \R^n$.  Moreover, since our lifting approach relies on univariate Gaussian representations of the factors,  it is a direct analogue of half-quadratic splitting \cite{Nikolova:2005}.  Our approach and variational lifting approaches like half-quadratic splitting will generally result in different models.  Interestingly, however, our lifting approach and the half-quadratic splitting approach from \cite{Chambolle:1997} lead to the same lifted model in case of isotropic TV regularizers. 

\subsection{Contributions} The main contributions of our work are:
\begin{itemize}
	\item A unifying perspective on many existing sampling algorithms from the literature.
	\item General samplers for the \eqref{def_poe} model and its various specializations that lead to efficient prior and posterior sampling algorithms for many standard imaging models.
	\item A rigorous treatment of improper priors in the context of sampling. 
	\item Publicly available high-quality implementations of our sampling routines that scale well to high-dimensional problems.
	\item Detailed numerical experiments that demonstrate the efficiency and efficacy of our proposed sampling approach in the context of various Bayesian imaging problems.
\end{itemize}

\subsection{Outline} 
The remainder of the paper is organized as follows.
Important preliminaries are summarized in \Cref{sec:prelim}.
Our proposed sampling approach and various specializations that are of interest in the context of imaging are derived in \Cref{sec:approach}.
Experimental results that cover prior and posterior sampling for many standard imaging models, comparisons with existing sampling methods, and a few example applications are shown in \Cref{sec:results}.
Finally, concluding remarks are given in \Cref{sec:conclusions}. 

Our notational conventions are given in \cref{sec:notation}, and detailed proofs for all propositions are given in \cref{sec:proofs}. 
Additional examples, experimental results, and derivations of some illuminating examples are provided in the supplementary materials.

%% file: sections/preliminaries.tex
\section{Preliminaries}%
\label{sec:prelim}
We will briefly review some important preliminaries and introduce some basic terminology before proceeding with the algorithmic parts of the paper.
Specifically, we first show how we can sample from an improper~\eqref{def_poe} model by extending it to a proper ~\eqref{def_poe} model without changing its distribution when constrained to the linear subspace where it was originally defined.
Afterwards, we review latent variable models and two standard techniques for sampling from them:
Ancestral and Gibbs sampling.

\subsection{Handling Improper Densities}
The~\eqref{def_poe} model only defines a proper distribution on $\R^n$ when the linear operator $K$ has a trivial kernel. In contrast to that, the \eqref{def_poe} defines a proper distribution only on a linear subspace of $\R^n$ whenever $K$ has a nontrivial kernel. It is therefore an improper distribution on $\R^n$ in those case. This is formally stated in the following proposition.
\begin{proposition}%
	\label{prop:nullspace_improper_dist}
	Assume that the factors $\phi_i$ in the~\eqref{def_poe} model are bounded and let $N \ceq \ker(K)$. Then the following holds:
	\begin{enumerate}[label=\alph*)]
		\item $\int_{\R^n}\fx{x}\;\mathrm{d}x < \infty$ if and only if $N=\{0\}$.
		\item Regardless of the injectivity of $K$, it always holds that $\int_{N^\perp}\fx{x}\mathrm{d}x < \infty$.
	\end{enumerate}
\end{proposition}
\begin{proof}
	See \cref{proof:nullspace_improper_dist}.
\end{proof}
Improper distributions are important in the context of imaging priors since there the operator \( K \) typically has a non-trivial kernel.
As an example of this, priors are often explicitly constructed such that constant signals are in the kernel of $K$ to enforce invariances with respect to radiometric shifts.

We can complement $\fx$ with an arbitrary distribution on $N \coloneqq \ker(K)$ to avoid the need to directly work with the constraint that defines the linear subspace $N^\perp$ where the~\eqref{def_poe} model is a proper distribution.
This yields a proper distribution on $\R^n$ while leaving the distribution restricted to $N^\perp$ unchanged, as formally stated in the following proposition.
\begin{proposition}%
	\label{prop:tie_breaking_works}
	Consider the~\eqref{def_poe} model.
	Suppose that $K$ has a non-trivial kernel $N \ceq \ker(K)$ and let $P_{N^\perp}$ denote the orthogonal projection onto its complement $N^\perp \ceq \spanv{K^T}$. Define $f_{\bar{X}}$ via
	\[
	f_{\bar{X}} \br{x} \propto \fx{x} \cdot f_0(x),
	\]
	where the \emph{complementary density} $f_0$ integrates to $1$ on $N$ and satisfies that $f_0(u + v) = f_0(u)$ for any $(u,v)\in N \times N^\perp$. Then $f_{\bar{X}}$ is proper and with $\bar{X} \da f_{\bar{X}}$ the distribution of the random variable defined by the projection $X \ceq P_{N^\perp} \bar{X}$ admits the density $\fx$ on $N^\perp$.
\end{proposition}
\begin{proof}
	See \cref{proof:tie_breaking_works}.
\end{proof}
\begin{remark}
	A complementary density $f_0$ always exists. Indeed, we may simply define $f_0(x) \propto \exp(-\norm{P_{N}x}^2)$ with $P_N$ denoting the orthogonal projection onto $N$. Similar to the proof of \cref{prop:nullspace_improper_dist} it follows that $\int_{N} f_0(x)\mathrm{d} x<\infty$ and by the properties of the projection that $f(u+v)=f(u)$ for $(u,v)\in N\times N^\perp$.
\end{remark}
\begin{remark}%
	\label{remark_example_tie_breaking}
	\Cref{prop:tie_breaking_works} relies on the fact that the complementary density $f_0$ is constant on $N^\perp$ as in the converse case the distribution of $P_{N^\perp} \bar{X}$ might not follow the density $\fx$ on $N^\perp$.
	For instance, consider the setting $X\in\R^2$, $m=1$, $K : (x_1, x_2) \mapsto x_1$, $\phi_1(t)=\exp(-t^2)$, and the two complementary densities
	\begin{align*}
		f_0(x) &\propto \exp(-x_2^2), \\
		\bar{f}_{0} (x) &\propto \exp(-x^2_1-x^2_2),
	\end{align*}
	that lead to the complemented densities
	\begin{align*}
		f_{\bar{X}}(x) &\propto \exp(-x^2_1)\exp(-x^2_2), \\
		\bar{f}_{\bar{X}}(x) &\propto \exp(-x^2_1)\exp(-x^2_1-x^2_2).
	\end{align*}
	The complementary density $f_{0}$ satisfies the assumptions in~\Cref{prop:tie_breaking_works}, whereas $\bar{f}_{0}$ does not.
	The corresponding distributions of $X_1 = P_{N^\perp} \bar{X}$ are $f_{X_1}(x_1)\propto\exp(-x^2_1)$ and $\bar{f}_{X_1}(x_1)\propto\exp(-2 x^2_1)$ and, consequently, we conclude that the choice of the complementary distribution as $\bar{f}_{0}$ alters the desired distribution on $N^\perp$.
	Interestingly, the generalization of this counterexample to higher dimensions where the complementary density is defined as a zero-mean normal distribution $f_0(x) = \exp\Bigl(-\frac{\norm{x}^2}{2\varepsilon^2} \Bigr)$ is a popular choice in many works (\eg,~\cite{Weiss:2007, Schmidt:2010}).
	The variance parameter \( \epsilon^2 \) is typically chosen to be very large in order to minimize the influence of the complementary density on the desired density.
\end{remark}

\begin{remark}%
\label{remark_tie_breaking_sufficient}
	The condition on the complementary density in~\Cref{prop:tie_breaking_works} is sufficient but not necessary.
	For instance, for the example in~\Cref{remark_example_tie_breaking}, the definition of the complementary density as $f_0(x) = \exp\bigl(-(x_1 + x_2)^2\bigr)$ violates the condition in \Cref{prop:tie_breaking_works}, but still produces the desired distribution $f_{X_1}(x_1)\propto\exp(-x^2_1)$ on $N^\perp$.
\end{remark}

\Cref{prop:tie_breaking_works} is arguably the natural approach for extending a distribution that is defined on a linear subspace to the entire space.
However, it turns out that the~\eqref{def_poe} model can be extended under an even milder condition.
This somewhat counter-intuitive result is summarized in the following proposition.

\begin{proposition}%
	\label{prop:smart_tie_breaking_works_too}
	Consider the~\eqref{def_poe} model.
	Suppose that $K$ has a non-trivial kernel $N \ceq \ker(K)$ of dimension $r$ and let $P_{N^\perp}$ denote the orthogonal projection onto its complement $N^\perp \ceq \spanv{K^T}$.
	Assume that $\bar{X} \da f_{\bar{X}}$ with
	\[
	f_{\bar{X}} \br{x} \propto \fx{x} \cdot f_0(x),
	\]
	where
	\begin{align*}
		f_0(x) \propto \prod_{i = 1}^r \bar{\phi}_i\br{\br{\bar{K} x}_i},
	\end{align*}
	and $\seq{\bar{\phi_i}}^r_{i = 1}: \R \to \R$ are univariate densities.
	Under any choice of the matrix $\bar{K} \in \R^{r \times n}$ such that the matrix
	\begin{align*}
		\begin{bmatrix}
			\, K \, \\
			\, \bar{K} \,
		\end{bmatrix}
	\end{align*}
	has full rank, the distribution of the random variable defined by the projection $X \ceq P_{N^\perp} \bar{X}$ admits the density $\fx$ on $N^\perp$.
\end{proposition}
\begin{proof}
	See \cref{proof:smart_tie_breaking_works_too}.
\end{proof}
\begin{remark}
	To the best of our knowledge, \Cref{prop:smart_tie_breaking_works_too} is a novel result.
	However, we decided to place it in this section to not distract from the main results in \Cref{sec:approach}.
\end{remark}
\begin{remark}
	\Cref{prop:smart_tie_breaking_works_too} is a generalization of the example in \Cref{remark_tie_breaking_sufficient}.
\end{remark}
\begin{remark}
	\Cref{prop:smart_tie_breaking_works_too} trivially implies that the direct sampling algorithms proposed in~\cite{Bohra:2023} can also be used to sample from improper pairwise chain priors.
	These algorithms were derived by discretizing Lévy processes, which leads to an improper pairwise chain prior augmented by an initial condition.
	The augmentation ensures that the model is proper on the entire space, but since it also satisfies the assumptions from \Cref{prop:smart_tie_breaking_works_too} it follows that resulting model produces correct samples from the improper prior as well.
\end{remark}
In other words, the condition in \Cref{prop:tie_breaking_works} necessitates that the $r$ appended rows form a basis of $\ker(K)$, whereas the condition in \Cref{prop:smart_tie_breaking_works_too} shows that it suffices to pick them such that they are linearly independent from the rows of $K$ and together with the rows of $K$ span the entire space $\R^n$.
Depending on the particular form of an improper~\eqref{def_poe} model, it might be easier to produce one or the other for model extension purposes.

The definition of $\f_0$ like in \Cref{prop:smart_tie_breaking_works_too} is a particularly convenient choice, since the resulting extended model $f_{\bar{X}}$ is again a~\eqref{def_poe} model.
This is formally summarized in the following corollary.
\begin{corollary}
	The extended model $f_{\bar{X}}$ in \Cref{prop:smart_tie_breaking_works_too} is a~\eqref{def_poe} model with $m + r$ factors, whose corresponding $\br{m + r} \times n$ matrix is of full rank.
\end{corollary}

Consequently, throughout the remainder of this work, we can assume that $K$ is an $m \times n$ matrix with full rank and $m \geq n$, which ensures that~\eqref{def_poe} defines a proper distribution on $\R^n$.
Moreover, we will refer to the cases of $m = n$ and $m > n$ as complete and overcomplete\footnote{%
	This terminology is borrowed from~\cite{Lewicki:2000}, which offers additional insights into complete and overcomplete probabilistic models.
	In particular, Figure 2 in~\cite{Lewicki:2000} shows some illuminating geometric examples that contrast the modeling capabilities of complete and overcomplete probabilistic models.%
}, respectively.

\subsection{Latent Variable Models}%
As mentioned in~\Cref{intro:summary}, the proposed sampling approach is based on latent variable models~\cite{Robert:2011}, which are formally defined as follows.
\begin{definition}[Latent variable model]%
	\label{def_lvm}
	A distribution $\fxz$ on $\mathcal{X}\times\mathcal{Z}$ is a latent variable model for the distribution $\fx$ on $\mathcal{X}$, if it satisfies the marginalization property
	\begin{align}
		\fx{x} = \int_{\mathcal{Z}} \fxz{x, z}\,\mathrm{d}z. \label{def_mp} \tag{MP}
	\end{align}
	In other words,  $\fx$ is the marginal distribution of $\fxz$.
	In this case, we refer to $Z \da \fz$ as the latent variable, where $\fz$ is defined through the marginalization $\fz(z) = \int_{\mathcal{X}} \fxz{x, z}\,\mathrm{d}x$.
\end{definition}

A latent variable model $\fxz$ of a distribution $\fx$ opens up an alternative approach to sample from the distribution $\fx$.
Specifically, if $(X,Z)\sim\fxz$, then $X\sim \fx$ by construction.
Consequently, sampling from $\fx$ can be implemented by sampling from $\fxz$ and discarding $Z$.
This often leads to simpler and more efficient sampling algorithms in cases when $\fxz$ has favourable structure that can be exploited for sampling.

\subsection{Sampling from Latent Variable Models}
The two most popular approaches to sample from latent variable models are ancestral and Gibbs sampling. Ancestral sampling~\cite{Barber:2012} relies on the factorization
\begin{align*}
	\fxz{x, z} = \fxgz{\xgz} \cdot \fz{z}
\end{align*}
of the joint distribution $\fxz$ into the product of the prior distribution $\fz$ of the latent variable $Z$ and the conditional distribution $\fxgz$ that describes the dependence of the variable of interest $X$ on the latent variable $Z$.
Sampling $X \da \fx$ via ancestral sampling amounts to sampling the latents $z \da \fz$ and then sampling $X$ from the conditional distribution $\f_{X \mid Z = z}$.
Ancestral sampling is a direct sampling algorithm and, consequently is the preferred approach whenever we can obtain the required factorization and are able to efficiently sample from it.

In contrast, Gibbs sampling is a class of \gls{mcmc} algorithms~\cite{Casella:1992, Robert:2004} where the variables of the target distribution are partitioned into disjoint blocks, and sampling is performed by iteratively drawing from the conditional distributions of each block given the remaining variables.
In the context of latent variable models as in \cref{def_lvm}, the easiest form of Gibbs sampling is the two-block Gibbs sampler that is defined by the iterations
\begin{equation}\label{eq:gibbs_intro}
\begin{cases}
	z^k \da \f_{Z \mid X = x^{k - 1}} \\
	x^k \da \f_{X \mid Z = z^k}
\end{cases}
\end{equation}
for $k = 1, 2, \ldots$ and a given initial condition $x^0$.
It can be shown that~\eqref{eq:gibbs_intro} admits $\fxz$ as a stationary measure~\cite[Theorem 10.6]{Robert:2004}.
Moreover, Harris recurrence and ergodicity of the Markov chain can be obtained under rather mild conditions---such as absolute continuity with respect to the underlying measure---on the transition kernel $K((x,z),(x',z')) = \f_{Z\mid X}(z'|x)\f_{X\mid Z}(x'|z')$ corresponding to~\eqref{eq:gibbs_intro} (see, e.g.~\cite[Theorem 9.6, Lemma 10.9, Theorem 10.10, Corollary 10.12]{Robert:2004}).

%% file: sections/approach.tex
\section{Proposed Approach}%
\label{sec:approach}
We now introduce \glspl{glm} and show that, under \cref{assumption_fmp}, a \gls{glm} is a latent variable model of the~\eqref{def_poe} model.
We then derive the corresponding distributions $\fxgz$, $\fz$, and $\fzgx$ that allow us to adapt the standard sampling approaches from latent variable models to \glspl{glm}, thereby providing a sampler for~\eqref{def_poe}.
In particular, we focus on Gibbs sampling since it leads to tractable subproblems in the general case.
We also show that a minor modification of the Gibbs sampling approach can be used to handle some specific nonlinear extensions of the \eqref{def_poe} model.
Finally, we show that efficient direct sampling algorithms can be obtained for complete~\eqref{def_poe} models and relate them to ancestral sampling in corresponding \glspl{glm}.

\subsection{Gaussian Latent Machines}
\glspl{glm} are a particular type of latent variable model, formally defined as follows.
\begin{definition}[Gaussian latent machine]%
	\label{def:glm}
	\glsreset{glm}
	Let $X \in \R^n$ and $Z \coloneqq \tuple{Z_1, Z_2, \ldots, Z_m} \in \mathcal{Z}_1 \times \mathcal{Z}_2 \times \ldots \times \mathcal{Z}_m \eqc \mathcal{Z}$ be random variables where $\mathcal{Z}_1, \ldots, \mathcal{Z}_m \subseteq \mathbb{R}$.
	A joint distribution $\fxz$ over the random variables $X$ and $Z$ of the form
	\begin{align}
		\fxz{x, z} \propto \prod_{i = 1}^m g_i\br*{\br{Kx}_i, z_i} \cdot \f{i}{z_i} \label{def_glm} \tag{GLM}
	\end{align}
	is called a \emph{\gls{glm}}, where $f_1: \mathcal{Z}_1 \to \R, \ldots, f_m: \mathcal{Z}_m \to \R$ are univariate distributions, $K$ is an $m \times n$ ($m \geq n$) matrix with full rank, and $g_1: \R \times \mathcal{Z}_1 \to \R, \ldots, g_m: \R \times \mathcal{Z}_m \to \R$ are univariate Gaussian factors of the form
	\begin{align*}
		g_i\br*{t, z_i} \coloneqq \nd*{t; \mu_i\br*{z_i}, \sigma^2_i\br*{z_i}},
	\end{align*}
	where $\mu_1: \mathcal{Z}_1 \to \R, \ldots, \mu_m: \mathcal{Z}_m \to \R$ and $\sigma^2_1: \mathcal{Z}_1 \to \Rpp, \ldots, \sigma^2_m: \mathcal{Z}_m \to \Rpp$.
\end{definition}

\begin{remark}
	As before, we refer to the cases of $m = n$ and $m > n$ as complete and overcomplete, respectively.
\end{remark}

\begin{remark}
	The notation $f_1, \ldots, f_m$ for the univariate distributions was chosen deliberately to emphasize that they differ from the marginal distributions $f_{Z_1}, \dotsc, f_{Z_m}$ of the univariate random variables $Z_1, \dotsc, Z_m$ in the~\eqref{def_glm} model in general.
	More precisely, they coincide in the complete case (see \Cref{prop:complete_independent_latents} for more details) but in general differ in the overcomplete case.
	The general form of the latent distribution $\fz$ is derived later in \Cref{prop:latent_distribution} and an example where they differ can be found in Section SM2 of the supplementary materials.
\end{remark}

\begin{remark}
	We refer to the functions $\mu_1, \ldots, \mu_m$ and $\sigma^2_1, \ldots, \sigma^2_m$ as \emph{latent mappings}.
\end{remark}

\Cref{def:glm} implies that \glspl{glm} are products of $m$ univariate \glspl{gmm} in their latent variable representation, where the $i$th Gaussian mixture acts on the scalar variable defined by the inner product of the \( i \)th row of \( K \) with the argument.
The \eqref{def_poe} model admits a \gls{glm} as a latent variable model under~\Cref{assumption_fmp}.
The following proposition states that the local marginalization property~\eqref{def_fmp} from \cref{assumption_fmp} on the univariate factors $\phi_i$ in the~\eqref{def_poe} model is sufficient to imply the global marginalization property~\eqref{def_mp} in~\cref{def_lvm}, thereby allowing the~\eqref{def_poe} model to be represented as a \gls{glm}.
As mentioned earlier, the factor marginalization property~\eqref{def_fmp} means that the univariate factors $\phi_i$ can be exactly represented by univariate Gaussian mixtures.
This is a mild condition since most univariate factors of practical interest for imaging can be either exactly described as Gaussian mixtures or closely approximated by them.
\begin{proposition}%
	\label{prop:marginalization}
	Suppose that \cref{assumption_fmp} holds.
	Then there exists a \gls{glm} $\fxz$ such that
	\begin{align*}
		\int_{\mathcal{Z}} \fxz{x, z}\,\mathrm{d}z \propto \prod_{i = 1}^{m} \phi_i\br*{\br*{K x}_i}. 
	\end{align*}
\end{proposition}
\begin{proof}
	See \cref{proof:marginalization}.
\end{proof}

\begin{remark}
	In principle, non-Gaussian mixtures could be used in the factor marginalization property~\eqref{def_fmp}.
	However, Gaussian mixtures are preferred as they result in significantly simpler algorithms for both sampling and factor representation/approximation.
	We will further elaborate on this point throughout the remainder of this section.
\end{remark}

\subsection{Deriving \texorpdfstring{$\fxgz$}{f\_\{X | Z\}}, \texorpdfstring{$\fz$}{f\_\{Z\}} and \texorpdfstring{$\fzgx$}{f\_\{Z | X\}}}
We now derive the analytical forms of the distributions $\fxgz$, $\fz$, and $\fzgx$ to adapt the standard sampling algorithms of latent variable models---ancestral sampling and Gibbs sampling---to \glspl{glm}.
These distributions will also provide additional insights into the modeling capabilities of \glspl{glm}.

\subsubsection{Deriving \texorpdfstring{$\fxgz$}{f\_\{X | Z\}}}
\begin{figure}[!t]
	\centering
	\begin{subfigure}{0.225\textwidth}
		\centering
		\includegraphics[width=\textwidth]{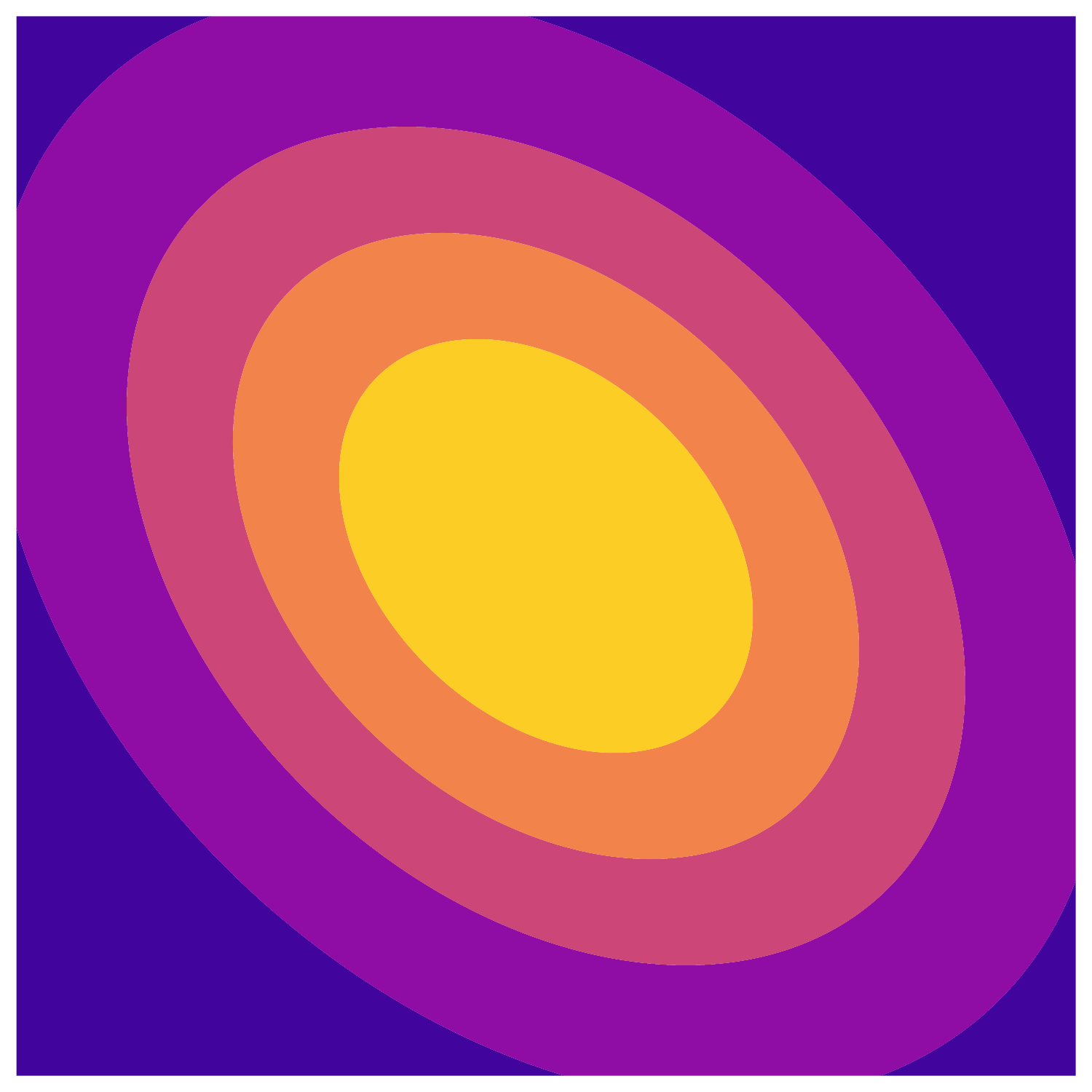}
		\caption{$\rho = \br*{\,\cdot\,}^2$.}
	\end{subfigure}
	\hspace{2cm}
	\begin{subfigure}{0.225\textwidth}
		\centering
		\includegraphics[width=\textwidth]{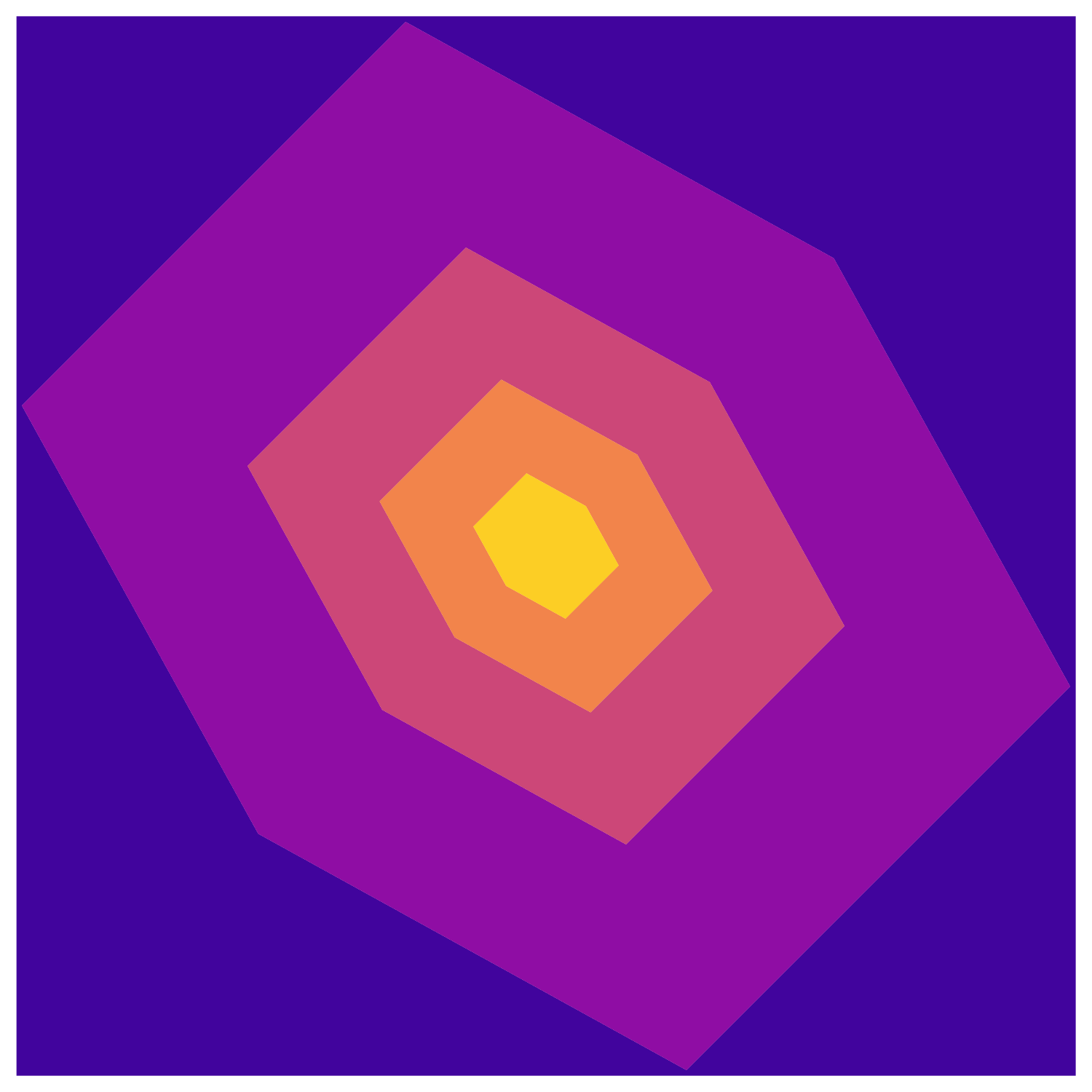}
		\caption{$\rho = \abs*{\,\cdot\,}$.}
	\end{subfigure}
	\caption{Contour plots that illustrate the closure property of overcomplete Gaussian models in $\R^2$. The contours are of the overcomplete model $\exp\br{-\sum_{i = 1}^3 \rho\br*{\ip{k_i}{x}}}$, where $k_1 \ceq \tuple{1, -1}$, $k_2 \ceq \tuple{\sqrt{3} - 1, \sqrt{3} + 1}$, and $k_3 \ceq \tuple{\sqrt{3} + 1, \sqrt{3} - 1}$. For $\rho = \br*{\,\cdot\,}^2$ the model is an overcomplete multivariate Gaussian on $\R^2$ that can be equivalently represented as a complete multivariate Gaussian on $\R^2$. For $\rho = \abs*{\,\cdot\,}$ the model is an overcomplete Laplace distribution on $\R^2$, which however cannot be represented as a complete Laplace distribution on $\R^2$ as the distribution clearly has more than two principle directions.}
	\label{fig:closure}
\end{figure}

The joint distribution $\fxz$ in~\eqref{def_glm} can be written as
\begin{align}
	\fxz{x, z} &\propto \prod_{i = 1}^m g_i\br*{\br{Kx}_i, z_i} \cdot \f{i}{z_i} \nonumber = \prod_{i = 1}^m \nd*{\br*{Kx}_i; \mu_i\br*{z_i}, \sigma^2_i\br*{z_i}} \cdot \f{i}{z_i} \nonumber \\
		&= \nd*{Kx; \mu_0\br*{z}, \Sigma_0\br*{z}} \cdot \prod_{i = 1}^m \f{i}{z_i} \label{eq_ocg},
\end{align}
where
\begin{align*}
	\mu_0\br*{z} \ceq \tuple*{\mu_1\br*{z_1}, \mu_2\br*{z_2}, \ldots, \mu_m\br*{z_m}} \quad \text{and} \quad \Sigma_0\br*{z} \ceq \diag \tuple*{\sigma^2_1\br*{z_1}, \sigma^2_2\br*{z_2}, \ldots, \sigma^2_m\br*{z_m}}.
\end{align*}
Consequently, the conditional distribution $\f_{X \mid Z = z}(x)$ at some point \( x \in \R^n \) is proportional to $\nd*{Kx; \mu_0\br*{z}, \Sigma_0\br*{z}}$. This is an overcomplete representation of a Gaussian distribution on $\R^n$ since $K$ is an $m \times n$ matrix of full rank and $m \geq n$ univariate Gaussian factors are used to define the distribution.
A crucial property of Gaussian distributions is that they are \enquote{closed under overcompleteness} in the sense of that we can always replace an overcomplete representation with an equivalent complete representation.
An example that illustrates the presence of this property for Gaussian distributions and the absence of this property for Laplace distributions is given  \cref{fig:closure}.

This closure property under overcompleteness implies that for some given latent vector \( z \in \mathcal{Z} \) the conditional distribution $\f_{X\mid Z = z}$ is a multivariate Gaussian distribution $\nd*{\mu\br*{z}, \Sigma\br*{z}}$ with appropriate mean vector $\mu\br*{z} \in \R^n$ and covariance matrix $\Sigma\br*{z} \in \symmpp^n$.
This is formally stated in the following proposition.
\begin{proposition}%
	\label{prop:xgz_mvg}
	Let $\fxz$ be a \gls{glm}.
	Then 
	\begin{align*}
		f_{X \mid Z = z} = \nd*{\mu\br*{z}, \Sigma\br*{z}},
	\end{align*}
	where
	\begin{align*}
		\mu\br*{z} \ceq \Sigma\br*{z} K^\top \Sigma^{-1}_0\br*{z} \mu_0\br*{z} \quad \text{and} \quad \Sigma\br*{z} \ceq \br{K^\top \Sigma^{-1}_0\br*{z} K}^{-1}.
	\end{align*}
\end{proposition}
\begin{proof}
	See \cref{proof:xgz_mvg}.
\end{proof}

\begin{remark}
	We again refer to the functions $\mu_0$, $\Sigma_0$, $\mu$, and $\Sigma$ as latent mappings.
\end{remark}

\begin{remark}%
	\label{remark_cov}
	The principle axes of the covariance matrix $\Sigma\br*{z}$ in \Cref{prop:xgz_mvg} might change depending on the entries of $\Sigma_0\br*{z} = \diag \tuple*{\sigma^2_1\br*{z_1}, \sigma^2_2\br*{z_2}, \ldots, \sigma^2_m\br*{z_m}}$.
	In other words,  the covariance matrix $\Sigma\br*{z}$ might be differently oriented depending on the value $z$ of the latent variable $Z$.
\end{remark}

\begin{remark}
	\Cref{prop:xgz_mvg} partially motivates our choice of using Gaussian instead of non-Gaussian mixtures to \enquote{lift} the factors $\phi_i$ in the~\eqref{def_poe} model to latent representations, as the conditional distribution $\fxgz$ of the resulting latent variable model is a multivariate Gaussian distribution, for which efficient sampling algorithms exist even in the high-dimensional setting.
	In contrast, sampling overcomplete models with non-Gaussian factors in high-dimensional spaces is in general a nontrivial problem.
\end{remark}

\subsubsection{Deriving \texorpdfstring{$\fz$}{f\_\{Z\}}}
We state the form of the prior distribution $\fz$ of the latent variable $Z$ from~\eqref{eq_ocg} in the following proposition.
\begin{proposition}%
	\label{prop:latent_distribution}
	Let $\fxz$ be a \gls{glm}.
	Then
	\begin{align*}
		\fz{z} \propto \g{z} \cdot \prod_{i = 1}^m \frac{\f{i}{z_i}}{\sigma_i\br*{z_i} \cdot \exp\br*{\frac{1}{2} \frac{\mu^2_i\br*{z_i}}{\sigma^2_i\br*{z_i}}}},
	\end{align*}
	where
	\begin{align*}
		\g{z} \ceq \sqrt{\det \Sigma\br*{z}} \cdot \exp\br*{\frac{1}{2} \norm{\mu\br*{z}}^2_{\Sigma^{-1}\br*{z}}},
	\end{align*}
	\begin{align*}
		\mu\br*{z} \ceq \Sigma\br*{z} K^\top \Sigma^{-1}_0\br*{z} \mu_0\br*{z}, \quad \text{and} \quad \Sigma\br*{z} \ceq \br{K^\top \Sigma^{-1}_0\br*{z} K}^{-1}.
	\end{align*}
\end{proposition}
\begin{proof}
	See \cref{proof:latent_distribution}.
\end{proof}

Unfortunately, the coupling term $g$ in this proposition makes the components of $Z$ dependent of each other\footnote{%
	An example of this can be found in Section SM3 of the supplementary materials.%
}, which in turn makes sampling from $\fz$ difficult.
Consequently, ancestral sampling for \glspl{glm} is difficult in general, which makes Gibbs sampling the preferred approach.
However, later at the end of this section, we will show that direct sampling is feasible and efficient in the case of complete~\eqref{def_poe} models. 

\cref{prop:xgz_mvg,prop:latent_distribution} imply that \glspl{glm} are implicitly defined \glspl{gmm} on $\R^n$, which is formally stated in the following corollary.
\begin{corollary}%
	\label{coro:glms_and_gmms}
	Let $\fxz$ be a \gls{glm}.
	Then $\fx$ is a \gls{gmm} on $\R^n$ of the form
	\begin{align*}
		\fx{x} = \int_{\mathcal{Z}} \fz\br*{z} \cdot \nd*{x; \mu\br*{z}, \Sigma\br*{z}}\,\mathrm{d}z,
	\end{align*}
	where
	\begin{align*}
		\mu\br*{z} \ceq \Sigma\br*{z} K^\top \Sigma^{-1}_0\br*{z} \mu_0\br*{z} \quad \text{and} \quad \Sigma\br*{z} \ceq \br{K^\top \Sigma^{-1}_0\br*{z} K}^{-1}.
	\end{align*}
\end{corollary}

\begin{remark}
	\Cref{remark_cov} implies that the resulting Gaussian mixture model in \Cref{coro:glms_and_gmms} is, in general, heteroscedastic with non-axis-aligned covariance matrices.
\end{remark}

\begin{remark}
	While the \gls{gmm} representation is conceptually insightful, obtaining or directly working with it is generally intractable.
	The number of components in the resulting \gls{gmm} either grows exponentially in \( m \), is uncountable, or a combination of both. 
	For instance, if all latent components $Z_1, \ldots, Z_m$ can only take two possible values,  then we obtain a \gls{gmm} on $\R^n$ with $2^m$ components.
	This quickly becomes intractable even for moderately large $m$ and prohibits the evaluation of the mixture weights $\fz$ or the storage of the \gls{gmm} in nearly all cases of practical interest.
\end{remark}

\begin{remark}
	\glspl{glm} closely resemble the decoder in a variational autoencoder~\cite{Kingma:2013} since both are Gaussian latent variable models. However, there are a few notable differences: The variational autoencoder relies on a low-dimensional\footnote{In the sense that the dimension of the latent $Z$ is considerably smaller than the dimension of $X$.} standard multivariate Gaussian latent distribution $\fz = \nd{0, I}$ and a Gaussian conditional distribution $f_{\XGZ = z} = \nd{\mu\br{z}, \Sigma\br{z}}$ whose latent mappings $\mu$ and $\Sigma$ are defined through a neural network.  In contrast, \glspl{glm} relies on a high-dimensional latent variable $Z$ distributed using a general latent distribution $\fz$ and a Gaussian conditional distribution $f_{\XGZ = z} = \nd{\mu\br{z}, \Sigma\br{z}}$ whose latent mappings $\mu$ and $\Sigma$ are defined through the latent mappings $\mu_0$ and $\Sigma_0$ and the linear operator $K$.  The components of the latent variable $Z$ in a variational autoencoder are independent from each other, whereas that is typically not the case in \glspl{glm}.  For tractability reasons, the covariance matrix $\Sigma\br{z}$ in variational autoencoders is typically assumed to be axis aligned, whereas that is not the case in \glspl{glm} in nearly all cases of practical interest.
\end{remark}

\subsubsection{Deriving \texorpdfstring{$\fzgx$}{f\_\{Z | X\}}}
Finally, we show that $f_{Z \mid X}$ decomposes into $m$ independent univariate distributions, as formally stated in the following proposition.
\begin{proposition}%
	\label{prop:fz_decomposes}
	Let $\fxz$ be a \gls{glm}.
	Then
	\begin{align*}
		\fzgx{\zgx} = \prod_{i = 1}^m \f_{Z_i \mid X}\br*{z_i \mid x},
	\end{align*}
	where
	\begin{align*}
		\f_{Z_i \mid X}\br*{z_i \mid x} \propto g_i\br*{\br{Kx}_i,  z_i} \cdot \f{i}{z_i} \quad \text{for} \quad i = 1, \ldots, m.
	\end{align*}
\end{proposition}
\begin{proof}
	See \cref{proof:fz_decomposes}.
\end{proof}
Efficient direct sampling from these univariate distributions is possible in many cases of practical interest, as summarized in the following subsection.
Among these, the most notable one is when a factor $\phi_i$ is a \gls{gmm} with a finite amount of components.
This implies that this subproblem can be tractably approximated in general, simply by approximating a factor $\phi_i$ with a finite component \gls{gmm}.

\subsection{Gibbs Sampling for Gaussian Latent Machines}%
\label{sec:glm_sampling}
We now focus on two-block Gibbs sampling for \glspl{glm} since the corresponding subproblems can be efficiently solved in general.
The Markov network of~\eqref{def_glm}, along with the corresponding two-block Gibbs sampling subproblems, is summarized in~\cref{fig:gibbs}.
As earlier shown in~\Cref{prop:xgz_mvg,prop:fz_decomposes}, sampling from $\fzgx$ reduces to sampling a multivariate Gaussian distribution with an implicitly defined mean vector and covariance matrix, while sampling from $\fzgx$ reduces to independently sampling from $m$ univariate distributions\footnote{%
	Note that the conditional independence property from~\Cref{prop:fz_decomposes} is directly implied from the Markov network structure of~\eqref{def_glm} as shown in \cref{fig:gibbs}, since $\set{X}$ is the Markov blanket~\cite{Koller:2009} for each latent component $Z_1, \ldots, Z_m$.%
} whose exact form depends on the latent distributions and mappings.
The following two subsections focus on efficient methods for solving these subproblems in standard Bayesian imaging models.
Specifically, we discuss techniques for sampling from a multivariate Gaussian distribution in a high-dimensional, matrix-free setting\footnote{%
	Matrix-free means that we can evaluate the forward and adjoint operations of a linear operator without the need to store it explicitly as a matrix.
	For example, if $A$ is a linear operator from $\R^n$ to $\R^m$, we assume that we can compute the forward operation $Ax$ for any vector $x \in \R^n$ and the adjoint operation $A^T y$ for any vector $y \in \R^m$ without storing $A$ as an $m \x n$ matrix and performing matrix-vector operations with it.
	This is unavoidable in high-dimensional settings since storing and operating with matrices quickly becomes intractable.
}, as well as efficient strategies for conditional latent sampling of various standard factors $\phi_i$ that are commonly found in imaging models.

\begin{figure}[!t]
	\centering
	\begin{subfigure}{0.45\textwidth}
		\centering
		\begin{tikzpicture}[framed]
			\node[circle, draw=black, thick, minimum size=1cm] (x)  at (0, 0) {$X$};
			%\node[above=0.05cm of x] {Markov network of $X, Z$};
			\node[circle, draw=black, thick, minimum size=1cm] (z1) at (-2.25, -2)  {$Z_1$};
			\node[circle, draw=black, thick, minimum size=1cm] (z2) at (-0.75, -2)  {$Z_2$};
			\node                                                     at ( 0.75, -2)  {$\ldots$};
			\node[circle, draw=black, thick, minimum size=1cm] (z3) at ( 2.25, -2)  {$Z_m$};  
			\foreach \from/\to in {x/z1,x/z2,x/z3} \draw[line width=1pt] (\from) -- (\to);
		\end{tikzpicture}
		\caption{Markov network that corresponds to~\eqref{def_glm}.}%
		\label{fig:gibbs-xgz-markov}
	\end{subfigure}
	\\[0.5em]
	\begin{subfigure}{0.45\textwidth}
		\centering
		\begin{tikzpicture}[framed]
			\node[circle, draw=black, thick, minimum size=1cm] (x)  at (0, 0) {$X$};
			%\node[above=0.05cm of x] {Markov network of $X \mid Z = z$};
			\node[circle, draw=black, thick, minimum size=1cm, fill=RedViolet!5!gray!25] (z1) at (-2.25, -2)  {$Z_1$};
			\node[circle, draw=black, thick, minimum size=1cm, fill=RedViolet!5!gray!25] (z2) at (-0.75, -2)  {$Z_2$};
			\node                                                     at ( 0.75, -2)  {$\ldots$};
			\node[circle, draw=black, thick, minimum size=1cm, fill=RedViolet!5!gray!25] (z3) at ( 2.25, -2)  {$Z_m$};  
			\foreach \from/\to in {x/z1,x/z2,x/z3} \draw[line width=1pt] (\from) -- (\to);
			\node[below=2.25cm of x] {$X \mid Z = z \da \nd*{\mu\br*{z}, \Sigma\br*{z}}$};
		\end{tikzpicture}
		\caption{Sampling $X \mid Z = z$.}%
		\label{fig:gibbs-xgz}
	\end{subfigure}
	\begin{subfigure}{0.45\textwidth}
		\centering
		\begin{tikzpicture}[framed]
			\node[circle, draw=black, thick, minimum size=1cm, fill=RedViolet!5!gray!25] (x)  at (0, 0) {$X$};
			%\node[above=0.05cm of x] {Markov network of $Z \mid X = x$};
			\node[circle, draw=black, thick, minimum size=1cm] (z1) at (-2.25, -2)  {$Z_1$};
			\node[circle, draw=black, thick, minimum size=1cm] (z2) at (-0.75, -2)  {$Z_2$};
			\node                                                     at ( 0.75, -2)  {$\ldots$};
			\node[circle, draw=black, thick, minimum size=1cm] (z3) at ( 2.25, -2)  {$Z_m$};  
			\foreach \from/\to in {x/z1,x/z2,x/z3} \draw[line width=1pt] (\from) -- (\to);
			\node[below=2.25cm of x] {$Z_i \mid X = x \da f_{Z_i \mid X = x}$};
		\end{tikzpicture}
		\caption{Sampling $Z \mid X = x$.}%
		\label{fig:gibbs-zgx}
	\end{subfigure}
	\caption{%
		Markov network of the~\eqref{def_glm} model and the corresponding Gibbs sampling subproblems.
		Sampling $X \mid Z = z$ reduces to sampling from a multivariate Gaussian distribution, while sampling $Z \mid x = x$ reduces to sampling from $m$ univariate distributions.%
	}%
	\label{fig:gibbs}
\end{figure}

\subsubsection{Sampling from \texorpdfstring{$\fxgz$}{f\_\{X | Z\}}}
The standard method for sampling from a multivariate Gaussian distribution $\nd*{\mu, \Sigma}$ relies on computing the Cholesky factorization of the covariance matrix $\Sigma$ or its inverse (commonly referred to as the precision matrix).
However, in high-dimensional settings, this approach is prohibitively expensive in terms of both computation and memory requirements~\cite{Vono:2022}.
Therefore, we discuss an alternative direct sampling approach that is specifically tailored to the structure of the multivariate Gaussian distributions $\fxgz$ that arise as subproblems in the Gibbs sampler for \glspl{glm}.
It is based on the following simple technical result that mirrors the structure of the multivariate Gaussian distribution from \cref{prop:xgz_mvg}.
\begin{proposition}%
	\label{prop:mvg_perturb_map}
	Let $Y \da \nd*{\mu_0, \Sigma_0}$, where $\mu_0 \in \R^m$ and $\Sigma_0 \in \symmpp^n$ is a real symmetric positive definite matrix of size ${m \times m}$. Suppose that $K \in \R^{m \times n}$ with $m \geq n$ and $\rank(K) = n$. Define the random variable $X$ through the linear transformation
	\begin{align*}
		X = \br{K^\top \Sigma^{-1}_0 K}^{-1} K^\top \Sigma^{-1}_0 Y.
	\end{align*}
	Then
	\begin{align*}
		X \da \nd*{\mu, \Sigma}, \quad \text{where} \quad \mu \ceq \Sigma K^\top \Sigma^{-1}_0 \mu_0 \quad \text{and} \quad \Sigma \ceq \br{K^\top \Sigma^{-1}_0 K}^{-1}.
	\end{align*}
\end{proposition}
\begin{proof}
	See \cref{proof:mvg_perturb_map}.
\end{proof}
In other words, to draw a sample $X$ from the multivariate Gaussian distribution $\nd*{\mu, \Sigma}$, we can first draw a sample $Y$ from the multivariate Gaussian distribution $\nd*{\mu_0, \Sigma_0}$ and then solve the linear system $K^\top \Sigma^{-1}_0 K X = K^\top \Sigma^{-1}_0 Y$.
Fortunately, both subproblems can be solved efficiently:
The first subproblem reduces to independently sampling from $m$ univariate Gaussian distributions since $\Sigma_0$ is always a diagonal matrix in our setting.
The second subproblem can be efficiently solved in a matrix-free fashion by iterative methods such as conjugate gradient.
The resulting algorithm to sample from the multivariate Gaussian distribution $\fxgz$ is summarized in \Cref{alg:perturb_and_map}.

\begin{algorithm}[!t]
	\caption{Direct sampling algorithm for $f_{\XGZ = z} = \nd{\mu\br{z}, \Sigma\br{z}}$}%
	\label{alg:perturb_and_map}
	\begin{algorithmic}[1]
		\Require{Latent mappings $\mu_0$ and $\Sigma_0$, and latent realization $z$}
		\Ensure{A sample $X \da f_{\XGZ = z}$}
		\State{Draw $Y_i \da \nd{\mu_i\br{z_i}, \sigma^2_i\br{z_i}}$ for $i = 1, \ldots, m$}\Comment{Sample $Y \da \nd{\mu_0\br{z}, \Sigma_0\br{z}}$}
		\State{Find \( X \) such that $K^\top \Sigma^{-1}_0 K X = K^\top \Sigma^{-1}_0 Y$}\Comment{Sample $X \da \nd{\mu\br{z}, \Sigma\br{z}}$}
	\end{algorithmic}
\end{algorithm}

\begin{remark}
	\Cref{prop:mvg_perturb_map,alg:perturb_and_map} have been rediscovered many times throughout the literature (see~\cite{Bardsley:2012, Bardsley:2014, Gilavert:2015, Orieux:2012, Papandreou:2010, Papandreou:2016, Rue:2001, Rue:2005, Schmidt:2010, Vono:2022} and the references therein) and are known under names such as Perturb-and-MAP~\cite{Papandreou:2010, Papandreou:2016}, Perturbation-Optimization~\cite{Orieux:2012}, and Randomize-Then-Optimize~\cite{Bardsley:2012, Bardsley:2014}.
	The terms \enquote{MAP} and \enquote{optimize/optimization} are used since solving the linear system in the second subproblem is equivalent to computing the \gls{map} estimate in a Gaussian linear inverse problem which, in turn, is equivalent to solving a least squares optimization problem.
\end{remark}

Due to the assumption that $K \in \mathbb{R}^{m \times n}$ with full rank and $m \geq n$, the linear system in line 2 of \Cref{alg:perturb_and_map} is always well-posed, but might be ill-conditioned depending on the structure of $K$ and $\Sigma_0$. Therefore, the diagonally preconditioned conjugate gradient method \cite[Algorithm 5.3]{Nocedal:2006} is a suitable choice for its resolution. The algorithm, adapted to our notation and setting, is summarized in \Cref{alg:precond_cg}\footnote{The algorithm solves the system $Ax = b$ by introducing the change of variable $\bar{X} = D^{\frac{1}{2}} X$, where $D \ceq \diag d$ is the so-called preconditioner. This leads to the modified system $D^{-\frac{1}{2}} A D^{-\frac{1}{2}} \bar{X} = D^{-\frac{1}{2}} b$, which is more appealing to solve if $D^{-\frac{1}{2}} A D^{-\frac{1}{2}}$ has a lower condition number than $A$. Note also that $r$ refers to the residual of the original system $Ax = b$.}.

\begin{remark}
	Strictly speaking, the corresponding multivariate Gaussian samples produced by \Cref{alg:perturb_and_map} and \ref{alg:precond_cg} will be slightly biased since the linear system is only approximately solved by \Cref{alg:precond_cg}. It was shown in \cite{Gilavert:2015} that, in a Gibbs sampling context, this bias can be compensated by adding a Metropolis-Hastings step at the end of \Cref{alg:perturb_and_map}. However, we omitted the correction step since the bias vanishes when the tolerance parameter in \Cref{alg:precond_cg} is sufficiently small.
\end{remark}

Executing \cref{alg:precond_cg} using double-precision floating-point numbers and without preconditioning (\ie, $d = 1_n$) was sufficient for most of our experiments. The poor conditioning in the remaining experiments was considerably improved by the standard choice $d = \diag \br{K^T \Sigma_0 K}$. The diagonal entries of $K^T \Sigma_0 K$ can be efficiently computed in a matrix-free fashion if $K$ models a collection of convolution operator, as formally summarized in the following proposition.
\begin{proposition}%
	\label{prop:precond}
	Let $K \in \R^{m \times n}$ and $\Sigma_0 \in \R^{m \times m}$ be block matrices defined as
	\begin{align*}
		K = \begin{bmatrix}
			K_1 \\
			K_2 \\
			\vdots \\
			K_k 
		\end{bmatrix}
		\quad \text{and} \quad
		\inv{\Sigma}_0 =
		\begin{bmatrix}
			\Sigma^{-1}_1 & 0 & \cdots & 0 \\
			0 & \Sigma^{-1}_2 & \cdots & 0 \\
			\vdots & \vdots & \ddots & \vdots \\
			0 & 0 & \cdots & \Sigma^{-1}_k
		\end{bmatrix},		
	\end{align*}
	where $K_i \in \R^{m_i \times n}$ and $\Sigma^{-1}_i \in \R^{m_i \times m_i}$ for $i = 1, \ldots, k$, and $m = \sum^k_{i = 1} m_i$. Then the following statements are true:
	\begin{enumerate}[label=\alph*)]
		\item $\diag\br{K^\top \Sigma^{-1}_0 K} = \sum_{i = 1}^k \diag{K^\top_i \Sigma^{-1}_i K_i}$. 
		\item If $\Sigma^{-1}_i$ is a diagonal matrix for any \( i = 1, \dotsc, k \), then 
		\[\diag{K^\top_i \Sigma^{-1}_i K_i} = \br{K^{\circ 2}_i}^\top \diag \Sigma^{-1}_i,\] 
		where \( K^{\circ 2}_i \) denotes the element-wise squaring of the entries in \( K_i \).
	\end{enumerate}
\end{proposition}
\begin{proof}
	See \cref{proof:precond}.
\end{proof}

\begin{algorithm}[!t]
	\caption{Diagonally preconditioned conjugate gradient algorithm for finding \( X \) such that $K^\top \Sigma^{-1}_0 K X = K^\top \Sigma^{-1}_0 Y$.}%
	\label{alg:precond_cg}
	\begin{algorithmic}[1]
		\Require{Linear operator $A \ceq K^\top \circ \Sigma^{-1}_0 \circ K$, vector $b \ceq K^\top \Sigma^{-1}_0 Y$, initial solution $X_0 \in \R^n$, preconditioning vector $d \in \Rpp^n$, tolerance parameter $\epsilon > 0$}
		\Ensure{$X$ such that $\norm{K^\top \Sigma^{-1}_0 K X - K^\top \Sigma^{-1}_0 Y} < \epsilon$}
		\State{$X \ceq X_0$}
		\State{$r \ceq b - A X$}\Comment{Compute initial residual}
		\State{$z \ceq r \oslash d$}\Comment{Compute initial search direction}
		\State{$p \ceq z$}
		\While{$\norm{r} > \epsilon$}\Comment{Iterate until desired tolerance is reached}
		\State{$\alpha \ceq \ip{r}{z} / \ip{p}{Ap}$}
		\State{$X \ceq X + \alpha \, p$ }\Comment{Update solution}
		\State{$\bar{r} \ceq r$ }\Comment{Update residual}
		\State{$r \ceq r - \alpha \, Ap$}
		\State{$\bar{z} \ceq z$ }\Comment{Update search direction}
		\State{$z \ceq r \oslash d$}
		\State{$\beta \ceq \ip{r}{z} / \ip{\bar{r}}{\bar{z}}$}
		\State{$p \ceq z + \beta \, p$}
		\EndWhile%
	\end{algorithmic}
\end{algorithm}

\begin{remark}
	Note that the linear operator $\br{K^{\circ 2}_i}^\top$ in part b of \Cref{prop:precond} can be easily computed whenever $K_i$ represents a local linear operator like a convolution. This can be achieved either by squaring the weights of the convolutional filter and then forming its adjoint operator, or by first forming the adjoint operator of the convolutional filter and then squaring its weights. Consequently, in many applications of practical interest, the diagonal entries of the matrix $K^T \Sigma_0 K$ can be computed in a matrix-free fashion with the same complexity required for a single iteration of conjugate gradient. Therefore, in such cases, computing the diagonal preconditioner is basically a for-free operation, making this form of diagonal preconditioning worth trying, as it may accelerate the sampling or improve its accuracy. On the other hand, this approach is sometimes not feasible when $K_i$ represents a global linear operator (\eg, discrete cosine transform), since obtaining the linear operator $\br{K^{\circ 2}_i}^\top$ might be nontrivial or even intractable in a matrix-free setting.
\end{remark}

The matrix-free approach for this subproblem is easily parallelizable on modern graphics processing units, both for a single system and for batches of systems (which is beneficial when running multiple Markov chains in parallel). Futhermore, in the context of Gibbs sampling, the solution from the previous iteration can serve as an initial guess to warm-start the conjugate gradient algorithm for the current iteration, which typically results in substantial runtime improvements for the Gibbs sampler.

\begin{remark}
	The efficiency of various algorithms for sampling high-dimensional multivariate Gaussian distributions is largely determined by the structure of the covariance matrix. The previous discussion focused on approaches that work well in the imaging setting considered in this work. However, given the broad applicability of our proposed approach beyond imaging problems, we refer to \cite{Vono:2022} for a recent survey on this topic. Additionally, the stochastic matrix-free equilibration method from \cite{Diamond:2017} might be of interest since it can efficiently compute diagonal preconditioners for general linear operators. In our experiments, we have instead used the result from \Cref{prop:precond}, as it is easier to implement and considerably faster to compute while leading to nearly identical behaviour as the approach from \cite{Diamond:2017}.
\end{remark}

\subsubsection{Sampling from \texorpdfstring{$\fzgx$}{f\_\{Z | X\}}}
\Cref{prop:fz_decomposes} states that the conditional distribution $\fzgx$ factors as the product of the individual conditional latent distributions $\f_{Z_i \mid X = x}$.
Consequently, sampling from $\fzgx$ amounts to independently sampling each conditional latent distribution.
Specifically, each conditional latent distribution is of the form $\f_{Z_i \mid X}\br*{z_i \mid x} \propto g_i\br*{\br{Kx}_i,  z_i} \cdot \f{i}{z_i}$, which means that its exact form is determined by its corresponding latent distribution $f_i$ and by the latent mappings $\mu_i$ and $\sigma^2_i$ used in its conditional Gaussian distribution $g_i$ (or, in other words, what kind of factor $\phi_i$ they represent as per \Cref{assumption_fmp}).
\Cref{tab:latents} lists several factors $\phi_i$ commonly found in imaging models, along with their corresponding factor latent distributions, latent mappings, and conditional latent distributions\footnote{%
	A summary of our notation for univariate distributions is provided in \Cref{tab:distributions} of \Cref{sec:notation}, and derivations of the conditional latent distributions $\f_{Z_i \mid X = x}$ in \Cref{tab:latents} can be found in Section SM5 of the supplementary materials.%
}. 
{
\renewcommand{\arraystretch}{2.}
\begin{table}[!t]
\centering
\begin{adjustbox}{width=\columnwidth,center}
\begin{tabular}{lllll}
	\toprule
	Factor type & Factor distribution $\phi_i$ & Factor latent distribution $f_i$& Latent mappings & Conditional latent distribution $\f_{Z_i \mid X = x}$ \\ 
	\midrule	
	Laplace & $\LaplaceDist\br{b}$ & $\ExpDist\br*{\frac{1}{2 b^2}}$ & $\mu_i\br*{z_i} = 0, \sigma^2_i\br*{z_i} = z_i$ & $\GIGDist\br*{\frac{1}{b^2}, \br{Kx}^2_i, \frac{1}{2}}$ \\
	Student-t & $\tDist\br{\nu}$ & $\GammaDist\br*{\frac{\nu}{2}, \frac{\nu}{2}}$ & $\mu_i\br*{z_i} = 0, \sigma^2_i\br*{z_i} = \frac{1}{z_i}$ & $\GammaDist\br*{\frac{\nu + 1}{2}, \frac{\nu + \br{Kx}^2_i}{2}}$ \\ 
	Symmetrized Gamma & $\SymGammaDist\br{\alpha, \beta}$ & $\GammaDist\br*{\alpha, \beta}$ & $\mu_i\br*{z_i} = 0, \sigma^2_i\br*{z_i} = z_i$ & $\GIGDist\br*{2 \beta, \br{Kx}^2_i, \alpha - \frac{1}{2}}$ \\
	Gaussian mixture model & $\GMMDist\br{w,  \mu,  \sigma^2}$ & $\CatDist\br*{w}$ & $\mu_i\br*{z_i} = \mu_{z_i}, \sigma^2_i\br*{z_i} = \sigma^2_{z_i}$ & $\CatDist\br*{\bar{w}}$ with $\bar{w}_j \ceq \frac{w_j \cdot \nd{\br{Kx}^2_i; \mu_j, \sigma^2_j}}{\sum_{k = 1}^d w_k \cdot \nd{\br{Kx}^2_i; \mu_k, \sigma^2_k}}$ for $j = 1, \ldots, d$ \\
    \bottomrule
   \end{tabular}
   \end{adjustbox}
\caption{Factor distributions and their corresponding factor latent distributions,  latent mappings and conditional latent distributions.}%
\label{tab:latents}
\end{table}
}

The representations of the first three factors in~\Cref{tab:latents} are examples of \glspl{gsm}, which are Gaussian mixtures as defined in \Cref{assumption_fmp} with a latent mapping $\mu_i \equiv 0$.
Such representations are well established in the statistics literature.
For instance, the \gls{gsm} representations of Laplace and Student-t factors are derived in~\cite{Andrews:1974}, while the symmetric Gamma distribution is constructed through its characteristic function as briefly described in~\cite{Wainwright:1999}.
The last row in \Cref{tab:latents} follows directly from the standard latent variable representation of a finite-component \gls{gmm}.

The key difficulty in sampling from the conditional distribution $\fzgx$ lies, therefore, in having access to efficient samplers for the various conditional latent distributions $\f_{Z_i \mid X = x}$ that arise in our applications.
Fortunately, efficient samplers exist for all distributions in \Cref{tab:latents}.
More specifically, samplers for the Gamma distribution---required for Student-t factors---are readily available in standard libraries such as \texttt{PyTorch}~\cite{Paszke:2019}.
Sampling from generalized inverse Gaussian distributions---required for Laplace and symmetrized Gamma factors---can be carried out by the efficient rejection sampling algorithms from~\cite{Devroye:2014, Hoermann:2014}.
Finally, sampling from categorical distributions---required for Gaussian mixture factors---is a standard sampling problem that is implemented in many libraries like \texttt{PyTorch}, but custom implementations might be preferred for memory efficiency. 

\begin{remark}
	To allow our implementations to scale to high-dimensional inverse problems, we implemented custom \texttt{CUDA}~\cite{Nickolls:2008} kernels for the efficient sampling from generalized inverse Gaussian and categorical distributions. For the former, following the example of~\cite{Bohra:2023} we ported the rejection sampling algorithm from~\cite{Devroye:2014}. For the latter, our \texttt{CUDA} kernel for categorical distributions is considerably more memory efficient than a standard implementation in \texttt{PyTorch} that relies on broadcasting and results in prohibitive memory demands even for relative small problems.
\end{remark}

\begin{remark}
	The second parameter of the generalized inverse Gaussian distributions in \cref{tab:latents} for the conditional latent distribution of Laplace and symmetrized Gamma factors is given by $\br{Kx}^2_i$.
	This parameter should be positive, but might be zero depending on the entries of $K$ and the value $x$.
	It is almost surely positive during the iterations of the Gibbs sampler for any reasonable initialization strategy since such isolated points have measure zero, but various initialization strategies, such as zero initialization, can lead to violations of this constraint.
	To protect against such cases of undefined behaviour, it is advised in implementations to replace $\br{Kx}^2_i$ with standard numerical safeguards such as $\max\set{\br{Kx}^2_i, \varepsilon}$, where $\varepsilon$ is a small positive constant (\eg, $\num{1e-7}$).
\end{remark}

\begin{remark}
	More general families of factors, such as $\alpha$-stable distributions~\cite{Wainwright:1999} or the exponential power family of distributions~\cite{West:1987} of Box \& Tiao~\cite{Box:1992}, can be represented as \glspl{gsm}.
	The systematic cataloging of such factors (in the same spirit as was done for proximal operators~\cite{Chierchia}) is an interesting research direction, but such efforts are beyond the scope of this work.
	However, it remains unclear whether such factors are meaningful in the context of imaging models and whether they admit efficient conditional latent sampling.
\end{remark}

\subsection{Specific Nonlinear Extensions}%
It is possible to adapt our Gibbs sampling approach to trivially support some specific nonlinear extensions of the~\eqref{def_poe} model. More precisely, we can add nonlinear terms of the form
\begin{align}
	\phi_i\br*{\sqrt{\br*{K_1 x}^2_i + \cdots + \br*{K_d x}^2_i}}
	\label{def_nfactor}
\end{align}
into the ~\eqref{def_poe} model, where $K_1, \ldots, K_d \in \R^{m\times n}$ are linear operators and $\phi_i: \R \to \Rpp$ is a univariate distribution that admits a \gls{gsm} representation of the form \eqref{def_fmp}. In that case, it follows by the lifted form of the factor $\phi_i$ that
\begin{align}
	\phi_i\br*{\sqrt{\br*{K_1 x}^2_i + \cdots + \br*{K_d x}^2_i}} &= \nd*{\sqrt{\br*{K_1 x}^2_i + \cdots + \br*{K_d x}^2_i}; 0, \sigma^2_i\br{z_i}} \cdot f_i\br{z_i} \nonumber \\
	&= f_i\br{z_i} \cdot \prod_{j = 1}^d \nd*{\br*{K_j x}_i; 0, \sigma^2_i\br{z_i}}.
	\label{def_lifted_nfactor}	
\end{align}
Therefore, a nonlinear factor of the form \eqref{def_nfactor} induces $d$ univariate Gaussian factors in the corresponding \gls{glm} representation that share the same latent variable. Consequently, the first Gibbs subproblem again simplifies to sampling from a multivariate Gaussian distribution of similar form as in the linear case. Similarly, from the first line in \eqref{def_lifted_nfactor} it follows that the conditional latent distribution of the corresponding $Z_i$ is given by
\begin{align*}
	\f_{Z_i \mid X}\br*{z_i \mid x} \propto g_i\br*{\sqrt{\br*{K_1 x}^2_i + \cdots + \br*{K_d x}^2_i}, z_i} \cdot \f{i}{z_i},
\end{align*}
which immediately implies that the value $(Kx)_i$ in the conditional latent distribution is replaced by $\sqrt{\br*{K_1 x}^2_i + \ldots + \br*{K_d x}^2_i}$. Practically, this means that terms of the form $\br{Kx}^2_i$ in the last column of \Cref{tab:latents} are simply replaced by terms of the form $\br*{K_1 x}^2_i + \cdots + \br*{K_d x}^2_i$ for factors $\phi_i$ that admit \gls{gsm} representations. Consequently, the second Gibbs subproblem is a trivial modification of the one from the linear case. 

Extensions of this form are relevant in the context of imaging, since they allow us to sample from models with isotropic regularizers like total variation. A particular instance of this extension was already hinted in \cite{Papandreou:2010} for posterior sampling in image denoising problems with isotropic total variation regularizers. 

\subsection{Direct Sampling for Complete Product of Experts Models}%
\label{sec:direct_sampling}
Direct sampling for the~\eqref{def_poe} model is possible whenever $K$ is an invertible $n \times n$ matrix, either directly or after model extension.
This is a trivial consequence of the change of variables formula and is formally stated in the following proposition. 
\begin{proposition}%
	\label{prop:direct_sampling}
	Consider the~\eqref{def_poe} model. When $K\in\R^{n\times n}$ is invertible, $U\coloneqq KX$ admits the density $\f{U}{u} = \prod_{i = 1}^n \f{U_i}{u_i}$, where $f_{U_i} = \phi_i$ for $ i = 1, \dots, n$.
\end{proposition}
\begin{proof}
	See \cref{proof:direct_sampling}.
\end{proof}
This immediately motivates \cref{alg:complete} for sampling in the complete case.

\begin{algorithm}[!t]
	\caption{Direct sampling algorithm for complete~\eqref{def_poe} models}%
	\label{alg:complete}
	\begin{algorithmic}[1]
		\Require{Invertible matrix $K \in \R^{n \times n}$ and univariate distributions $\phi_1: \R \to \R, \ldots, \phi_n: \R \to \R$}
		\Ensure{A sample from~\eqref{def_poe}}
		\State{Draw $U_i \da \phi_i$ for $i = 1, \ldots, n$ independently}\Comment{Sample $U \da \f_U$}
		\State{Find \( X \) such that \( KX = U \)}\Comment{Sample $X \da \fx$}
	\end{algorithmic}
\end{algorithm}
\begin{remark}
	\Cref{prop:direct_sampling} and the resulting \Cref{alg:complete} generalize the direct prior sampling algorithms of~\cite{Bohra:2023} from chain topologies to complete~\eqref{def_poe} models.
\end{remark}

For certain structures of \( K \) that frequently arise in imaging applications, it is possible to efficiently find an \( X \) such that $KX = U$ for some given \( U \in \R^n \).
In particular, this is the case when the~\eqref{def_poe} model describes certain improper priors defined on chain or tree graphs\footnote{%
	Such as the Gibbs distribution counterparts of the variational models considered in~\cite{Kolmogorov:2016, Kuric:2024} that cover chain and tree variants of standard image priors such as total variation~\cite{Rudin:1992} and total generalized variation~\cite{Bredies:2010}.%
}.
For instance, consider improper priors on $\R^n$ of the form
\begin{align}
	\fx{x} \propto \prod_{ij \in \edge} \phi_{ij}\br{x_j - x_i},
	\label{def_tp} \tag{TP}
\end{align}
defined over a tree graph $\tuple{\node, \edge}$, where $n \ceq \abs{\mathcal{V}}$ and $\phi_{ij}: \R \to \R$ are univariate distributions for $ij \in \edge$.
The~\eqref{def_tp} model is a special case of the~\eqref{def_poe} model, where the corresponding operator $K$ is an $\br{n - 1} \times n$ finite difference matrix.
Since $\ker(K) = \set{\lambda \cdot 1}{\lambda \in \R}$\footnote{Here $1$ refers to the vector in $\R^n$ whose entries are all equal to $1$.}, we can extend the~\eqref{def_tp} model as per~\Cref{prop:smart_tie_breaking_works_too} by adding an appropriately chosen row to $K$.
For instance, adding the constant vector leads to the extended model
\begin{align*}
	\fx{x} \propto \phi_0\br*{\sum_{i \in \node} x_i} \cdot \prod_{ij \in \edge} \phi_{ij}\br{x_j - x_i},
\end{align*}
where $\phi_0: \R \to \R$ is an arbitrary univariate distribution.
Then, for this extended model and for any \( U \) we can efficiently find \( X \) such that $KX = U$, as formally stated in the following proposition. 
\begin{proposition}%
	\label{prop:tree_efficient_sampler}
	Let $\tuple{\node, \edge}$ be a directed tree with nodes $V = \set{1, \ldots, n}$, where $n \geq 1$, and all edges are oriented away from the root node $1$.  Let $u_0$,  $\seq{u_{ij}}_{ij \in \edge}$ and $\seq{x_{ij}}_{ij \in \edge}$ denote a collection of real-valued scalars.  Define $z_i$ as the total distance of the path from the root node to an arbitrary node $i \in \node$,  given by
	\begin{align*}
			z_i \ceq \begin{cases}
				0,  &\text{if } i = 1, \\
				\sum_{ij \in \br{1 \to i}} u_{ij},  &\text{otherwise},
			\end{cases}
	\end{align*}
	where $\br{1 \to i}$ denotes the set of directed edges along the unique path from the root node to node $i$. 

	Then the solution of the linear system
	\begin{align*}
		\sum_{i = 1}^n x_i &= u_0 \\
		x_j - x_i &= u_{ij} \quad \text{for all} \quad ij \in \edge
	\end{align*}
	is given by
	\begin{align*}
		x_1 &= \frac{1}{n}\br*{u_0 - \sum_{i = 1}^n z_i} \\
		x_i &= x_1 + z_i \quad \text{for all} \quad i \in \set{2, \ldots, n}.
	\end{align*}
\end{proposition}
\begin{proof}
	See \cref{proof:tree_efficient_sampler}.
\end{proof}

This immediately leads to \cref{alg:trees}, where $\mathcal{E}_{\text{preorder}}$ denotes the tree edges sorted in preorder. \cref{alg:trees} can efficiently solve the system in with linear time and memory complexity. Consequently, this implies that samples from chain and tree priors of the form~\eqref{def_tp} can be drawn in linear time and memory under the reasonable assumption that drawing each individual component of $u$ can be done in constant time and memory.

\Cref{prop:direct_sampling} also implies that the marginal distribution of $U_i = \br{KX}_i$ is given by the factor $\phi_i$.
This trivially follows by the marginalization over the components of $U$ in $f_U$ in \Cref{prop:direct_sampling}, and is summarized in the following corollary. 
\begin{corollary}%
	\label{cor_marginals}
	It follows from~\Cref{prop:direct_sampling} that the random variable $U_i \ceq \br{KX}_i$ is distributed as $\f_{U_i} = \phi_i$ for $i = 1, \ldots, n$.  
\end{corollary}
In other words, the factors $\phi_i$ are equivalent to the marginal distributions $\f_{U_i}$ (where the marginal $U_i$ is defined through the linear transformation $U_i = \br{KX}_i$ for $i = 1, \ldots, n$) in complete~\eqref{def_poe} models.
Consequently, it follows by~\Cref{prop:smart_tie_breaking_works_too} that the marginal distribution of edges $(X_j - X_i)$ in the~\eqref{def_tp} model is given by $\phi_{ij}$ for all $ij \in \edge$.

\begin{remark}
	Unfortunately, \Cref{cor_marginals} does not hold for overcomplete~\eqref{def_poe} models, as is shown through various examples in~\Cref{sec:results}. 
\end{remark}

Finally, the following proposition formalized that complete~\eqref{def_poe} models are a nontrivial case where the latent variables are independent from each other in the corresponding \gls{glm}.
\begin{proposition}%
	\label{prop:complete_independent_latents}
	Let $\fxz$ be a \gls{glm} and suppose that $K$ is an invertible $n \times n$ matrix. Then $\fz{z} = \prod_{i = 1}^n f_{Z_i}$, where $f_{Z_i} = f_i$ for $i = 1, \ldots, n$.
\end{proposition}
\begin{proof}
	See \cref{proof:complete_independent_latents}.
\end{proof}

\begin{algorithm}[!t]
	\caption{Efficient direct sampling algorithm for tree priors}%
	\label{alg:trees}
	\begin{algorithmic}[1]
		\Require{Directed tree $\tuple{\node,  \edge}$ whose edges are oriented toward the leaf nodes,  univariate distributions $\phi_0,  \seq{\phi_{ij}}_{ij \in \edge}$.}
		\Ensure{A sample from~\eqref{def_tp}}.
		\State{Draw $u_0 \da \phi_0$ and $u_{ij} \da \phi_{ij}$ for $ij \in \edge$ independently}\Comment{Sample $u \da \f_U$}
		\State{$z_1 \ceq 0$}\Comment{Compute the distances of each node to the root node}
		\For{$ij \in \mathcal{E}_{\text{preorder}}$}\Comment{Sample $x \da \fx$}
		\State{$z_j \ceq z_i + u_{ij}$}
		\EndFor{}
		\State{$x_1 \ceq \frac{1}{n} \br*{u_0 - \sum_{i = 1}^n z_i}$}\Comment{Efficiently solve $Kx = u$}
		\For{$i \ceq 2$ to $\abs{\mathcal{V}}$}
		\State{$x_i \ceq x_1 + z_i$}
		\EndFor{}
	\end{algorithmic}
\end{algorithm}

Therefore, the direct sampling algorithms~\Cref{alg:complete,alg:trees} are special cases of ancestral sampling, except that they operate directly on the factors $\phi_i$ of the~\eqref{def_poe} model instead of their lifted representations.
The sampling of the $U$ components could be alternatively carried out via ancestral sampling by considering the lifted factors (in cases when~\Cref{assumption_fmp} is satisfied).
However, it is unclear if such a reformulation provides any advantage when sampling from complete~\eqref{def_poe} models.

%% file: sections/results.tex
\section{Experimental Results}
\label{sec:results}
We now present experimental results that demonstrate the efficacy of our proposed Gibbs sampling approach. We focus mainly on prior and posterior sampling applications in the context of Bayesian imaging since they allow us to evaluate how well our proposed approach performs in high-dimensional settings. The source code required to reproduce all experimental results is available online\footnote{The source code is available at \href{https://github.com/mkuric/GLM}{https://github.com/mkuric/GLM}}.

\subsection{Prior Sampling Experiments} We first outline our evaluation protocol, then present simple baseline experiments before moving on to more complicated prior sampling experiments.

\subsubsection{Evaluation Protocol} 
We adhere to a unified evaluation protocol in our prior sampling experiments. 
We run $\num{10000}$ chains in parallel for $N$ iterations, all initialized with the same initial condition. 
The total number of iterations $N$ is selected such that a representative empirical marginal distribution---formed by taking marginal samples from all chains at a given iteration---does not change throughout the iterations $\ci{N/2}{N}$. 
The change is quantified through visual inspection of the empirical CDFs of the marginal distributions, which corresponds to achieving a predefined Wasserstein distance threshold. 

We rely on one-dimensional distributions as representative marginals for practical reasons since accurately evaluating statistical distances in high-dimensional spaces is computationally intractable, and storing the samples across the iterates for all chains becomes quickly infeasible\footnote{For instance, storing scalar samples in single-precision floating point format of $\num{10000}$ chains with $\num{10000}$ samples per chain requires already nearly 1/3 GB of memory.}. Details on how these representative marginals were chosen are provided in each experiment.

We compare our proposed Gibbs sampling approach against the \gls{mala}~\cite{Roberts:1996}. We have chosen \gls{mala} since it is the most popular general sampling approach in imaging and it is guaranteed to converge to the ground-truth distribution in all cases considered in our experiments. 
Since the performance of \gls{mala} heavily depends on the step-sizes, we spent roughly one week to tune \gls{mala} step-sizes in all of our experiments. 
In cases where the ground-truth marginal cannot be computed directly, we simply take the empirical marginal distribution at the final iteration of Gibbs as an approximation of it. Note that this particular choice is arbitrary since we ensure that both Gibbs and \gls{mala} have converged. 

After a sufficiently large number of iterations, the samples from all chains at a given iteration will be approximately independent and identically distributed samples from the ground-truth distribution. 
Therefore, this evaluation protocol allows us to quantify the length of the initialization phase (often referred to as burn-in) by tracking the distance of the empirical marginal distribution over the iterations with respect to the corresponding ground-truth marginal. 
For measuring the statistical distance between distributions, we use the Wasserstein-1 distance \cite{Villani:2009}, as it can be easily computed in the univariate case both between two sets of samples and between a set of samples and a univariate distribution.

In addition to quantifying the duration of the initialization phase, we assess the correlation between successive samples within the chains. Specifically, we calculate the autocorrelation function \cite{Gelman:2013} for each chain and report the mean and standard deviation of the autocorrelation function across all chains. 
Samples from the first half of the iterations are discarded to ensure the chains have reached the stationary regime, thereby allowing an accurate computation of the autocorrelation function. 
The rate at which the autocorrelation function decays to zero reflects the strength of the correlation between successive samples in a chain. 
A faster decay indicates weaker correlations, with an autocorrelation function immediately dropping to zero, signifying that successive samples are practically independent. 

This can be further summarized into a scalar $\gamma$ that lies in $\ci{0}{1}$ and quantifies the sampling efficiency, where a higher number means better sampling efficiency.
It is formally defined as
\begin{align*}
	\gamma \ceq \frac{1}{1 + 2 \sum_{k = 1}^K \rho_k},
\end{align*}
where $\rho_k$ denotes the autocorrelation function at lag $k$. 
For example, drawing $1000$ samples sequentially from a Markov chain in stationary regime with a sampling efficiency of $\gamma = 0.25$ is roughly equivalent to drawing $250$ independent samples. 
We choose the lag upper bound $K$ such that the autocorrelation function has dropped below $\num{0.05}$, following standard practice \cite{Pereyra:2023}.
Again, we report the mean and standard deviation of the sampling efficiency across all chains. 

The faster a sampler's convergence rate for a specific problem is, the shorter the initialization phase is. Similarly, the faster the mixing of a sampler is, the less correlated successive samples are. Therefore, our overall evaluation protocol consists of running these two types of tests on a suite of different sampling problems and comparing Gibbs against \gls{mala}. 

\begin{remark} We intentionally skew the evaluation in favour of \gls{mala} by providing high-quality initial conditions at the modes of the target distributions in each prior sampling experiment. This hides the general limitation of \gls{mala} that its performance is highly sensitive to initial conditions, and that the identification of suitable ones is a challenging problem in itself.
\end{remark}

\subsubsection{Baseline Experiments} We begin with a set of simple baseline experiments for which representative ground-truth marginals can be computed explicitly. These experiments serve several purposes.  They allow us to establish the performance of both Gibbs and \gls{mala} before transitioning to similar problems in higher dimensions. Futhermore, they allow to easily demonstrate the difference in behaviour between Gibbs and \gls{mala}. Lastly,  they provide valuable insights into the marginal distributions of standard image priors. 

We show the four factor graph topologies considered in the baseline experiments in \cref{fig:baseline-graphs}. When the factors $\phi$ are proper univariate distributions, the factor and product topologies describe proper distributions on $\R$,  while the loop and grid topologies describe improper distributions on $\R^4$ and $\R^6$, respectively.  Since the factor and product topologies induce easily computable univariate ground-truth distributions (the factor itself and a product of the factor with itself five times), we can directly compare the samplers on the corresponding ground-truth distribution without the need for any marginalization.  The loop and grid topology represent small versions of standard image priors with pairwise factors defined on the image edges. The kernels of the corresponding linear operators in those priors consists of constant images so, per \Cref{prop:smart_tie_breaking_works_too}, we introduce a factor acting on the mean of the entries to obtain proper distributions. 

In the literature, much focus has been placed on the marginal distribution of edges as a mechanism for determining the form of pairwise factors in image priors. We therefore concentrate on edge distributions as representative marginals for the loop and grid topologies, to illustrate the relationship between the pairwise factors and the edge marginals they induce.  Under the assumption that the factors $\phi$ are symmetric, the edge marginals in the loop topology are identically distributed, while in the grid topology the outer six edges are identically distributed, and the inner edge follows a different distribution. We exploit this fact to effectively achieve \( 4 \times \) the number of chains in the loop case and \( 6 \times \) the number of chains for the outer marginals in the grid case. Detailed derivations of the edge marginals for these two cases are provided in Section SM6 of the supplementary materials. 

\begin{figure}
\centering
\begin{subfigure}{0.11\textwidth}
  \centering
  \begin{tikzpicture}[framed]
    \node[circle, draw=black, thick, minimum size=1cm] (x)  at (0, 1.265) {$X$};
    \node[circle, thick, minimum size=1cm]  at (0, -1.25) {};
    \node[rectangle, draw=black, thick, fill=black, minimum size=0.01em] (f)  at (0, -1.25) {};
    \draw[line width=1pt] (x) -- (f);
  \end{tikzpicture}
  \caption{Factor.}
  \label{fig:baseline-factor}
\end{subfigure}
\hspace{-0.075cm}
\begin{subfigure}{0.16725\textwidth}
  \centering
  \begin{tikzpicture}[framed]
    \node[circle, draw=black, thick, minimum size=1cm] (x)  at (0, 1.265) {$X$};
    \node[circle, thick, minimum size=1cm]  at (0, -1.25) {};
    \node[rectangle, draw=black, thick, fill=black, minimum size=0.01em] (f1)  at (-1, -1.25) {};
    \node[rectangle, draw=black, thick, fill=black, minimum size=0.01em] (f2)  at (-0.5, -1.25) {}; 
    \node[rectangle, draw=black, thick, fill=black, minimum size=0.01em] (f3)  at (0, -1.25) {};
    \node[rectangle, draw=black, thick, fill=black, minimum size=0.01em] (f4)  at (0.5, -1.25) {};
	\node[rectangle, draw=black, thick, fill=black, minimum size=0.01em] (f5)  at (1., -1.25) {};
    
    \draw[line width=1pt] (x) -- (f1);
    \draw[line width=1pt] (x) -- (f2);
    \draw[line width=1pt] (x) -- (f3);
    \draw[line width=1pt] (x) -- (f4);
    \draw[line width=1pt] (x) -- (f5);
  \end{tikzpicture}
  \caption{Product.}
  \label{fig:baseline-product}
\end{subfigure}
\hspace{0.11cm}
\begin{subfigure}{0.2475\textwidth}
  \centering
  \begin{tikzpicture}[framed]
    \node[circle, draw=black, thick, minimum size=1cm] (x1)  at (-1.25, 1.25) {$X_1$};
    \node[circle, draw=black, thick, minimum size=1cm] (x2)  at (1.25, 1.25) {$X_2$}; 
    \node[circle, draw=black, thick, minimum size=1cm] (x3)  at (-1.25, -1.25) {$X_3$};
    \node[circle, draw=black, thick, minimum size=1cm] (x4)  at (1.25, -1.25) {$X_4$};
    
    \draw[line width=1pt] (x1) -- (x2) node[rectangle, midway, draw=black, thick, fill=black, minimum size=0.01em]{};
    \draw[line width=1pt] (x2) -- (x4) node[rectangle, midway, draw=black, thick, fill=black, minimum size=0.01em]{};
    \draw[line width=1pt] (x4) -- (x3) node[rectangle, midway, draw=black, thick, fill=black, minimum size=0.01em]{};
    \draw[line width=1pt] (x3) -- (x1) node[rectangle, midway, draw=black, thick, fill=black, minimum size=0.01em]{};
  \end{tikzpicture}
  \caption{Loop.}
  \label{fig:baseline-loop}
\end{subfigure}
\hspace{0cm}
\begin{subfigure}{0.425\textwidth}
  \centering
  \begin{tikzpicture}[framed]
    \node[circle, draw=black, thick, minimum size=1cm] (x1)  at (-2.5, 1.25) {$X_1$};
    \node[circle, draw=black, thick, minimum size=1cm] (x2)  at (0, 1.25) {$X_2$}; 
    \node[circle, draw=black, thick, minimum size=1cm] (x3)  at (2.5, 1.25) {$X_3$};
    \node[circle, draw=black, thick, minimum size=1cm] (x4)  at (-2.5, -1.25) {$X_4$};
    \node[circle, draw=black, thick, minimum size=1cm] (x5)  at (0, -1.25) {$X_5$};
    \node[circle, draw=black, thick, minimum size=1cm] (x6)  at (2.5, -1.25) {$X_6$};
    
    \draw[line width=1pt] (x1) -- (x2) node[rectangle, midway, draw=black, thick, fill=black, minimum size=0.01em]{};
    \draw[line width=1pt] (x2) -- (x3) node[rectangle, midway, draw=black, thick, fill=black, minimum size=0.01em]{};
    \draw[line width=1pt] (x3) -- (x6) node[rectangle, midway, draw=black, thick, fill=black, minimum size=0.01em]{};
    \draw[line width=1pt] (x6) -- (x5) node[rectangle, midway, draw=black, thick, fill=black, minimum size=0.01em]{};
    \draw[line width=1pt] (x5) -- (x4) node[rectangle, midway, draw=black, thick, fill=black, minimum size=0.01em]{};
    \draw[line width=1pt] (x4) -- (x1) node[rectangle, midway, draw=black, thick, fill=black, minimum size=0.01em]{};
    \draw[line width=1pt] (x2) -- (x5) node[rectangle, midway, draw=black, thick, fill=black, minimum size=0.01em]{};
  \end{tikzpicture}
  \caption{Grid.}
  \label{fig:baseline-grid}
\end{subfigure}
\caption{Factor graph topologies used in the baseline experiments. Each node defines a scalar random variable and each factor is of the same functional form $\phi$. The pairwise factors in (c) and (d) are of the form $\phi_{ij}\br*{x_i, x_j} := \phi\br*{x_j - x_i}$ and are defined for edges $ij$ such that $i < j$.}
\label{fig:baseline-graphs}
\end{figure}

All variations of the baseline experiments were run for $\num{15000}$ iterations, initialized with all zeros since this represents a mode in every test case.
Since runtime is not that relevant in these small-scale experiments, we plot the Wasserstein-1 distances over the iterations instead of the runtime in all baseline experiments.

\subsubsection{Baseline Performance}
The ground-truth negative log marginals for the considered topologies and four different types of factors (normal,  Laplace, Student-t, and \gls{gmm}) are shown in \cref{fig:baseline-gt}. An important observation is that the marginals in the overcomplete models differ from the factors. In particular, they are generally not even from the same parametric family, nor are they necessarily translational invariant as seen from the edge marginals in the grid topology. The belief that factors and edge marginals are the same is a common misconception that can be often found in publications, even though the seminal works \cite{Zhu:1998, Zhu:1997a} already established that the factors are not marginals themselves but rather dual variables in a corresponding maximum entropy problem that ensure that the marginals of the model match some target marginals. Notice also that overcomplete models reduce the variance of the marginals in comparison to their prescribed factors, which is the reason why all the marginal distributions in the product, loop and grid models are narrower than what is prescribed by the factors\footnote{It is rather intuitive that overcompleting probabilistic models tends to decrease variance. The easiest way to see this is by considering models with a product topology. For instance, if we consider such a model with Gaussian factors $\phi = \nd{0, 1}$ then it follows that $\fx = \nd{0, 1/5}$. Hence, adding more and more factors into an overcomplete model tends to reduce the variance along different directions.}.

We show the Wasserstein-1 distances over iterations in the baseline experiments in \Cref{fig:baseline-w1}. The Gibbs sampler is a direct sampling approach in all baseline experiments with normal factors. Consequently, in those cases, it converges to the ground-truth distribution within a single iteration, as seen in the first column of \Cref{fig:baseline-w1}. For all the other cases,  Gibbs sampling requires at most 20 iterations to converge to ground-truth distribution. In contrast, \gls{mala} struggles even in very simple cases---for instance, the \gls{gmm} potential in the grid case requires close to $\num{6000}$ iterations to reach the representative stationary distribution\footnote{In particular, note that it is a rather nontrivial problem to tune a constant step-size for multichain variants of \gls{mala} when trying to sample from factors whose log density has a wide range of different curvature values.}. The Wasserstein-1 distance saturates at some low but nonzero value due to the finite sample size.

We show in \Cref{fig:baseline-acf} the corresponding autocorrelation functions in the baseline experiments. As mentioned before, the Gibbs sampler is a direct sampling approach in all baseline experiments with normal factors and, consequently, produces independent samples along each chain in those cases. Unsurprisingly,  the autocorrelation functions computed from Gibbs samples are equal to zero in those cases,  as seen in the first column of \Cref{fig:baseline-acf}. However, the autocorrelation functions computed from Gibbs samples are also practically equal to zero in all remaining cases, even though the Gibbs sampler is not a direct sampling algorithm in those cases. Therefore,  Gibbs sampling effectively produces independent samples along each chain in all baseline experiments.
In contrast,  \gls{mala} tends to yield highly correlated samples, even in these small-scale problems. This is further corroborated by the corresponding sampling efficiencies shown in \Cref{tab:baseline_samp_eff}. Gibbs achieves a sampling efficiency of $1$ on average in nearly all baseline experiments, while \gls{mala} achieves poor efficiencies close to $0$ in some of the baseline experiments.

In summary, Gibbs sampling achieves near direct sampling performance in all considered baseline experiments, while \gls{mala} behaves rather poorly in some of them despite the fact that our step-size tuning in the baseline experiments is likely close to optimal. Furthermore, Gibbs does not require any kind of hyperparameter tuning, while inferring suitable step-sizes that allow \gls{mala} to explore the target distribution quickly is rather tedious, even in these small-scale problems. 

% Baseline
\begin{figure}
\centering
 \begin{tikzpicture}
 	\def\basepathhh{./fig/data/ground-truth-csvs/ground-truth_}
 		\begin{groupplot}[marginal group plot]
 			\pgfplotsforeachungrouped \component/\mytitle in {1/Normal, 2/Laplace, 3/Student-t, 4/GMM} {
 				\edef\tmp{%
 				\noexpand\nextgroupplot[title={\mytitle}, ylabel={\ifnum\component=1 Factor\fi}]%
 				\noexpand\addplot [line width=1pt, maincolor] table[col sep=comma,x=x,y=y]{\basepathhh1_\component.csv};%
 			}\tmp}
 			\pgfplotsforeachungrouped \component in {1, ..., 4} {
 				\edef\tmp{%
 				\noexpand\nextgroupplot[ylabel={\ifnum\component=1 Product\fi}]%
 				\noexpand\addplot [line width=1pt, maincolor] table[col sep=comma,x=x,y=y]{\basepathhh2_\component.csv};%
 			}\tmp}
 			\pgfplotsforeachungrouped \component in {1, ..., 4} {
 				\edef\tmp{%
 				\noexpand\nextgroupplot[ylabel={\ifnum\component=1 Loop\fi}]%
 				\noexpand\addplot [line width=1pt, maincolor] table[col sep=comma,x=x,y=y]{\basepathhh3_\component.csv};%
 			}\tmp}
 			\pgfplotsforeachungrouped \component in {1, ..., 4} {
 				\edef\tmp{%
 				\noexpand\nextgroupplot[ylabel={\ifnum\component=1 Grid\fi}]%
 				\noexpand\addplot [line width=1pt, maincolor] table[col sep=comma,x=x,y=y]{\basepathhh4_\component.csv};%
 				\noexpand\addplot [line width=1pt, secondarycolor] table[col sep=comma,x=x,y=y]{\basepathhh4_\component_orange.csv};%
 			}\tmp}
 		\end{groupplot}
 \end{tikzpicture}
\caption[]{%
\tikzexternaldisable
Ground-truth negative $\log$ marginals in the baseline experiments for
\protect\tikz[baseline=-\the\dimexpr\fontdimen22\textfont2\relax]\protect\draw [maincolor, ultra thick] (0,0) -- (.5, 0); the stationary distribution for the factor and product topologies, the edge marginal for the loop topology, and the inner edge marginal in the grid topology, and \protect\tikz[baseline=-\the\dimexpr\fontdimen22\textfont2\relax]\protect\draw [secondarycolor, ultra thick] (0,0) -- (.5, 0); the outer edge marginal in the grid topology.
\tikzexternalenable
}%
\label{fig:baseline-gt}
\end{figure}

\begin{figure}
\centering
 \begin{tikzpicture}
 	\def\basepathhh{./fig/data/baseline-wasserstein/baseline_wasserstein}
 		\begin{groupplot}[wasserstein group plot,/tikz/line join=bevel]
 			\pgfplotsforeachungrouped \component/\mytitle in {1/Normal, 2/Laplace, 3/Student-t, 4/\gls{gmm}} {
 				\edef\tmp{%
 				\noexpand\nextgroupplot[title={\mytitle}, ylabel={\ifnum\component=1 Factor\fi}]%
 				\noexpand\addplot [line width=1pt, maincolor] table[col sep=comma,x=x,y=y]{\basepathhh_gibbs_1_\component.csv};%
 				\noexpand\addplot [line width=1pt, maincolor, dashed] table[col sep=comma,x=x,y=y]{\basepathhh_mala_1_\component.csv};%
 			}\tmp}
 			\pgfplotsforeachungrouped \component in {1, ..., 4} {
 				\edef\tmp{%
 				\noexpand\nextgroupplot[ylabel={\ifnum\component=1 Product\fi}]%
 				\noexpand\addplot [line width=1pt, maincolor] table[col sep=comma,x=x,y=y]{\basepathhh_gibbs_2_\component.csv};%
 				\noexpand\addplot [line width=1pt, maincolor, dashed] table[col sep=comma,x=x,y=y]{\basepathhh_mala_2_\component.csv};%
 			}\tmp}
 			\pgfplotsforeachungrouped \component in {1, ..., 4} {
 				\edef\tmp{%
 				\noexpand\nextgroupplot[ylabel={\ifnum\component=1 Loop\fi}]%
 				\noexpand\addplot [line width=1pt, maincolor] table[col sep=comma,x=x,y=y]{\basepathhh_gibbs_3_\component.csv};%
 				\noexpand\addplot [line width=1pt, maincolor, dashed] table[col sep=comma,x=x,y=y]{\basepathhh_mala_3_\component.csv};%
 			}\tmp}
 			\pgfplotsforeachungrouped \component in {1, ..., 4} {
 				\edef\tmp{%
 				\noexpand\nextgroupplot[ylabel={\ifnum\component=1 Grid\fi}]%
 				\noexpand\addplot [line width=1pt, maincolor] table[col sep=comma,x=x,y=y]{\basepathhh_inner_gibbs_4_\component.csv};%
 				\noexpand\addplot [line width=1pt, maincolor, dashed] table[col sep=comma,x=x,y=y]{\basepathhh_inner_mala_4_\component.csv};%
 				\noexpand\addplot [line width=1pt, secondarycolor] table[col sep=comma,x=x,y=y]{\basepathhh_outer_gibbs_4_\component.csv};%
 				\noexpand\addplot [line width=1pt, secondarycolor, dashed] table[col sep=comma,x=x,y=y]{\basepathhh_outer_mala_4_\component.csv};%
 			}\tmp}
 		\end{groupplot}
 \end{tikzpicture}
\caption[]{%
\tikzexternaldisable
$\log_{10}$ Wasserstein-1 distances of Gibbs (solid, \protect\tikz[baseline=-\the\dimexpr\fontdimen22\textfont2\relax]\protect\draw [black, ultra thick] (0,0) -- (.5, 0);) and \gls{mala} (dashed, \protect\tikz[baseline=-\the\dimexpr\fontdimen22\textfont2\relax]\protect\draw [black, ultra thick, dashed] (0,0) -- (.5, 0);)  over iterations in the baseline experiments for \protect\tikz[baseline=-\the\dimexpr\fontdimen22\textfont2\relax]\protect\draw [maincolor, ultra thick] (0,0) -- (.5, 0); the factor and product topologies, the edge marginal of the loop topology, and the inner edge marginal of the grid topology and \protect\tikz[baseline=-\the\dimexpr\fontdimen22\textfont2\relax]\protect\draw [secondarycolor, ultra thick] (0,0) -- (.5, 0); the outer edge marginal of the grid topology. 
\tikzexternalenable
}%
\label{fig:baseline-w1}
\end{figure}

\begin{figure}
\centering
 \begin{tikzpicture}
 	\def\basepathhh{./fig/data/bacf/acf}
 		\begin{groupplot}[marginal group plot, xmin=1, xmax=30, yticklabel style={/pgf/number format/fixed,/pgf/number format/precision=3}]
 			\pgfplotsforeachungrouped \component/\mytitle in {1/Normal, 2/Laplace, 3/Student-t, 4/\gls{gmm}} {
 				\edef\tmp{%
 				\noexpand\nextgroupplot[title={\mytitle}, ylabel={\ifnum\component=1 Factor\fi}]%
 				\noexpand\addplot [name path=upper, draw=none] table [col sep=comma, x=x, y=upper] {\basepathhh_gibbs_1_\component.csv};%
 				\noexpand\addplot [name path=lower, draw=none] table [col sep=comma, x=x, y=lower] {\basepathhh_gibbs_1_\component.csv};%
 				\noexpand\addplot [maincolor, fill opacity=0.5] fill between[of=upper and lower];
 				\noexpand\addplot [line width=1pt, maincolor, mark=*, mark size=1pt, mark options={fill=white, draw=maincolor}] table[col sep=comma,x=x,y=y]{\basepathhh_gibbs_1_\component.csv};%
 				\noexpand\addplot [name path=upper, draw=none] table [col sep=comma, x=x, y=upper] {\basepathhh_mala_1_\component.csv};%
 				\noexpand\addplot [name path=lower, draw=none] table [col sep=comma, x=x, y=lower] {\basepathhh_mala_1_\component.csv};%
 				\noexpand\addplot [maincolor, fill opacity=0.5] fill between[of=upper and lower];
 				\noexpand\addplot [line width=1pt, maincolor, mark=*, mark size=1pt, mark options={fill=maincolor, draw=maincolor}] table[col sep=comma,x=x,y=y]{\basepathhh_mala_1_\component.csv};%
 			}\tmp}
 			\pgfplotsforeachungrouped \component in {1, ..., 4} {
 				\edef\tmp{%
 				\noexpand\nextgroupplot[ylabel={\ifnum\component=1 Product\fi}]%
 				\noexpand\addplot [name path=upper, draw=none] table [col sep=comma, x=x, y=upper] {\basepathhh_gibbs_2_\component.csv};%
 				\noexpand\addplot [name path=lower, draw=none] table [col sep=comma, x=x, y=lower] {\basepathhh_gibbs_2_\component.csv};%
 				\noexpand\addplot [maincolor, fill opacity=0.5] fill between[of=upper and lower];
 				\noexpand\addplot [line width=1pt, maincolor, mark=*, mark size=1pt, mark options={fill=white, draw=maincolor}] table[col sep=comma,x=x,y=y]{\basepathhh_gibbs_2_\component.csv};%
 				\noexpand\addplot [name path=upper, draw=none] table [col sep=comma, x=x, y=upper] {\basepathhh_mala_2_\component.csv};%
 				\noexpand\addplot [name path=lower, draw=none] table [col sep=comma, x=x, y=lower] {\basepathhh_mala_2_\component.csv};%
 				\noexpand\addplot [maincolor, fill opacity=0.5] fill between[of=upper and lower];
 				\noexpand\addplot [line width=1pt, maincolor, mark=*, mark size=1pt, mark options={fill=maincolor, draw=maincolor}] table[col sep=comma,x=x,y=y]{\basepathhh_mala_2_\component.csv};%
 			}\tmp}
 			\pgfplotsforeachungrouped \component in {1, ..., 4} {
 				\edef\tmp{%
 				\noexpand\nextgroupplot[ylabel={\ifnum\component=1 Loop\fi}]%
 				\noexpand\addplot [name path=upper, draw=none] table [col sep=comma, x=x, y=upper] {\basepathhh_gibbs_3_\component.csv};%
 				\noexpand\addplot [name path=lower, draw=none] table [col sep=comma, x=x, y=lower] {\basepathhh_gibbs_3_\component.csv};%
 				\noexpand\addplot [maincolor, fill opacity=0.5] fill between[of=upper and lower];
 				\noexpand\addplot [line width=1pt, maincolor, mark=*, mark size=1pt, mark options={fill=white, draw=maincolor}] table[col sep=comma,x=x,y=y]{\basepathhh_gibbs_3_\component.csv};%
 				\noexpand\addplot [name path=upper, draw=none] table [col sep=comma, x=x, y=upper] {\basepathhh_mala_3_\component.csv};%
 				\noexpand\addplot [name path=lower, draw=none] table [col sep=comma, x=x, y=lower] {\basepathhh_mala_3_\component.csv};%
 				\noexpand\addplot [maincolor, fill opacity=0.5] fill between[of=upper and lower];
 				\noexpand\addplot [line width=1pt, maincolor, mark=*, mark size=1pt, mark options={fill=maincolor, draw=maincolor}] table[col sep=comma,x=x,y=y]{\basepathhh_mala_3_\component.csv};%
 			}\tmp}
 			\pgfplotsforeachungrouped \component in {1, ..., 4} {
 				\edef\tmp{%
 				\noexpand\nextgroupplot[ylabel={\ifnum\component=1 Grid\fi}]%
 				\noexpand\addplot [name path=upper, draw=none] table [col sep=comma, x=x, y=upper] {\basepathhh_gibbs_inner_4_\component.csv};%
 				\noexpand\addplot [name path=lower, draw=none] table [col sep=comma, x=x, y=lower] {\basepathhh_gibbs_inner_4_\component.csv};%
 				\noexpand\addplot [maincolor, fill opacity=0.5] fill between[of=upper and lower];
 				\noexpand\addplot [line width=1pt, maincolor, mark=*, mark size=1pt, mark options={fill=white, draw=maincolor}] table[col sep=comma,x=x,y=y]{\basepathhh_gibbs_inner_4_\component.csv};%
 				\noexpand\addplot [name path=upper, draw=none] table [col sep=comma, x=x, y=upper] {\basepathhh_mala_inner_4_\component.csv};%
 				\noexpand\addplot [name path=lower, draw=none] table [col sep=comma, x=x, y=lower] {\basepathhh_mala_inner_4_\component.csv};%
 				\noexpand\addplot [maincolor, fill opacity=0.5] fill between[of=upper and lower];
 				\noexpand\addplot [line width=1pt, maincolor, mark=*, mark size=1pt, mark options={fill=maincolor, draw=maincolor}] table[col sep=comma,x=x,y=y]{\basepathhh_mala_inner_4_\component.csv};%
 				\noexpand\addplot [name path=upper, draw=none] table [col sep=comma, x=x, y=upper] {\basepathhh_gibbs_outer_4_\component.csv};%
 				\noexpand\addplot [name path=lower, draw=none] table [col sep=comma, x=x, y=lower] {\basepathhh_gibbs_outer_4_\component.csv};%
 				\noexpand\addplot [secondarycolor, fill opacity=0.5] fill between[of=upper and lower];
 				\noexpand\addplot [line width=1pt, secondarycolor, mark=*, mark size=1pt, mark options={fill=white, draw=secondarycolor}] table[col sep=comma,x=x,y=y]{\basepathhh_gibbs_outer_4_\component.csv};%
 				\noexpand\addplot [name path=upper, draw=none] table [col sep=comma, x=x, y=upper] {\basepathhh_mala_outer_4_\component.csv};%
 				\noexpand\addplot [name path=lower, draw=none] table [col sep=comma, x=x, y=lower] {\basepathhh_mala_outer_4_\component.csv};%
 				\noexpand\addplot [secondarycolor, fill opacity=0.5] fill between[of=upper and lower];
 				\noexpand\addplot [line width=1pt, secondarycolor, mark=*, mark size=1pt, mark options={fill=secondarycolor, draw=secondarycolor}] table[col sep=comma,x=x,y=y]{\basepathhh_mala_outer_4_\component.csv};%
 			}\tmp}
 		\end{groupplot}
 \end{tikzpicture}
\caption[]{%
\tikzexternaldisable
Autocorrelation functions of Gibbs (open circles, \protect\tikz[baseline=-\the\dimexpr\fontdimen22\textfont2\relax]\protect\draw [black, ultra thick] (0,0) circle (1mm);) and \gls{mala} (closed circles, \protect\tikz[baseline=-\the\dimexpr\fontdimen22\textfont2\relax]\protect\fill [black, ultra thick] (0,0) circle (1mm);) in the baseline experiments for \protect\tikz[baseline=-\the\dimexpr\fontdimen22\textfont2\relax]\protect\draw [maincolor, ultra thick] (0,0) -- (.5, 0); the factor and product topologies, the edge marginal for the loop topology, and the inner edge marginal for the grid topology and \protect\tikz[baseline=-\the\dimexpr\fontdimen22\textfont2\relax]\protect\draw [secondarycolor, ultra thick] (0,0) -- (.5, 0); the outer edge marginal for the grid topology.
The shaded areas around the means denote $\pm$ one standard deviation.
\tikzexternalenable
}%
\label{fig:baseline-acf}
\end{figure}

\begin{table}
    \centering
    \begin{subtable}{0.48\textwidth}
        \centering
        \resizebox{\textwidth}{!}{ % Resizing the table
        \begin{tabular}{lcccc}
        		\toprule
        		& Normal & Laplace & Student-t & \gls{gmm} \\
        		\midrule
        		Factor & $1.000 \pm 0.002$ & $1.000 \pm 0.004$ & $1.000 \pm 0.003$ & $1.000 \pm 0.006$ \\
        		Product & $1.000 \pm 0.002$ & $1.000 \pm 0.003$ & $1.000 \pm 0.002$ & $0.857 \pm 0.157$  \\
        		Loop & $1.000 \pm 0.001$ & $1.000 \pm 0.002$ & $1.000 \pm 0.001$ & $1.000 \pm 0.007$ \\
        		Grid (inner edge) & $1.000 \pm 0.001$ & $1.000 \pm 0.002$ & $1.000 \pm 0.001$ & $1.000 \pm 0.007$  \\
        		Grid (outer edge) & $1.000 \pm 0.002$ & $1.000 \pm 0.002$ & $1.000 \pm 0.002$ & $1.000 \pm 0.007$  \\
        		\bottomrule
    		\end{tabular}
        }
		\caption{Gibbs.}
    \end{subtable}
    \hfill
    \begin{subtable}{0.48\textwidth}
        \centering
        \resizebox{\textwidth}{!}{ % Resizing the table
    		\begin{tabular}{lcccc}
        		\toprule
        		& Normal & Laplace & Student-t & \gls{gmm} \\
        		\midrule
        		Factor & $0.737 \pm 0.036$ & $0.464 \pm 0.037$ & $0.578 \pm 0.082$ & $0.002 \pm 0.001$ \\
        		Product & $0.737 \pm 0.036$ & $0.453 \pm 0.036$ & $0.746 \pm 0.038$ & $0.388 \pm 0.275$  \\
        		Loop & $0.315 \pm 0.018$ & $0.111 \pm 0.015$ & $0.273 \pm 0.019$ & $0.002 \pm 0.001$ \\
        		Grid (inner edge) & $0.236 \pm 0.016$ & $0.096 \pm 0.013$ & $0.223 \pm 0.017$ & $0.002 \pm 0.001$  \\
        		Grid (outer edge) & $0.173 \pm 0.034$ & $0.064 \pm 0.013$ & $0.154 \pm 0.031$ & $0.002 \pm 0.001$  \\
        		\bottomrule
    		\end{tabular}
        }
		\caption{\gls{mala}.}
    \end{subtable}
    \caption{Mean $\pm$ standard deviation of the sampling efficiency $\gamma$ in the baseline experiments.}
    \label{tab:baseline_samp_eff}
\end{table}

\subsubsection{Sensitivity to Initial Condition}
\gls{mala} relies solely on local information about the distribution at each step, whereas Gibbs sampling leverages a global representation. As a result, \gls{mala} may perform poorly when initialized far away from the modes of the distribution. These regions might have gradients near zero, which forces \gls{mala} to rely on a Gaussian random walk to escape these regions, or they might have quite extreme gradient values, which would result in either high proposal rejection rates or require very small step sizes. Gibbs sampling, on the other hand, is agnostic to these problems since its global representation allows it to escape such regions in one step.  

To illustrate the dependence on the initial condition, we rerun the same baseline experiment using a Student-t factor and place the initial condition at $x_0 = c \cdot 1$, where $c > 0$ is a scalar that allows us to shift away the initial condition from the corresponding mode of the distribution. Due to the heavy-tailedness of the Student-t factor, the resulting baseline models will also be similarly heavy-tailed\footnote{Similarly to the variance decreasing effect, overcompleting probabilistic models also tends to decrease the degree of heavy-tailedness in comparison to the factors.}. Therefore, the more we shift away from the modes of the models, the more we place the initial condition in regions where gradients are near zero. The Wasserstein-1 distances over iterations for various choices of the scalar \( c \) such that the initial condition has a certain norm is shown in \cref{fig:baseline-init-w1}. The results clearly demonstrate that the convergence rate of \gls{mala} can be made arbitrarily poor by putting the initialization in a low gradient region, while Gibbs sampling remains practically unaffected.

% Initialization
\begin{figure}
\centering
\begin{tikzpicture}
	\def\basepathhh{./fig/data/init-wasserstein/initialization_wasserstein}
		\begin{groupplot}[marginal group plot,/tikz/line join=bevel]
			\pgfplotsforeachungrouped \component/\mytitle in {1/{\norm{x_0} = 1}, 2/{\norm{x_0} = 5}, 3/{\norm{x_0} = 10}, 4/{\norm{x_0} = 15}} {
				\edef\tmp{%
				\noexpand\nextgroupplot[title={\mytitle}, ylabel={\ifnum\component=1 Factor\fi}]%
				\noexpand\addplot [line width=1pt, maincolor] table[col sep=comma,x=x,y=y]{\basepathhh_gibbs_1_\component.csv};%
				\noexpand\addplot [line width=1pt, maincolor, dashed] table[col sep=comma,x=x,y=y]{\basepathhh_mala_1_\component.csv};%
			}\tmp}
			\pgfplotsforeachungrouped \component in {1, ..., 4} {
				\edef\tmp{%
				\noexpand\nextgroupplot[ylabel={\ifnum\component=1 Product\fi}]%
				\noexpand\addplot [line width=1pt, maincolor] table[col sep=comma,x=x,y=y]{\basepathhh_gibbs_2_\component.csv};%
				\noexpand\addplot [line width=1pt, maincolor, dashed] table[col sep=comma,x=x,y=y]{\basepathhh_mala_2_\component.csv};%
			}\tmp}
			\pgfplotsforeachungrouped \component in {1, ..., 4} {
				\edef\tmp{%
				\noexpand\nextgroupplot[ylabel={\ifnum\component=1 Loop\fi}]%
				\noexpand\addplot [line width=1pt, maincolor] table[col sep=comma,x=x,y=y]{\basepathhh_gibbs_3_\component.csv};%
				\noexpand\addplot [line width=1pt, maincolor, dashed] table[col sep=comma,x=x,y=y]{\basepathhh_mala_3_\component.csv};%
			}\tmp}
			\pgfplotsforeachungrouped \component in {1, ..., 4} {
				\edef\tmp{%
				\noexpand\nextgroupplot[ylabel={\ifnum\component=1 Grid\fi}]%
				\noexpand\addplot [line width=1pt, maincolor] table[col sep=comma,x=x,y=y]{\basepathhh_inner_gibbs_4_\component.csv};%
				\noexpand\addplot [line width=1pt, maincolor, dashed] table[col sep=comma,x=x,y=y]{\basepathhh_inner_mala_4_\component.csv};%
				\noexpand\addplot [line width=1pt, secondarycolor] table[col sep=comma,x=x,y=y]{\basepathhh_outer_gibbs_4_\component.csv};%
				\noexpand\addplot [line width=1pt, secondarycolor, dashed] table[col sep=comma,x=x,y=y]{\basepathhh_outer_mala_4_\component.csv};%
			}\tmp}
		\end{groupplot}
\end{tikzpicture}
\caption[]{%
\tikzexternaldisable
Effect of the initialization on the sampling performance:
$\log_{10}$ Wasserstein-1 distances of Gibbs (solid, \protect\tikz[baseline=-\the\dimexpr\fontdimen22\textfont2\relax]\protect\draw [black, ultra thick] (0,0) -- (.5, 0);) and \gls{mala} (dashed, \protect\tikz[baseline=-\the\dimexpr\fontdimen22\textfont2\relax]\protect\draw [black, ultra thick, dashed] (0,0) -- (.5, 0);) over iterations for the initialization baseline experiments for \protect\tikz[baseline=-\the\dimexpr\fontdimen22\textfont2\relax]\protect\draw [maincolor, ultra thick] (0,0) -- (.5, 0); the factor and product topologies, the edge marginal of the loop topology, and the inner edge marginal of the grid topology and \protect\tikz[baseline=-\the\dimexpr\fontdimen22\textfont2\relax]\protect\draw [secondarycolor, ultra thick] (0,0) -- (.5, 0); the outer edge marginal of the grid topology.
\tikzexternalenable
}%
\label{fig:baseline-init-w1}
\end{figure}

% Uniform approximation 
\begin{figure}
\centering
 \begin{tikzpicture}
 	\def\basepathhh{./fig/data/uniform-approx-ground-truth-cdfs-csvs/ground-truth-cdfs}
 		\begin{groupplot}[asmarginal group plot]
 			\pgfplotsforeachungrouped \component/\mytitle in {1/$L = 512$, 2/$L = 1024$, 3/$L = 1536$, 4/$L = 2048$} {
 				\edef\tmp{%
 				\noexpand\nextgroupplot[title={\mytitle}, xmin=-0.515, xmax=0.515, ymin=-0.03, ymax=1.03, ylabel={\ifnum\component=1 Factor\fi}]%
 				\noexpand\addplot [line width=1pt, maincolor] table[col sep=comma,x=x,y=y,each nth point=10]{\basepathhh_gibbs_1_\component.csv};%
 				\noexpand\addplot [line width=1pt, dashed, maincolor] table[col sep=comma,x=x,y=y,each nth point=100]{\basepathhh_mala_1_\component.csv};%
 			}\tmp}
 			\pgfplotsforeachungrouped \component in {1, ..., 4} {
 				\edef\tmp{%
 				\noexpand\nextgroupplot[xmin=-0.515, xmax=0.515, ymin=-0.03, ymax=1.03, ylabel={\ifnum\component=1 Product\fi}]%
 				\noexpand\addplot [line width=1pt, maincolor] table[col sep=comma,x=x,y=y,each nth point=10]{\basepathhh_gibbs_2_\component.csv};%
 				\noexpand\addplot [line width=1pt, dashed, maincolor] table[col sep=comma,x=x,y=y,each nth point=100]{\basepathhh_mala_2_\component.csv};%
 			}\tmp}
 			\pgfplotsforeachungrouped \component in {1, ..., 4} {
 				\edef\tmp{%
 				\noexpand\nextgroupplot[xmin=-0.515, xmax=0.515, ymin=-0.03, ymax=1.03, ylabel={\ifnum\component=1 Loop\fi}]%
 				\noexpand\addplot [line width=1pt, maincolor] table[col sep=comma,x=x,y=y,each nth point=10]{\basepathhh_gibbs_3_\component.csv};%
 				\noexpand\addplot [line width=1pt, dashed, maincolor] table[col sep=comma,x=x,y=y,each nth point=100]{\basepathhh_mala_3_\component.csv};%
 			}\tmp}
 			\pgfplotsforeachungrouped \component in {1, ..., 4} {
 				\edef\tmp{%
 				\noexpand\nextgroupplot[xmin=-0.515, xmax=0.515, ymin=-0.03, ymax=1.03, ylabel={\ifnum\component=1 Grid\fi}]%
 				\noexpand\addplot [line width=1pt, maincolor] table[col sep=comma,x=x,y=y,each nth point=10]{\basepathhh_gibbs_inner_4_\component.csv};%
 				\noexpand\addplot [line width=1pt, dashed, maincolor] table[col sep=comma,x=x,y=y,each nth point=100]{\basepathhh_mala_inner_4_\component.csv};%
 				\noexpand\addplot [line width=1pt, secondarycolor] table[col sep=comma,x=x,y=y,each nth point=10]{\basepathhh_gibbs_outer_4_\component.csv};%
 				\noexpand\addplot [line width=1pt, dashed, secondarycolor] table[col sep=comma,x=x,y=y,each nth point=100]{\basepathhh_mala_outer_4_\component.csv};%
 			}\tmp}
 		\end{groupplot}
 \end{tikzpicture}
\caption[]{%
\tikzexternaldisable
Effect of the number of components \( L \) when approximating a factor with a Gaussian mixture:
The reference marginal CDF (dashed, \protect\tikz[baseline=-\the\dimexpr\fontdimen22\textfont2\relax]\protect\draw [black, ultra thick, dashed] (0,0) -- (.5, 0);) compared to the empirical marginal CDF after \num{15000} iterations of Gibbs (solid, \protect\tikz[baseline=-\the\dimexpr\fontdimen22\textfont2\relax]\protect\draw [black, ultra thick] (0,0) -- (.5, 0);) for \protect\tikz[baseline=-\the\dimexpr\fontdimen22\textfont2\relax]\protect\draw [maincolor, ultra thick] (0,0) -- (.5, 0); the factor and product topologies, the edge marginal of the loop topology, and the inner edge marginal of the grid topology and \protect\tikz[baseline=-\the\dimexpr\fontdimen22\textfont2\relax]\protect\draw [secondarycolor, ultra thick] (0,0) -- (.5, 0); the outer edge marginal for the grid topology. 
\tikzexternalenable
}%
\label{baseline-uniform-approx-cdf}
\end{figure}

\subsubsection{Sensitivity to \texorpdfstring{\gls{gmm}}{GMM} Parametrization}
The proposed Gibbs sampling approach can, in principle, be applied to any type of continuous factor by obtaining suitable \gls{gmm} approximations. However, the specific \gls{gmm} parameterization can significantly affect the convergence rate of the Gibbs sampler. 

To illustrate this, we conducted an experiment where we approximated a Laplace factor using two different \gls{gmm} parametrizations. In the first parametrization, the means of the \gls{gmm} are uniformly spaced on the interval $[-0.5, 0.5]$, with each Gaussian having the same variance,  $\sigma^2 = 1 / \br{L - 1}$, where $L$ denotes the number of components used in the approximation. To approximate a Laplace distribution with parameter $b$,  the weights are set as $w_i \propto \exp\br*{-\abs{\mu_i} / b}$ for $i \in \set{1, \ldots, L}$.  In the second parametrization,  we directly discretize the GSM representation of the Laplace distribution, resulting in a \gls{gmm} where all components have zero mean but vary in weights and variances. If the number of components in the approximations is sufficiently large, then increasing the number of components leads practically to the same approximate factor. Therefore, ideally, the convergence rate of a sampler should remain unaffected by the number of components.

The Gibbs sampler showed significantly different behaviors when applied to the \gls{gmm} and the \gls{gsm} parametrizations.
In fact, in the former it failed to converge to the stationary distribution after \num{15000} iterations for all considered number of components and topologies.
Therefore, for the \gls{gmm} parametrization we illustrate the influence of the number of components on the performance of the Gibbs sampler by comparing the empirical marginal CDF after \num{15000} iterations of Gibbs to the reference marginal CDF (which is synonymous to the empirical marginal CDF after \num{15000} iterations of \gls{mala}, since \gls{mala} converged for all considered number of components and topologies) in \Cref{baseline-uniform-approx-cdf}.
The results show that increasing the number of components in the \gls{gmm} parametrization considerably slows down the convergence rate of Gibbs.
In particular, increasing the number of components compresses the value range of the empirical marginals computed from the Gibbs samples.
This is because the Gaussian components of the parametrization become narrower by increasing the number of components in the approximation, which limits the exploration rate of the Gibbs sampler.
In contrast, \gls{mala} relies on gradient information, which is practically the same when increasing the number of components. 

For the \gls{gsm} representation, both methods converged and we show the Wasserstein-1 distance over the iterations and the autocorrelation functions in \Cref{baseline-gsm-approx-w1,baseline-gsm-approx-acf}.
In these experiments, both Gibbs and \gls{mala} are practically unaffected by the increase in number of components.
Informally speaking, this suggests that we would want to obtain \gls{gmm} parametrizations whose components have the largest variance values possible in order to allow the Gibbs sampler to efficiently explore the space.
However, obtaining a precise notion of what a good parametrization entails is beyond the scope of this work.

We included this set of experiments to make readers aware that this phenomenon exists. However, we believe it is highly unlikely to occur in practice, as it requires deliberate efforts to construct such poorly behaved \gls{gmm} representations. From our experience, representations that allow for varying weights and variances across the \gls{gmm} components lead to well-behaved representations. Furthermore, this phenomenon is trivial to diagnose by checking if the univariate conditional latent distributions become extremely narrow relative to their support, and can be easily mitigated through serial or parallel tempering \cite{Geyer:2011} on the variances of the univariate \glspl{gmm}. 

% Approximation GSM
\begin{figure}
\centering
\begin{tikzpicture}
	\def\basepathhh{./fig/data/gsm-apx/approximation-gsm_wasserstein}
		\begin{groupplot}[marginal group plot,/tikz/line join=bevel]
			\pgfplotsforeachungrouped \component/\mytitle in {1/$L = 256$, 2/$L = 512$, 3/$L = 1024$, 4/$L = 2048$} {
				\edef\tmp{%
				\noexpand\nextgroupplot[title={\mytitle}, ylabel={\ifnum\component=1 Factor\fi}]%
				\noexpand\addplot [line width=1pt, maincolor] table[col sep=comma,x=x,y=y]{\basepathhh_gibbs_1_\component.csv};%
				\noexpand\addplot [line width=1pt, maincolor, dashed] table[col sep=comma,x=x,y=y]{\basepathhh_mala_1_\component.csv};%
			}\tmp}
			\pgfplotsforeachungrouped \component in {1, ..., 4} {
				\edef\tmp{%
				\noexpand\nextgroupplot[ylabel={\ifnum\component=1 Product\fi}]%
				\noexpand\addplot [line width=1pt, maincolor] table[col sep=comma,x=x,y=y]{\basepathhh_gibbs_2_\component.csv};%
				\noexpand\addplot [line width=1pt, maincolor, dashed] table[col sep=comma,x=x,y=y]{\basepathhh_mala_2_\component.csv};%
			}\tmp}
			\pgfplotsforeachungrouped \component in {1, ..., 4} {
				\edef\tmp{%
				\noexpand\nextgroupplot[ylabel={\ifnum\component=1 Loop\fi}]%
				\noexpand\addplot [line width=1pt, maincolor] table[col sep=comma,x=x,y=y]{\basepathhh_gibbs_3_\component.csv};%
				\noexpand\addplot [line width=1pt, maincolor, dashed] table[col sep=comma,x=x,y=y]{\basepathhh_mala_3_\component.csv};%
			}\tmp}
			\pgfplotsforeachungrouped \component in {1, ..., 4} {
				\edef\tmp{%
				\noexpand\nextgroupplot[ylabel={\ifnum\component=1 Grid\fi}]%
				\noexpand\addplot [line width=1pt, maincolor] table[col sep=comma,x=x,y=y]{\basepathhh_inner_gibbs_4_\component.csv};%
				\noexpand\addplot [line width=1pt, maincolor, dashed] table[col sep=comma,x=x,y=y]{\basepathhh_inner_mala_4_\component.csv};%
				\noexpand\addplot [line width=1pt, secondarycolor] table[col sep=comma,x=x,y=y]{\basepathhh_outer_gibbs_4_\component.csv};%
				\noexpand\addplot [line width=1pt, secondarycolor, dashed] table[col sep=comma,x=x,y=y]{\basepathhh_outer_mala_4_\component.csv};%
			}\tmp}
		\end{groupplot}
\end{tikzpicture}
\caption[]{%
\tikzexternaldisable
$\log_{10}$ Wasserstein-1 distances of Gibbs (solid, \protect\tikz[baseline=-\the\dimexpr\fontdimen22\textfont2\relax]\protect\draw [black, ultra thick] (0,0) -- (.5, 0);) and \gls{mala} (dashed, \protect\tikz[baseline=-\the\dimexpr\fontdimen22\textfont2\relax]\protect\draw [black, ultra thick, dashed] (0,0) -- (.5, 0);) over iterations for the \( L \)-component GSM approximation baseline experiments for \protect\tikz[baseline=-\the\dimexpr\fontdimen22\textfont2\relax]\protect\draw [maincolor, ultra thick] (0,0) -- (.5, 0); the factor and product topologies, the edge marginal of the loop topology, and the inner edge marginal of the grid topology and \protect\tikz[baseline=-\the\dimexpr\fontdimen22\textfont2\relax]\protect\draw [secondarycolor, ultra thick] (0,0) -- (.5, 0); the outer edge marginal of the grid topology.
\tikzexternalenable
}%
\label{baseline-gsm-approx-w1}
\end{figure}

\begin{figure}
\centering
\begin{tikzpicture}
	\def\basepathhh{./fig/data/gsmacf/acf}
			\begin{groupplot}[marginal group plot, xmin=1, xmax=30, yticklabel style={/pgf/number format/fixed,/pgf/number format/precision=3}]
			\pgfplotsforeachungrouped \component/\mytitle in {1/$L=256$, 2/$L=512$, 3/$L=2024$, 4/$L=2048$} {
				\edef\tmp{%
				\noexpand\nextgroupplot[title={\mytitle}, ylabel={\ifnum\component=1 Factor\fi}]%
				\noexpand\addplot [name path=upper, draw=none] table [col sep=comma, x=x, y=upper] {\basepathhh_gibbs_1_\component.csv};%
				\noexpand\addplot [name path=lower, draw=none] table [col sep=comma, x=x, y=lower] {\basepathhh_gibbs_1_\component.csv};%
				\noexpand\addplot [maincolor, fill opacity=0.5] fill between[of=upper and lower];
				\noexpand\addplot [line width=1pt, maincolor, mark=*, mark size=1pt, mark options={fill=white, draw=maincolor}] table[col sep=comma,x=x,y=y]{\basepathhh_gibbs_1_\component.csv};%
				\noexpand\addplot [name path=upper, draw=none] table [col sep=comma, x=x, y=upper] {\basepathhh_mala_1_\component.csv};%
				\noexpand\addplot [name path=lower, draw=none] table [col sep=comma, x=x, y=lower] {\basepathhh_mala_1_\component.csv};%
				\noexpand\addplot [maincolor, fill opacity=0.5] fill between[of=upper and lower];
				\noexpand\addplot [line width=1pt, maincolor, mark=*, mark size=1pt, mark options={fill=maincolor, draw=maincolor}] table[col sep=comma,x=x,y=y]{\basepathhh_mala_1_\component.csv};%
			}\tmp}
			\pgfplotsforeachungrouped \component in {1, ..., 4} {
				\edef\tmp{%
				\noexpand\nextgroupplot[ylabel={\ifnum\component=1 Product\fi}]%
				\noexpand\addplot [name path=upper, draw=none] table [col sep=comma, x=x, y=upper] {\basepathhh_gibbs_2_\component.csv};%
				\noexpand\addplot [name path=lower, draw=none] table [col sep=comma, x=x, y=lower] {\basepathhh_gibbs_2_\component.csv};%
				\noexpand\addplot [maincolor, fill opacity=0.5] fill between[of=upper and lower];
				\noexpand\addplot [line width=1pt, maincolor, mark=*, mark size=1pt, mark options={fill=white, draw=maincolor}] table[col sep=comma,x=x,y=y]{\basepathhh_gibbs_2_\component.csv};%
				\noexpand\addplot [name path=upper, draw=none] table [col sep=comma, x=x, y=upper] {\basepathhh_mala_2_\component.csv};%
				\noexpand\addplot [name path=lower, draw=none] table [col sep=comma, x=x, y=lower] {\basepathhh_mala_2_\component.csv};%
				\noexpand\addplot [maincolor, fill opacity=0.5] fill between[of=upper and lower];
				\noexpand\addplot [line width=1pt, maincolor, mark=*, mark size=1pt, mark options={fill=maincolor, draw=maincolor}] table[col sep=comma,x=x,y=y]{\basepathhh_mala_2_\component.csv};%
			}\tmp}
			\pgfplotsforeachungrouped \component in {1, ..., 4} {
				\edef\tmp{%
				\noexpand\nextgroupplot[ylabel={\ifnum\component=1 Loop\fi}]%
				\noexpand\addplot [name path=upper, draw=none] table [col sep=comma, x=x, y=upper] {\basepathhh_gibbs_3_\component.csv};%
				\noexpand\addplot [name path=lower, draw=none] table [col sep=comma, x=x, y=lower] {\basepathhh_gibbs_3_\component.csv};%
				\noexpand\addplot [maincolor, fill opacity=0.5] fill between[of=upper and lower];
				\noexpand\addplot [line width=1pt, maincolor, mark=*, mark size=1pt, mark options={fill=white, draw=maincolor}] table[col sep=comma,x=x,y=y]{\basepathhh_gibbs_3_\component.csv};%
				\noexpand\addplot [name path=upper, draw=none] table [col sep=comma, x=x, y=upper] {\basepathhh_mala_3_\component.csv};%
				\noexpand\addplot [name path=lower, draw=none] table [col sep=comma, x=x, y=lower] {\basepathhh_mala_3_\component.csv};%
				\noexpand\addplot [maincolor, fill opacity=0.5] fill between[of=upper and lower];
				\noexpand\addplot [line width=1pt, maincolor, mark=*, mark size=1pt, mark options={fill=maincolor, draw=maincolor}] table[col sep=comma,x=x,y=y]{\basepathhh_mala_3_\component.csv};%
			}\tmp}
			\pgfplotsforeachungrouped \component in {1, ..., 4} {
				\edef\tmp{%
				\noexpand\nextgroupplot[ylabel={\ifnum\component=1 Grid\fi}]%
				\noexpand\addplot [name path=upper, draw=none] table [col sep=comma, x=x, y=upper] {\basepathhh_gibbs_inner_4_\component.csv};%
				\noexpand\addplot [name path=lower, draw=none] table [col sep=comma, x=x, y=lower] {\basepathhh_gibbs_inner_4_\component.csv};%
				\noexpand\addplot [maincolor, fill opacity=0.5] fill between[of=upper and lower];
				\noexpand\addplot [line width=1pt, maincolor, mark=*, mark size=1pt, mark options={fill=white, draw=maincolor}] table[col sep=comma,x=x,y=y]{\basepathhh_gibbs_inner_4_\component.csv};%
				\noexpand\addplot [name path=upper, draw=none] table [col sep=comma, x=x, y=upper] {\basepathhh_mala_inner_4_\component.csv};%
				\noexpand\addplot [name path=lower, draw=none] table [col sep=comma, x=x, y=lower] {\basepathhh_mala_inner_4_\component.csv};%
				\noexpand\addplot [maincolor, fill opacity=0.5] fill between[of=upper and lower];
				\noexpand\addplot [line width=1pt, maincolor, mark=*, mark size=1pt, mark options={fill=maincolor, draw=maincolor}] table[col sep=comma,x=x,y=y]{\basepathhh_mala_inner_4_\component.csv};%
				\noexpand\addplot [name path=upper, draw=none] table [col sep=comma, x=x, y=upper] {\basepathhh_gibbs_outer_4_\component.csv};%
				\noexpand\addplot [name path=lower, draw=none] table [col sep=comma, x=x, y=lower] {\basepathhh_gibbs_outer_4_\component.csv};%
				\noexpand\addplot [secondarycolor, fill opacity=0.5] fill between[of=upper and lower];
				\noexpand\addplot [line width=1pt, secondarycolor, mark=*, mark size=1pt, mark options={fill=white, draw=secondarycolor}] table[col sep=comma,x=x,y=y]{\basepathhh_gibbs_outer_4_\component.csv};%
				\noexpand\addplot [name path=upper, draw=none] table [col sep=comma, x=x, y=upper] {\basepathhh_mala_outer_4_\component.csv};%
				\noexpand\addplot [name path=lower, draw=none] table [col sep=comma, x=x, y=lower] {\basepathhh_mala_outer_4_\component.csv};%
				\noexpand\addplot [secondarycolor, fill opacity=0.5] fill between[of=upper and lower];
				\noexpand\addplot [line width=1pt, secondarycolor, mark=*, mark size=1pt, mark options={fill=secondarycolor, draw=secondarycolor}] table[col sep=comma,x=x,y=y]{\basepathhh_mala_outer_4_\component.csv};%
			}\tmp}
		\end{groupplot}
\end{tikzpicture}
\caption[]{%
\tikzexternaldisable
Autocorrelation functions of Gibbs (open circles, \protect\tikz[baseline=-\the\dimexpr\fontdimen22\textfont2\relax]\protect\draw [black, ultra thick] (0,0) circle (1mm);) and \gls{mala} (closed circles, \protect\tikz[baseline=-\the\dimexpr\fontdimen22\textfont2\relax]\protect\draw [black, ultra thick, fill] (0,0) circle (1mm);) when approximating a Laplace factor with an \( L \)-component GSM for \protect\tikz[baseline=-\the\dimexpr\fontdimen22\textfont2\relax]\protect\draw [maincolor, ultra thick] (0,0) -- (.5, 0); the factor and product topologies, the edge marginal of the loop topology, and the inner edge marginal of the grid topology and \protect\tikz[baseline=-\the\dimexpr\fontdimen22\textfont2\relax]\protect\draw [secondarycolor, ultra thick] (0,0) -- (.5, 0); the outer edge marginal of the grid topology. 
The shaded areas around the means denote $\pm$ one standard deviation.
\tikzexternalenable
}%
\label{baseline-gsm-approx-acf}
\end{figure}

\subsubsection{Image Prior Sampling Experiments}
\label{sec:image_prior_sampling}
We now turn to higher-dimensional problems to evaluate how our proposed approach scales with problem size. To that end, we consider the task of sampling from improper image priors of the form
\begin{align}
	\fx{x} \propto \prod_{i = 1}^{2n} \phi\br*{\br{Kx}_i},
	\label{def_image_prior}
\end{align}
where $x \in \R^n$, $K \in \R^{2n \times n}$ and $\phi: \R \to \R$ is an arbitrary univariate distribution. The linear operator $K$ is a finite difference operator with circular boundary conditions\footnote{Circular boundary conditions were chosen deliberately, as they induce translation invariant edge marginals. We do not, however, prove this fact, since it is not of central importance for the proposed sampling approach.} that computes the horizontal and vertical image gradients. The problem size is the total number of pixels $n$. Note that \eqref{def_image_prior} includes the improper prior counterpart of the standard anisotropic total variational regularizer \cite{Rudin:1992} with circular boundary conditions and its extensions from Laplace factors to other types of factors. 

For sampling purposes, we again, as per \Cref{prop:smart_tie_breaking_works_too}, introduce a factor that acts on the mean of the entries to obtain proper distributions. To quantify how well our proposed approach scales with problem size, we consider images of sizes $12 \times 12$, $24 \times 24$, $48 \times 48$, and $96 \times 96$ (\ie, $n \in \set{12^2, 24^2, 48^2, 96^2}$). We use $\num{15000}$ iterations for Gibbs and $\num{3000000}$ iterations for \gls{mala} to accommodate for the fact that a single iteration of Gibbs is more expensive than a single iteration of \gls{mala}. This, however, forces us to only run $1000$ chains in parallel due to the prohibitive memory demands that arise from storing the \gls{mala} iterations. For the sake of simplicity, we again consider the same four factor types as in the baseline experiments. Note that some of the corresponding image prior sampling experiments are quite difficult sampling problems, since the target distributions are nonsmooth, heavy-tailed, or non-log-concave. These difficulties are further exacerbated with increase in image size. 

We choose the inner product $\ip{v}{x}$ as representative marginal, where $x$ denotes a sample and $v$ denotes the eigenvector of $K^\top K$ associated with the second smallest eigenvalue of $K^\top K$. This particular choice 
is an approximation of the direction along which the improper prior \eqref{def_image_prior} has the highest variance since the eigenvector associated with the smallest eigenvalue of $K^\top K$ would correspond to the kernel. It is a sensible choice of approximate worst-case direction and marginal in our setting due to the symmetric structure of our improper prior. Therefore, our evaluation protocol is based on the sliced Wasserstein-1 distance along this estimated worst-case direction $v$. Since \gls{mala} exhibited extremely slow convergence rates in these experiments,  we only checked that \gls{mala} converged within the last $7500$ iterations to the representative stationary distribution and based our autocorrelation function and sampling efficiency computations only on these last $7500$ iterations of each chain.

The Wasserstein-1 distances over time in seconds for the image prior sampling experiments in \Cref{fig:prior-w1} show that the Gibbs sampler converged to the representative ground-truth distribution within at most $200$ seconds,  while \gls{mala} converged extremely slowly in all test cases. \Cref{fig:prior-w1} is missing the Wasserstein-1 distances over time of \gls{mala} for priors of size $48 \times 48$ and $96 \times 96$ with Laplace and \gls{gmm} factors, since \gls{mala} either did not converge in these cases within reasonable time or was computationally too expensive to run.  In particular,  we did not run \gls{mala} for priors of size $96 \times 96$ with \gls{gmm} factors since it would require more than $640$ hours of runtime. The corresponding runtimes of the image prior sampling experiments are given in \Cref{tab:prior_runtimes}. These correspond to the time to run $1000$ parallel chains for $\num{15000}$ Gibbs iterations and $\num{3000000}$ MALA iterations. To reiterate, these do not reflect the time required for Gibbs to converge, and the extremely fast convergence times for Gibbs in each of the prior sampling experiments can be read off from \Cref{fig:prior-w1}. MALA on the other hand did not even reach convergence in some of the experiments, despite the excessive runtimes reported in  \Cref{tab:prior_runtimes}.

\Cref{fig:prior-acf} shows the autocorrelation functions in the image prior sampling experiments,  while \Cref{tab:prior_samp_eff} shows the corresponding sampling efficiencies. \gls{mala} results for priors of size $48 \times 48$ and $96 \times 96$ with Laplace and \gls{gmm} factors are again not shown since \gls{mala} either did not converge to the representative stationary distribution or was computationally too expensive to compute.  Akin to the baseline experiments,  Gibbs produces practically independent samples,  while \gls{mala} produces extremely correlated samples.  

% Priors
\begin{figure}
\centering
\begin{tikzpicture}
	\def\basepathhh{./fig/data/wprior/prior_wasserstein}
		\begin{groupplot}[wasserstein group plot, enlarge x limits={value=0.0075,lower}, enlarge y limits={value=0.0075,upper}]
			\pgfplotsforeachungrouped \component/\mytitle in {1/Normal, 2/Laplace, 3/Student-t, 4/\gls{gmm}} {
				\edef\tmp{%
				%  \else\ifnum\component=3
				\noexpand\nextgroupplot[title={\mytitle}, ylabel={\ifnum\component=1 $12 \times 12$\fi}, xmax={\ifnum\component=1 0.4\else\ifnum\component=2 2\else\ifnum\component=3 0.5\else\ifnum\component=4 2\fi\fi\fi\fi}]%
				\noexpand\addplot [line width=1pt, maincolor] table[col sep=comma,x=x,y=y]{\basepathhh_gibbs_1_\component.csv};%
				\noexpand\addplot [line width=1pt, maincolor, dashed] table[col sep=comma,x=x,y=y]{\basepathhh_mala_1_\component.csv};%
			}\tmp}
			\pgfplotsforeachungrouped \component in {1, ..., 4} {
				\edef\tmp{%
				\noexpand\nextgroupplot[ylabel={\ifnum\component=1 $24 \times 24$\fi}, xmax={\ifnum\component=1 0.4\else\ifnum\component=2 4\else\ifnum\component=3 1\else\ifnum\component=4 10\fi\fi\fi\fi}]%
				\noexpand\addplot [line width=1pt, maincolor] table[col sep=comma,x=x,y=y]{\basepathhh_gibbs_2_\component.csv};%
				\noexpand\addplot [line width=1pt, maincolor, dashed] table[col sep=comma,x=x,y=y]{\basepathhh_mala_2_\component.csv};%
			}\tmp}
			\pgfplotsforeachungrouped \component in {1, ..., 4} {
				\edef\tmp{%
				\noexpand\nextgroupplot[ylabel={\ifnum\component=1 $48 \times 48$\fi}, , xmax={\ifnum\component=1 4\else\ifnum\component=2 30\else\ifnum\component=3 4\else\ifnum\component=4 45\fi\fi\fi\fi}]%
				\noexpand\addplot [line width=1pt, maincolor] table[col sep=comma,x=x,y=y]{\basepathhh_gibbs_3_\component.csv};%
				\ifnum\component=1\relax
    				\noexpand\addplot [line width=1pt, maincolor, dashed] table[col sep=comma,x=x,y=y]{\basepathhh_mala_3_\component.csv};%				
    				\fi
    				\ifnum\component=3\relax
    				\noexpand\addplot [line width=1pt, maincolor, dashed] table[col sep=comma,x=x,y=y]{\basepathhh_mala_3_\component.csv};%				
    				\fi
			}\tmp}
			\pgfplotsforeachungrouped \component in {1, ..., 4} {
				\edef\tmp{%
				\noexpand\nextgroupplot[ylabel={\ifnum\component=1 $96 \times 96$\fi}, xmax={\ifnum\component=1 40\else\ifnum\component=2 250\else\ifnum\component=3 40\else\ifnum\component=4 500\fi\fi\fi\fi}]%
				\noexpand\addplot [line width=1pt, maincolor] table[col sep=comma,x=x,y=y]{\basepathhh_gibbs_4_\component.csv};%
				\ifnum\component=1\relax
    				\noexpand\addplot [line width=1pt, maincolor, dashed] table[col sep=comma,x=x,y=y]{\basepathhh_mala_4_\component.csv};%				
    				\fi
    				\ifnum\component=3\relax
    				\noexpand\addplot [line width=1pt, maincolor, dashed] table[col sep=comma,x=x,y=y]{\basepathhh_mala_4_\component.csv};%				
    				\fi
			}\tmp}
		\end{groupplot}
\end{tikzpicture}
\caption[]{%
\tikzexternaldisable
$\log_{10}$ Wasserstein-1 distances of \protect\tikz[baseline=-\the\dimexpr\fontdimen22\textfont2\relax]\protect\draw [maincolor, ultra thick] (0,0) -- (.5, 0); Gibbs and \protect\tikz[baseline=-\the\dimexpr\fontdimen22\textfont2\relax]\protect\draw [maincolor, ultra thick, dashed] (0,0) -- (.5, 0); \gls{mala} over time in seconds for the image prior sampling experiments. 
\tikzexternalenable
}%
\label{fig:prior-w1}
\end{figure}

\begin{figure}
\centering
\begin{tikzpicture}
	\def\basepathhh{./fig/data/pacf/acf}
		\begin{groupplot}[asmarginal group plot, xmin=1, xmax=500, ymin=0, ymax=1, xtick={250,500}, yticklabel style={/pgf/number format/fixed,/pgf/number format/precision=3}]
			\pgfplotsforeachungrouped \component/\mytitle in {1/Normal, 2/Laplace, 3/Student-t, 4/\gls{gmm}} {
				\edef\tmp{%
				\noexpand\nextgroupplot[title={\mytitle}, ylabel={\ifnum\component=1 $12 \times 12$\fi}]%
				\noexpand\addplot [name path=upper, draw=none, each nth point=20] table [col sep=comma, x=x, y=upper] {\basepathhh_gibbs_1_\component.csv};%
				\noexpand\addplot [name path=lower, draw=none, each nth point=20] table [col sep=comma, x=x, y=lower] {\basepathhh_gibbs_1_\component.csv};%
				\noexpand\addplot [maincolor, fill opacity=0.5] fill between[of=upper and lower];
				\noexpand\addplot [line width=1pt, maincolor, mark=*, mark size=1pt, mark options={fill=white, draw=maincolor}, each nth point=20] table[col sep=comma,x=x,y=y]{\basepathhh_gibbs_1_\component.csv};%
				\noexpand\addplot [name path=upper, draw=none, each nth point=20] table [col sep=comma, x=x, y=upper] {\basepathhh_mala_1_\component.csv};%
				\noexpand\addplot [name path=lower, draw=none, each nth point=20] table [col sep=comma, x=x, y=lower] {\basepathhh_mala_1_\component.csv};%
				\noexpand\addplot [maincolor, fill opacity=0.5] fill between[of=upper and lower];
				\noexpand\addplot [line width=1pt, maincolor, mark=*, mark size=1pt, mark options={fill=maincolor, draw=maincolor}, each nth point=20] table[col sep=comma,x=x,y=y]{\basepathhh_mala_1_\component.csv};%
			}\tmp}
			\pgfplotsforeachungrouped \component in {1, ..., 4} {
				\edef\tmp{%
				\noexpand\nextgroupplot[ylabel={\ifnum\component=1 $24 \times 24$\fi}]%
				\noexpand\addplot [name path=upper, draw=none, each nth point=20] table [col sep=comma, x=x, y=upper] {\basepathhh_gibbs_2_\component.csv};%
				\noexpand\addplot [name path=lower, draw=none, each nth point=20] table [col sep=comma, x=x, y=lower] {\basepathhh_gibbs_2_\component.csv};%
				\noexpand\addplot [maincolor, fill opacity=0.5] fill between[of=upper and lower];
				\noexpand\addplot [line width=1pt, maincolor, mark=*, mark size=1pt, mark options={fill=white, draw=maincolor}, each nth point=20] table[col sep=comma,x=x,y=y]{\basepathhh_gibbs_2_\component.csv};%
				\noexpand\addplot [name path=upper, draw=none, each nth point=20] table [col sep=comma, x=x, y=upper] {\basepathhh_mala_2_\component.csv};%
				\noexpand\addplot [name path=lower, draw=none, each nth point=20] table [col sep=comma, x=x, y=lower] {\basepathhh_mala_2_\component.csv};%
				\noexpand\addplot [maincolor, fill opacity=0.5] fill between[of=upper and lower];
				\noexpand\addplot [line width=1pt, maincolor, mark=*, mark size=1pt, mark options={fill=maincolor, draw=maincolor}, each nth point=20] table[col sep=comma,x=x,y=y]{\basepathhh_mala_2_\component.csv};%
			}\tmp}
			\pgfplotsforeachungrouped \component in {1, ..., 4} {
				\edef\tmp{%
				\noexpand\nextgroupplot[ylabel={\ifnum\component=1 $48 \times 48$\fi}]%
				\noexpand\addplot [name path=upper, draw=none, each nth point=20] table [col sep=comma, x=x, y=upper] {\basepathhh_gibbs_3_\component.csv};%
				\noexpand\addplot [name path=lower, draw=none, each nth point=20] table [col sep=comma, x=x, y=lower] {\basepathhh_gibbs_3_\component.csv};%
				\noexpand\addplot [maincolor, fill opacity=0.5] fill between[of=upper and lower];
				\noexpand\addplot [line width=1pt, maincolor, mark=*, mark size=1pt, mark options={fill=white, draw=maincolor}, each nth point=20] table[col sep=comma,x=x,y=y]{\basepathhh_gibbs_3_\component.csv};%
				\ifnum\component=1\relax
				\noexpand\addplot [name path=upper, draw=none, each nth point=20] table [col sep=comma, x=x, y=upper] {\basepathhh_mala_3_\component.csv};%
				\noexpand\addplot [name path=lower, draw=none, each nth point=20] table [col sep=comma, x=x, y=lower] {\basepathhh_mala_3_\component.csv};%
				\noexpand\addplot [maincolor, fill opacity=0.5] fill between[of=upper and lower];
				\noexpand\addplot [line width=1pt, maincolor, mark=*, mark size=1pt, mark options={fill=maincolor, draw=maincolor}, each nth point=20] table[col sep=comma,x=x,y=y]{\basepathhh_mala_3_\component.csv};%
				\fi
				\ifnum\component=3\relax
				\noexpand\addplot [name path=upper, draw=none, each nth point=20] table [col sep=comma, x=x, y=upper] {\basepathhh_mala_3_\component.csv};%
				\noexpand\addplot [name path=lower, draw=none, each nth point=20] table [col sep=comma, x=x, y=lower] {\basepathhh_mala_3_\component.csv};%
				\noexpand\addplot [maincolor, fill opacity=0.5, each nth point=20] fill between[of=upper and lower];
				\noexpand\addplot [line width=1pt, maincolor, mark=*, mark size=1pt, mark options={fill=maincolor, draw=maincolor}, each nth point=20] table[col sep=comma,x=x,y=y]{\basepathhh_mala_3_\component.csv};%
				\fi
			}\tmp}
			\pgfplotsforeachungrouped \component in {1, ..., 4} {
				\edef\tmp{%
				\noexpand\nextgroupplot[ylabel={\ifnum\component=1 $96 \times 96$\fi}]%
				\noexpand\addplot [name path=upper, draw=none, each nth point=20] table [col sep=comma, x=x, y=upper] {\basepathhh_gibbs_4_\component.csv};%
				\noexpand\addplot [name path=lower, draw=none, each nth point=20] table [col sep=comma, x=x, y=lower] {\basepathhh_gibbs_4_\component.csv};%
				\noexpand\addplot [maincolor, fill opacity=0.5] fill between[of=upper and lower];
				\noexpand\addplot [line width=1pt, maincolor, mark=*, mark size=1pt, mark options={fill=white, draw=maincolor}, each nth point=20] table[col sep=comma,x=x,y=y]{\basepathhh_gibbs_4_\component.csv};%
				\ifnum\component=1\relax				
				\noexpand\addplot [name path=upper, draw=none, each nth point=20] table [col sep=comma, x=x, y=upper] {\basepathhh_mala_4_\component.csv};%
				\noexpand\addplot [name path=lower, draw=none, each nth point=20] table [col sep=comma, x=x, y=lower] {\basepathhh_mala_4_\component.csv};%
				\noexpand\addplot [maincolor, fill opacity=0.5] fill between[of=upper and lower];
				\noexpand\addplot [line width=1pt, maincolor, mark=*, mark size=1pt, mark options={fill=maincolor, draw=maincolor}, each nth point=20] table[col sep=comma,x=x,y=y]{\basepathhh_mala_4_\component.csv};%
				\fi
				\ifnum\component=3\relax				
				\noexpand\addplot [name path=upper, draw=none, each nth point=20] table [col sep=comma, x=x, y=upper] {\basepathhh_mala_4_\component.csv};%
				\noexpand\addplot [name path=lower, draw=none, each nth point=20] table [col sep=comma, x=x, y=lower] {\basepathhh_mala_4_\component.csv};%
				\noexpand\addplot [maincolor, fill opacity=0.5] fill between[of=upper and lower];
				\noexpand\addplot [line width=1pt, maincolor, mark=*, mark size=1pt, mark options={fill=maincolor, draw=maincolor}, each nth point=20] table[col sep=comma,x=x,y=y]{\basepathhh_mala_4_\component.csv};%
				\fi
			}\tmp}
		\end{groupplot}
\end{tikzpicture}
\caption[]{%
\tikzexternaldisable
Autocorrelation functions of \protect\tikz[baseline=-\the\dimexpr\fontdimen22\textfont2\relax]\protect\draw [maincolor, ultra thick] (0,0) circle (1mm); Gibbs and \protect\tikz[baseline=-\the\dimexpr\fontdimen22\textfont2\relax]\protect\draw [maincolor, ultra thick,fill] (0,0) circle (1mm); \gls{mala} in the image prior sampling experiments.
The shaded areas around the means denote $\pm$ one standard deviation.
\tikzexternalenable
}%
\label{fig:prior-acf}
\end{figure}

\begin{figure}
\def\imwidth{1.9cm}
\def\vpad{1mm}
\def\hpad{1mm}
\centering
\begin{tikzpicture}
	\foreach [count=\ifactor] \annotation/\vmin/\vmax in {Normal/-1.63/1.53, Laplace/-1.41/1.06, Student-t/-3.73/3.87, \gls{gmm}/-0.50/0.46} {
		\pgfmathsetlengthmacro{\myy}{-\ifactor* (\imwidth + \vpad)}
		\node[rotate=90] at (.7, \myy) {\annotation};
		\pgfmathsetlengthmacro{\cbarx}{7 * (\imwidth + \hpad) + 1.1cm}
		\node at (\cbarx, \myy) {\drawcolorbar};
		\node at (\cbarx+4mm, \myy-.85cm) {\tiny\vmin};
		\node at (\cbarx+4mm, \myy+.85cm) {\tiny\vmax};
		\foreach \iindex in {1, ..., 7} {
			\pgfmathsetlengthmacro{\myx}{\iindex* (\imwidth + \hpad)}
			\node at (\myx, \myy) {\includegraphics[width=\imwidth]{./fig/data/prior-samples-collage/collage_\ifactor_\iindex.png}};
		}
	}
\end{tikzpicture}
\caption{Generated samples of size $96 \times 96$ in the image prior sampling experiments.}
\label{fig:prior-samples}
\end{figure}

\begin{table}
    \centering
    \begin{subtable}{0.48\textwidth}
        \centering
        \resizebox{\textwidth}{!}{ % Resizing the table
            \begin{tabular}{l r r r r }
                \toprule
                {} & Normal & Laplace & Student-t & \gls{gmm} \\
                \midrule
                $12 \times 12$   & 00:02:16 & 00:14:34 & 00:04:26 & 00:20:21 \\
                $24 \times 24$   & 00:04:54 & 00:40:59 & 00:08:12 & 00:57:22 \\
                $48 \times 48$   & 00:28:34 & 03:09:07 & 00:39:46 & 04:02:18 \\
                $96 \times 96$   & 06:16:28 & 36:39:14 & 08:04:04 & 44:31:45 \\
                \bottomrule
            \end{tabular}
        }
		\caption{Gibbs runtimes.}
    \end{subtable}
    \hfill
    \begin{subtable}{0.48\textwidth}
        \centering
        \resizebox{\textwidth}{!}{ % Resizing the table
            \begin{tabular}{l r r r r }
                \toprule
                {} & Normal & Laplace & Student-t & \gls{gmm} \\
                \midrule
                $12 \times 12$   & 06:06:06 & 05:46:22 & 05:10:59 & 10:25:47 \\
                $24 \times 24$   & 06:08:11 & 03:42:31 & 07:13:25 & 38:14:46 \\
                $48 \times 48$   & 06:12:42 & 06:02:56 & 07:38:47 & 161:35:16 \\
                $96 \times 96$   & 40:00:58 & 35:25:49 & 47:46:17 & NA \\
                \bottomrule
            \end{tabular}
        }
		\caption{\gls{mala} runtimes.}
    \end{subtable}
    \caption{Total runtimes of the image prior sampling experiments in hours:minutes:seconds. These correspond to the time to run $1000$ parallel chains for $\num{15000}$ Gibbs iterations and $\num{3000000}$ MALA iterations. Note that these do not reflect the time required for Gibbs to converge. Gibbs converged even in the most difficult experiment within at most $200$ seconds, while MALA failed to converge in some of the experiments despite these excessive runtimes.}
    \label{tab:prior_runtimes}
\end{table}

\begin{table}
    \centering
    \begin{subtable}{0.48\textwidth}
        \centering
        \resizebox{\textwidth}{!}{ % Resizing the table
        \begin{tabular}{lcccc}
        		\toprule
        		& Normal & Laplace & Student-t & \gls{gmm} \\
        		\midrule
        		$12 \times 12$ & $0.9998 \pm 0.0043$ & $0.9995 \pm 0.0065$ & $0.9997 \pm 0.0053$ & $0.9991 \pm 0.0094$ \\
        		$24 \times 24$ & $0.9995 \pm 0.0065$ & $0.9997 \pm 0.0051$ & $0.9995 \pm 0.0066$ & $0.9997 \pm 0.0051$ \\
        		$48 \times 48$ & $0.9994 \pm 0.0075$ & $0.9998 \pm 0.0041$ & $0.9996 \pm 0.0061$ & $0.9995 \pm 0.0065$ \\
        		$96 \times 96$ & $0.9998 \pm 0.0042$ & $0.9999 \pm 0.0031$ & $0.9996 \pm 0.0058$ & $0.9999 \pm 0.0029$ \\
        		\bottomrule
    		\end{tabular}
        }
		\caption{Gibbs.}
    \end{subtable}
    \hfill
    \begin{subtable}{0.48\textwidth}
        \centering
        \resizebox{\textwidth}{!}{ % Resizing the table
    		\begin{tabular}{lcccc}
        		\toprule
        		& Normal & Laplace & Student-t & \gls{gmm} \\
			\midrule
        		$12 \times 12$ & $0.0026 \pm 0.0008$ & $0.0010 \pm 0.0004$ & $0.0011 \pm 0.0005$ & $0.0011 \pm 0.0005$ \\
        		$24 \times 24$ & $0.0009 \pm 0.0004$ & $0.0007 \pm 0.0004$ & $0.0008 \pm 0.0004$ & $0.0008 \pm 0.0004$ \\
        		$48 \times 48$ & $0.0008 \pm 0.0004$ & NA & $0.0008 \pm 0.0004$ & NA \\
        		$96 \times 96$ & $0.0008 \pm 0.0004$ & NA & $0.0008 \pm 0.0004$ & NA \\
        		\bottomrule
    		\end{tabular}
        }
		\caption{\gls{mala}.}
    \end{subtable}
    \caption{Mean $\pm$ standard deviation of the sampling efficiency $\gamma$ in the image prior sampling experiments.}
    \label{tab:prior_samp_eff}
\end{table} 

Exemplary generated samples of size $96 \times 96$ are shown in \Cref{fig:prior-samples}.  The parameters for the normal, Laplace, and Student-t factors in these image prior sampling experiments were chosen arbitrarily. In contrast, the \gls{gmm} factors were learned beforehand in such a way that the resulting image prior \eqref{def_image_prior}, when applied to images of size $96 \times 96$, produces samples whose marginal distribution of image gradients matches the marginal distribution of image gradients found in natural images\footnote{The exact details on how this image prior was learned are not of central importance for the proposed sampling approach and are therefore omitted for brevity.}. This can be seen in \Cref{fig:gradient_marginals} that shows the negative logarithm of the marginal distributions of gradients obtained from samples of the BSDS500 dataset \cite{Arbelaez:2011} and from samples generated by the image prior \eqref{def_image_prior} for $n = 96^2$ and the four considered factors.  

\begin{figure}
\centering
\begin{tikzpicture}
	\begin{axis}[xmin=-1.24, xmax=1.24,/tikz/line join=bevel]
	\def\basepathhh{./fig/data/dx-marginals/dx_marginals}
		\addplot [ultra thick, maincolor] table[col sep=comma,x=x,y=y]{\basepathhh_dataset.csv};%
		\addplot [ultra thick, secondarycolor] table[col sep=comma,x=x,y=y]{\basepathhh_normal.csv};%
		\addplot [ultra thick, tertiarycolor] table[col sep=comma,x=x,y=y]{\basepathhh_laplace.csv};%
		\addplot [ultra thick, quaternarycolor] table[col sep=comma,x=x,y=y]{\basepathhh_student-t.csv};%
		\addplot [ultra thick, quinarycolor] table[col sep=comma,x=x,y=y]{\basepathhh_gmm.csv};%
	\end{axis}
\end{tikzpicture}
\caption[]{%
\tikzexternaldisable
Empirical negative $\log$ edge marginals of \protect\tikz[baseline=-\the\dimexpr\fontdimen22\textfont2\relax]\protect\draw [maincolor, ultra thick] (0,0) -- (.5, 0); the BSDS500 dataset and the considered image priors for size $96 \times 96$ with \protect\tikz[baseline=-\the\dimexpr\fontdimen22\textfont2\relax]\protect\draw [secondarycolor, ultra thick] (0,0) -- (.5, 0); normal, \protect\tikz[baseline=-\the\dimexpr\fontdimen22\textfont2\relax]\protect\draw [tertiarycolor, ultra thick] (0,0) -- (.5, 0); Laplace, \protect\tikz[baseline=-\the\dimexpr\fontdimen22\textfont2\relax]\protect\draw [quaternarycolor, ultra thick] (0,0) -- (.5, 0); Student-t, and \protect\tikz[baseline=-\the\dimexpr\fontdimen22\textfont2\relax]\protect\draw [quinarycolor, ultra thick] (0,0) -- (.5, 0); \gls{gmm} factors.
Each prior sample was rescaled in accordance to their optimal scale parameter $\lambda$ given in \Cref{sec:poster_sampling}.
The minimum value was subtracted from each curve.
\tikzexternalenable
}%
\label{fig:gradient_marginals}
\end{figure}

\subsection{Posterior Sampling Experiments}
\label{sec:poster_sampling}
We now turn to posterior sampling experiments that demonstrate some potential imaging applications. More specifically, we show how our proposed Gibbs sampling approach can be used to solve a large collection of inverse problems. 

\subsubsection{Inverse Problems}
To that end, we consider the inverse problems of denoising and \gls{dct} inpainting with improper image priors of the form \eqref{def_image_prior}. We briefly describe these inverse problems as follows.

In denoising, it is assumed that an observation $Y \in \R^n$ was generated from an unknown ground-truth image $X \in \R^n$ by the forward model 
\begin{align}
	Y = X + \eta,
	\label{forward_model_denoising}
\end{align}
where $\eta \da \nd{0, \sigma^2 \cdot I_n}$ with known variance parameter $\sigma^2$. In other words, it is assumed that $Y$ is a noisy image that was generated by adding i.i.d.~Gaussian noise to each pixel of an unknown clean ground-truth image $X$. This corresponds to a Gaussian likelihood term of the form
\begin{align}
	\fygx{\ygx} = \nd{y; x, \sigma^2 \cdot I_n}.
	\label{likelihood_denoising}
\end{align}

Similarly, in \gls{dct} inpainting it is assumed that an observation $Y \in \R^d$ was generated from an unknown ground-truth image $X \in \R^n$ by the forward model
\begin{align}
	Y = M D X + \eta,
	\label{forward_model_dct_inpainting}
\end{align}
where $M \in \R^{d \times n}$ is a linear operator that selects $d$ entries from an $n$-dimensional vector and $d < n$, $D \in \R^{n \times n}$ is the discrete cosine transform, and $\eta \da \nd{0, \sigma^2 \cdot I_n}$ with known variance parameter $\sigma^2$. In other words, $Y$ is generated by selecting a subset of \gls{dct} coefficients of an unknown ground-truth image $X$ and adding i.i.d.~Gaussian noise to them. Although this is a toy problem, it mimics the structure of the forward models found in many imaging applications like computed tomography and magnetic resonance imaging. The forward model \eqref{forward_model_dct_inpainting} corresponds to a linear Gaussian likelihood term of the form
\begin{align}
	\fygx{\ygx} = \nd{y; M D x, \sigma^2 \cdot I_d}.	
	\label{likelihood_dct_inpainting}
\end{align}

On a side note, these problems illustrate why prior information is crucial for solving inverse problems, since recovering $X$ from the likelihood term alone often leads to reconstructions that are not meaningful, are of poor quality, or not even uniquely defined. For instance, maximizing the denoising likelihood term \eqref{likelihood_denoising} for a given measurement realization $Y = y$ suggests that the most likely reconstruction is the noisy image itself. Similarly, maximizing the \gls{dct} inpainting likelihood term \eqref{likelihood_dct_inpainting} for a given measurement realization $Y = y$ suggests that there are uncountable many images that are candidates for the most likely reconstruction. Among those, the so-called zero-fill solution is often chosen, which is obtained as $D^T M^T y$ for a given measurement realization $Y = y$, or in other words by setting all unobserved \gls{dct} coefficients to zero and applying the inverse discrete cosine transform. While computationally convenient, such a reconstruction typically results in artifacts due to the noise in the observed \gls{dct} coefficients and the missing frequency information. 

\subsubsection{Posterior Distributions}
Therefore, we consider a Bayesian treatment of these inverse problems with improper image priors of the form \eqref{def_image_prior}. However, instead of considering posterior distributions of the form
\begin{align*}
	\fxgy{\xgy} \propto \fygx{\ygx} \cdot \fx{x},
\end{align*}
it is widespread practice in imaging applications to consider a modified version of the form
\begin{align}
	\fxgy{\xgy} \propto \fygx{\ygx} \cdot \fx^{\lambda}\br{x},
	\label{modified_posterior}
\end{align}
where $\lambda \in \Rpp$ is a tunable parameter. The parameter $\lambda$ can be interpreted as an additional degree of freedom that allows us, for instance, to compensate for modeling mismatches in cases when we do not have access to the correct prior, to compensate for approximate inference schemes, or to tweak end-to-end performance with respect to some performance metric of interest. Consequently, it is an important practical mechanism to squeeze out additional performance from imaging models. 

This modification has to be handled slightly differently in the context of Gaussian latent machines, since exponentiations of the form $\fx^\lambda$ require that our proposed approach can accommodate for exponentiated factors. This is possible for Gaussian, Laplace, and Student-t factors, but might result in some tedious implementation details such as relating the parameters of an exponentiated factor $\phi^{\lambda}_i$ to the parameters of the original factor $\phi_i$. However, it is not trivially possible for \gls{gmm} factors since exponentiated factors of the form $\phi^{\lambda}_i$ are not necessarily \gls{gmm} factors again. Given the generality of the proposed approach, there might exist other factor types besides \gls{gmm} factors for which this is not possible. For instance, it might be the case that there exist other factors for which assumption \Cref{assumption_fmp} holds, but which lead to corresponding exponentiated factors $\phi^{\lambda}_i$ for which \Cref{assumption_fmp} does not hold. It might also be the case that there exist factors $\phi_i$ for which the proposed approach is tractable, but which lead to corresponding exponentiated factors $\phi^{\lambda}_i$ for which it becomes intractable. Or it could happen that exponentation of the factors could lead to violations of other assumptions, like the assumption that the factors are univariate distributions. An example of this are Student-t factors, which after exponentiation with a positive exponent might not even yield a univariate distribution anymore.

To not have to deal with these technicalities in the context of Gaussian latent machines, we can instead of \eqref{modified_posterior} consider a similar posterior modification of the form
\begin{align}
	\fxgy{\xgy} \propto \fygx{\ygx} \cdot \fx\br{\lambda \cdot x}.
	\label{posterior_glm}
\end{align}
The parameter $\lambda$ in this modification can be interpreted as a scale parameter that controls the variance of the prior (or,  in other words, controls the expected intensity range of the prior). This modification can also be easily handled implementation-wise in our proposed approach, since this scale parameter can be trivially absorbed into the $K$ operator, even in a matrix-free fashion.

For the sake of simplicity, we again consider the same four factor types as in the baseline experiments in the improper prior \eqref{def_image_prior}. The scale parameter $\lambda$ in  \eqref{posterior_glm} for the four different factor types was chosen as follows. Both the scaling parameter $\lambda$ and the shape of the \gls{gmm} factors were fully learned via maximum likelihood. Only the scaling parameter was learned via maximum likelihood for the normal, Laplace, and Student-t factors. Notice that the scaling parameter $\lambda$ can be absorbed into the parameters $\sigma^2$ and $b$ of the normal and Laplace factors. Consequently, computing the optimal scaling $\lambda$ is equivalent to computing the optimal parameters of improper priors of the form \eqref{def_image_prior} with normal and Laplace factors in a maximum likelihood sense. That is not the case for Student-t factors, since the scaling $\lambda$ cannot be absorbed in the degree of freedom parameter of Student-t factors. Hence, this form of learning only infers the optimal scaling in improper priors of the form \eqref{def_image_prior} with Student-t factors for a particular degree of freedom. However, the exact details on how these scale parameters were learned are not of central importance for the proposed sampling approach and are therefore omitted for brevity.

\subsubsection{Posterior Sampling Results}
We run the posterior sampling experiments on $256$ ground-truth images of size $96 \times 96$ that were randomly extracted from the BSDS500 dataset. For each inverse problem and ground-truth image, a measurement realization $Y = y$ was synthesized by simulating the forward models \eqref{forward_model_denoising} and \eqref{forward_model_dct_inpainting}. The selection operator $M$ was randomized for each \gls{dct} inpainting problem by keeping $1/9$ of the \gls{dct} coefficients that correspond to low-frequency components (i.e., the structure of the image) and uniformly randomly removing  $25\%$ of the remaining coefficients. The variance parameter both for denoising and \gls{dct} inpainting was set to $\sigma^2 = 0.1^2$. 

For each problem instance and considered prior models, a set of $128$ samples from the posterior distribution \eqref{posterior_glm} was drawn by running $128$ parallel chains of Gibbs sampling and taking the last iteration of each chain as a sample. The effective number of samples could have been increased by relying on ergodicity, but we chose this variant instead for the sake of simplicity.  In case of the prior models with Laplace, Student-t, and \gls{gmm} factors,  the parallel chains were run for $3000$ iterations for the denoising instances and for $1000$ iterations for the \gls{dct} inpainting instances. These number of iterations were conservatively chosen and do not reflect the number of iterations required to reach stationarity in each problem instance. We did so to not have to explicitly check the convergence in all the considered posterior sampling problem instances, which would amount to more than a thousand checks. Considerably fewer iterations are needed to reach stationarity since these posterior sampling experiments are considerably easier sampling problems than the image prior sampling experiments due to the presence of the Gaussian likelihood terms.  We only run a single Gibbs iteration in case of the prior model with normal factors since our posterior sampling algorithms in that case are direct sampling algorithms. 

We do not need to extend the improper priors, since the posterior distributions in all our problem instances are proper. In fact, extending the improper priors would actually lead to undesired modifications in the posterior distributions. 

Note that running this experiment with \gls{mala} would require step-size tuning in each of the $2048$ posterior sampling problems. The denoising problem instances are probably similar enough that a single step-size could be chosen for all instances that use the same type of prior. However, separate step-size tuning is likely required in the DCT inpainting instances, since the likelihood term between the instances might vary considerably. 

Based on the samples, we compute point estimators for each problem instance and model to provide some summary statistics. Specifically, we compute the empirical mean for each resulting posterior \eqref{posterior_glm} as it is the point estimator that minimizes mean squared error. We also compute empirical pixel-wise standard deviation for each resulting posterior \eqref{posterior_glm} to provide some measure of uncertainty. 

Box plots that show the \gls{psnr} boost with respect to the naive reconstructions across the four considered models in the denoising and \gls{dct} inpainting posterior sampling experiments are shown in \Cref{psnr-boosts} and corresponding summary statistics are given in \Cref{tab-psnr-boosts}.  Interestingly, only the model with the learned \gls{gmm} prior improves over the \gls{psnr} of the naive reconstruction in all problem instances.  In other words, there are a few problem instances for which the other models perform worse than the naive reconstruction.  Furthermore, the learned \gls{gmm} prior performs best on average and achieves higher top-end performance than the other models. 

Exemplary denoising and \gls{dct} inpainting results are shown in \Cref{denoising-results,dct-inpainting-results}.  We see from these that the learned \gls{gmm} prior also provides visually the best reconstruction results. The performance in all models deteriorates as we go from test images with many constant regions to highly textured test images. This happens because the linear operator $K$ in the improper prior is only a collection of horizontal and vertical gradient filters and therefore cannot capture richer image statistics. Finally, the computed pixel-wise standard deviations in the models with \gls{gmm} and Laplace factors correlate strongly with the edges of the test images, which is considerably less the case for the other two models.  

Note that these experiments do not evaluate the best reconstruction performance that can be obtained from each model.  For instance, it is well-known that the reconstruction performance for the models with Laplace and Student-t factors can be improved by switching from computing the conditional mean to the conditional mode and by tweaking the trade-off parameter $\lambda$ accordingly.  This would bring their average \gls{psnr} boost much closer to the one obtained for the learned \gls{gmm} prior.  However,  this evaluation and the one given in \Cref{fig:gradient_marginals} do evaluate how suited the considered improper priors are as statistical models of natural images.  Therefore,  these experiments and the image prior sampling experiments suggest that total variation and fields-of-experts models with Student-t factors are poor priors for natural images.  

Strictly speaking,  our experiments only validated Student-t factors for a particular degree of freedom value. However,  the best reconstruction performance in classical fields-of-experts models with Student-t factors is achieved for very small or impermissible degree of freedom values.  The former would lead to prior distributions that are too heavy-tailed and thus produce samples considerably outside the $\ci{0}{1}$ intensity range,  while the latter do not even correspond to valid prior distributions.  Another argument against heavy-tailedness is our learned \gls{gmm} prior,  which in fact is not heavy-tailed at all,  but still matches the edge marginal statistics of natural images accurately, as shown in \Cref{fig:gradient_marginals}.  We conjecture that the shape of the Student-t distribution is not even sufficiently flexible to obtain the edge marginals as shown in  \Cref{fig:gradient_marginals}.  The edge marginals in natural images are invariant to the size of the images.  Therefore,  edge marginals would correspond to the factor for images of size $2 \times 1$ pixels, and a Student-t factor cannot adequately approximate the edge marginals as shown in \Cref{fig:gradient_marginals}.  These observations suggest that Student-t and other heavy-tailed factors are actually not good choices to model the statistics of natural images, even though this is widely believed in the field. 

\subsubsection{Empirical Convergence Rates in the Posterior Sampling Experiments}
To give an insight into the practicality of the proposed approach for solving inverse problems, we look at the achieved empirical convergence rates in the posterior sampling experiments. To that end we take the empirical mean and pixel-wise standard deviation computed across all chains at the final iteration of Gibbs as ground-truth and look how fast the Gibbs sampler converges over time to these estimated ground-truth quantities as measured in mean-squared error. 

The corresponding mean squared errors of Gibbs over time are shown in \cref{fig:posterior_times}. We can observe that both the mean and pixel-wise standard deviation in the worst case converge in the order of seconds across all experiments. Furthermore, the convergence to the posterior pixel-wise standard deviation requires roughly double the time necessary to converge in posterior mean. The results for the experiments with normal factors are not shown since Gibbs converges in a single iteration in those experiments. In those experiments, Gibbs converged on average in $0.026$ seconds (with a standard deviation of $0.009$ seconds) for a single denoising problem instance and in $0.051$ seconds (with a standard deviation of $0.025$ seconds) for a single \gls{dct} inpainting problem instance. Therefore, the proposed Gibbs sampler is both very practical and efficient in the context of sampling from posterior distributions that arise from inverse problems. 

For the sake of completeness, the total runtimes of the posterior sampling experiments are given in \Cref{tab:denoising_inpainting}.

\begin{figure}
\centering
\begin{tikzpicture}
\def\basepathden{./fig/data/denboost}
\def\basepathdct{./fig/data/dctboost}
\begin{groupplot}[
    group style={
        group size=2 by 1,
        horizontal sep=0.8cm,
        ylabels at=edge left,
        y descriptions at=edge left,
    },
    width=8cm,
    height=8.75cm,
    ylabel={\(\PSNR\br{\hat{x}, x_{\mathrm{gt}}} - \PSNR\br{\hat{x}_{\mathrm{naive}}, x_{\mathrm{gt}}} \; \mathrm{[dB]}\)},
    boxplot/draw direction=y,
    boxplot/box extend=0.475,
	boxplot/every box/.append style={line width=0.75pt},
    boxplot/every whisker/.append style={line width=0.75pt},
    boxplot/every median/.append style={line width=0.75pt},
    boxplot/every median/.append style={tertiarycolor},
	ymin=-3.75,
	ymax=17.75,
	ytick={-2.5, 0.0, 2.5, 5.0, 7.5, 10.0, 12.5, 15.0},
	scaled y ticks = false,
    y tick label style={
      /pgf/number format/.cd,
      fixed,
      precision=1,
      zerofill
    },
    	title style={text depth=0pt, font=\normalsize, yshift=-0.2cm, align=center},
    	ylabel style={text depth=0pt, font=\normalsize, yshift=-0.2cm, align=center},
	xlabel style={text depth=0pt, font=\tiny, align=center, yshift=0.5cm},
	ticklabel style={font=\tiny},
]

% Denoising
\nextgroupplot[
    title={Denoising},
    xtick={1,2,3,4},
    xticklabels={Normal,Laplace,Student-t,\gls{gmm}},
]

% Normal
\addplot+[
  boxplot,
  color=black,
] 
table[col sep=comma, y index=0] {\basepathden/normal.csv};

% Laplace
\addplot+[
  boxplot,
  color=black
] 
table[col sep=comma, y index=0] {\basepathden/laplace.csv};

% Student-t
\addplot+[
  boxplot,
  color=black
] 
table[col sep=comma, y index=0] {\basepathden/student-t.csv};

% GMM
\addplot+[
  boxplot,
  color=black
] 
table[col sep=comma, y index=0] {\basepathden/gmm.csv};

% DCT inpainting
\nextgroupplot[
    title={DCT inpainting},
    xtick={1,2,3,4},
    xticklabels={Normal,Laplace,Student-t,\gls{gmm}}
]

% Normal
\addplot+[
  boxplot,
  color=black
] 
table[col sep=comma, y index=0] {\basepathdct/normal.csv};

% Laplace
\addplot+[
  boxplot,
  color=black
] 
table[col sep=comma, y index=0] {\basepathdct/laplace.csv};

% Student-t
\addplot+[
  boxplot,
  color=black
] 
table[col sep=comma, y index=0] {\basepathdct/student-t.csv};

% GMM
\addplot+[
  boxplot,
  color=black
] 
table[col sep=comma, y index=0] {\basepathdct/gmm.csv};

\end{groupplot}
\end{tikzpicture}
\caption{Box plots that show the PSNR boost across the four considered models in the denoising and DCT inpainting posterior sampling experiments.  Here $x_{\mathrm{gt}}$ refers to a ground truth image, $\hat{x}_{\mathrm{naive}}$ to its noisy observation for denoising and to the zero-fill solution for DCT inpainting,  and $\hat{x}$ to the conditional mean of the posterior.}
\label{psnr-boosts}
\end{figure}

\begin{table}
    \centering
    \begin{subtable}{0.48\textwidth}
        \centering
        \resizebox{\textwidth}{!}{ % Resizing the table
            \begin{tabular}{l r r r r }
                \toprule
                Potential & Min & Median & Mean & Max \\
                \midrule
                Normal   & -2.14 & 5.54 & 5.63 & 9.56 \\
                Laplace  & -1.83 & 5.89 & 6.37 & 13.18 \\
                Student-t & -2.78 & 5.46 & 5.93 & 12.41 \\
                \gls{gmm}      &  1.98 & 6.76 & 7.54 & 16.22 \\
                \bottomrule
            \end{tabular}
        }
		\caption{Denoising}
    \end{subtable}
    \hfill
    \begin{subtable}{0.48\textwidth}
        \centering
        \resizebox{\textwidth}{!}{ % Resizing the table
            \begin{tabular}{l r r r r }
                \toprule
                Potential & Min & Median & Mean & Max \\
                \midrule
                Normal   & -1.32 & 4.50 & 4.73 & 9.14 \\
                Laplace  & -1.25 & 4.77 & 5.39 & 12.61 \\
                Student-t & -1.85 & 4.40 & 5.00 & 11.87 \\
                \gls{gmm}      &  1.10 & 5.60 & 6.45 & 15.44 \\
                \bottomrule
            \end{tabular}
        }
    	\caption{DCT inpainting}
    \end{subtable}
    \caption{Minimum, median, mean, and maximum PSNR boost across the four considered models in the denoising and DCT inpainting posterior experiments.}
    \label{tab-psnr-boosts}
\end{table}

% Denoising results
\begin{figure}
    \centering
    \begin{tikzpicture}
    		\def\basepathhh{./fig/data/denoising/denoising_}
        \begin{groupplot}[
            group style={
                group size=5 by 8,
                horizontal sep=0.58cm,
                vertical sep=1.2cm,
                xlabels at=all,
                group name=mygroup,
            },
            width=3.275cm, height=3.275cm,
            axis lines=box,
            xtick=\empty, 
            ytick=\empty,
            enlargelimits=false,
            clip=false,
            ticklabel style={font=\tiny},
            	ylabel style={text depth=0pt, font=\normalsize, yshift=-0.9cm, align=center},
            	xlabel style={text depth=0pt, font=\tiny, align=center, yshift=0.5cm},
			title style={text depth=0pt, font=\normalsize, yshift=-0.2cm, align=center},
        ]
        
        % Min performance top row
		\pgfplotsforeachungrouped \component/\mytitle in {1/{}, 2/{Normal}, 3/{Laplace}, 4/{Student-t}, 5/{\gls{gmm}}} {
		\edef\tmp{%
		\noexpand\nextgroupplot[xlabel={\ifnum\component=2 17.79 [dB] \else\ifnum\component=3 18.10 [dB] \else\ifnum\component=4 17.16 [dB] \else\ifnum\component=5 21.99 [dB] \else  \fi\fi\fi\fi}, title={\mytitle}, ylabel={\ifnum\component=1 $x_{\noexpand \mathrm{gt}}$ \else\ifnum\component=2 $\noexpand \hat{x}$ \fi\fi}]%
		\noexpand\addplot graphics [xmin=0, xmax=10, ymin=0, ymax=10] {\basepathhh1_\component.png};%
		\noexpand\draw[black, line width=0.25mm] (axis cs:-0.0775,-0.0775) rectangle (axis cs:10.0775,10.0775);%
		}\tmp}
		
		% Min performance bottom row
		\pgfplotsforeachungrouped \component in {1, ..., 5} {
		\edef\tmp{%
		\noexpand\nextgroupplot[yshift=0.65cm, xlabel={\ifnum\component=1 19.93 [dB]\fi}, ylabel={\ifnum\component=1 $\noexpand \hat{x}_{\noexpand \mathrm{naive}}$ \else\ifnum\component=2 $\noexpand \sigma_{\noexpand \hat{x}}$ \fi\fi}]%
		\noexpand\addplot graphics [xmin=0, xmax=10, ymin=0, ymax=10] {\basepathhh2_\component.png};%
		\noexpand\draw[black, line width=0.25mm] (axis cs:-0.0775,-0.0775) rectangle (axis cs:10.0775,10.0775);%
		}\tmp}
		
		% Median performance top row
		\pgfplotsforeachungrouped \component/\mytitle in {1/{}, 2/{26.18 [dB]}, 3/{26.32 [dB]}, 4/{26.20 [dB]}, 5/{26.79 [dB]}} {
		\edef\tmp{%
		\noexpand\nextgroupplot[xlabel={\mytitle}, ylabel={\ifnum\component=1 $x_{\noexpand \mathrm{gt}}$ \else\ifnum\component=2 $\noexpand \hat{x}$ \fi\fi}]%
		\noexpand\addplot graphics [xmin=0, xmax=10, ymin=0, ymax=10] {\basepathhh3_\component.png};%
		\noexpand\draw[black, line width=0.25mm] (axis cs:-0.0775,-0.0775) rectangle (axis cs:10.0775,10.0775);%
		}\tmp}
		
		% Median performance bottom row
		\pgfplotsforeachungrouped \component in {1, ..., 5} {
		\edef\tmp{%
		\noexpand\nextgroupplot[yshift=0.65cm, xlabel={\ifnum\component=1 19.92 [dB]\fi}, ylabel={\ifnum\component=1 $\noexpand \hat{x}_{\noexpand \mathrm{naive}}$ \else\ifnum\component=2 $\noexpand \sigma_{\noexpand \hat{x}}$ \fi\fi}]%
		\noexpand\addplot graphics [xmin=0, xmax=10, ymin=0, ymax=10] {\basepathhh4_\component.png};%
		\noexpand\draw[black, line width=0.25mm] (axis cs:-0.0775,-0.0775) rectangle (axis cs:10.0775,10.0775);%
		}\tmp}
		
		% Mean performance top row
		\pgfplotsforeachungrouped \component/\mytitle in {1/{}, 2/{25.06 [dB]}, 3/{25.60 [dB]}, 4/{24.68 [dB]}, 5/{27.50 [dB]}} {
		\edef\tmp{%
		\noexpand\nextgroupplot[xlabel={\mytitle}, ylabel={\ifnum\component=1 $x_{\noexpand \mathrm{gt}}$ \else\ifnum\component=2 $\noexpand \hat{x}$ \fi\fi}]%
		\noexpand\addplot graphics [xmin=0, xmax=10, ymin=0, ymax=10] {\basepathhh5_\component.png};%
		\noexpand\draw[black, line width=0.25mm] (axis cs:-0.0775,-0.0775) rectangle (axis cs:10.0775,10.0775);%
		}\tmp}
		
		% Mean performance bottom row
		\pgfplotsforeachungrouped \component in {1, ..., 5} {
		\edef\tmp{%
		\noexpand\nextgroupplot[yshift=0.65cm, xlabel={\ifnum\component=1 20.05 [dB]\fi}, ylabel={\ifnum\component=1 $\noexpand \hat{x}_{\noexpand \mathrm{naive}}$ \else\ifnum\component=2 $\noexpand \sigma_{\noexpand \hat{x}}$ \fi\fi}]%
		\noexpand\addplot graphics [xmin=0, xmax=10, ymin=0, ymax=10] {\basepathhh6_\component.png};%
		\noexpand\draw[black, line width=0.25mm] (axis cs:-0.0775,-0.0775) rectangle (axis cs:10.0775,10.0775);%
		}\tmp}
		
		% Max performance top row
		\pgfplotsforeachungrouped \component/\mytitle in {1/{}, 2/{28.58 [dB]}, 3/{31.00 [dB]}, 4/{30.22 [dB]}, 5/{33.77 [dB]}} {
		\edef\tmp{%
		\noexpand\nextgroupplot[xlabel={\mytitle}, ylabel={\ifnum\component=1 $x_{\noexpand \mathrm{gt}}$ \else\ifnum\component=2 $\noexpand \hat{x}$ \fi\fi}]%
		\noexpand\addplot graphics [xmin=0, xmax=10, ymin=0, ymax=10] {\basepathhh7_\component.png};%
		\noexpand\draw[black, line width=0.25mm] (axis cs:-0.0775,-0.0775) rectangle (axis cs:10.0775,10.0775);%
		}\tmp}
		
		% Max performance bottom row
		\pgfplotsforeachungrouped \component in {1, ..., 5} {
		\edef\tmp{%
		\noexpand\nextgroupplot[yshift=0.65cm, xlabel={\ifnum\component=1 20.01 [dB]\fi}, ylabel={\ifnum\component=1 $\noexpand \hat{x}_{\noexpand \mathrm{naive}}$ \else\ifnum\component=2 $\noexpand \sigma_{\noexpand \hat{x}}$ \fi\fi}]%
		\noexpand\addplot graphics [xmin=0, xmax=10, ymin=0, ymax=10] {\basepathhh8_\component.png};%
		\noexpand\draw[black, line width=0.25mm] (axis cs:-0.0775,-0.0775) rectangle (axis cs:10.0775,10.0775);%
		}\tmp}
        \end{groupplot}
        % Row 2
  		\foreach [count=\col from 2] \leftnum/\rightnum in {4.04/6.51, 4.24/6.60, 4.07/6.64, 4.25/6.60} {%
    		\node at ($(mygroup c\col r2.south)+(0,-0.245)$) {\drawcolorbarh{5.503cm}};
    		\node at ($(mygroup c\col r2.south)+(-0.65,-0.5)$) {\tiny\leftnum};
    		\node at ($(mygroup c\col r2.south)+(0.65,-0.5)$) {\tiny\rightnum};
  		}
        
        % Row 4
		\foreach [count=\col from 2] \leftnum/\rightnum in {3.21/9.19, 3.01/7.04, 3.14/8.26, 2.95/5.99} {%
    		\node at ($(mygroup c\col r4.south)+(0,-0.245)$) {\drawcolorbarh{5.503cm}};
    		\node at ($(mygroup c\col r4.south)+(-0.65,-0.5)$) {\tiny\leftnum};
    		\node at ($(mygroup c\col r4.south)+(0.65,-0.5)$) {\tiny\rightnum};
  		}
        
		% Row 6
        \foreach [count=\col from 2] \leftnum/\rightnum in {3.16/8.47, 3.36/5.67, 3.36/6.87, 3.20/5.54} {%
    		\node at ($(mygroup c\col r6.south)+(0,-0.245)$) {\drawcolorbarh{5.503cm}};
    		\node at ($(mygroup c\col r6.south)+(-0.65,-0.5)$) {\tiny\leftnum};
    		\node at ($(mygroup c\col r6.south)+(0.65,-0.5)$) {\tiny\rightnum};
  		}
        
		% Row 8
        \foreach [count=\col from 2] \leftnum/\rightnum in {2.19/14.90, 1.80/15.30, 1.88/15.60, 1.65/16.70} {%
    		\node at ($(mygroup c\col r8.south)+(0,-0.245)$) {\drawcolorbarh{5.503cm}};
    		\node at ($(mygroup c\col r8.south)+(-0.65,-0.5)$) {\tiny\leftnum};
    		\node at ($(mygroup c\col r8.south)+(0.65,-0.5)$) {\tiny\rightnum};
  		}
    \end{tikzpicture}
    \caption{Exemplary denoising results. Here $x_{\mathrm{gt}}$ refers to a ground truth image, $\hat{x}_{\mathrm{naive}}$ to its noisy observation, $\hat{x}$ to the conditional mean of the posterior, and $\sigma_{\hat{x}}$ to the pixel-wise standard deviation of the posterior. From top to bottom, the row pairs roughly correspond to the minimum, median, mean, and maximum performance for each model. The standard deviations are scaled by a factor of $100$, while the other images are clipped to the range $\ci{0}{1}$.}
\label{denoising-results}
\end{figure}

% DCT inpainting results
\begin{figure}
    \centering
    \begin{tikzpicture}
        	\def\basepathhh{./fig/data/dct/dct_}
        \begin{groupplot}[
            group style={
                group size=5 by 8,
                horizontal sep=0.58cm,
                vertical sep=1.2cm,
                xlabels at=all,
                group name=mygroup,
            },
            width=3.275cm, height=3.275cm,
            axis lines=box,
            xtick=\empty, 
            ytick=\empty,
            enlargelimits=false,
            clip=false,
            ticklabel style={font=\tiny},
            	ylabel style={text depth=0pt, font=\normalsize, yshift=-0.9cm, align=center},
            	xlabel style={text depth=0pt, font=\tiny, align=center, yshift=0.5cm},
			title style={text depth=0pt, font=\normalsize, yshift=-0.2cm, align=center},
        ]
        
        % Min performance top row
		\pgfplotsforeachungrouped \component/\mytitle in {1/{}, 2/{Normal}, 3/{Laplace}, 4/{Student-t}, 5/{\gls{gmm}}} {
		\edef\tmp{%
		\noexpand\nextgroupplot[xlabel={\ifnum\component=2 16.96 [dB] \else\ifnum\component=3 17.03 [dB] \else\ifnum\component=4 16.42 [dB] \else\ifnum\component=5 20.30 [dB] \else  \fi\fi\fi\fi}, title={\mytitle}, ylabel={\ifnum\component=1 $x_{\noexpand \mathrm{gt}}$ \else\ifnum\component=2 $\noexpand \hat{x}$ \fi\fi}]%
		\noexpand\addplot graphics [xmin=0, xmax=10, ymin=0, ymax=10] {\basepathhh1_\component.png};%
		\noexpand\draw[black, line width=0.25mm] (axis cs:-0.0775,-0.0775) rectangle (axis cs:10.0775,10.0775);%
		}\tmp}
		
		% Min performance bottom row
		\pgfplotsforeachungrouped \component in {1, ..., 5} {
		\edef\tmp{%
		\noexpand\nextgroupplot[yshift=0.65cm, xlabel={\ifnum\component=1 18.27 [dB]\fi}, ylabel={\ifnum\component=1 $\noexpand \hat{x}_{\noexpand \mathrm{naive}}$ \else\ifnum\component=2 $\noexpand \sigma_{\noexpand \hat{x}}$ \fi\fi}]%
		\noexpand\addplot graphics [xmin=0, xmax=10, ymin=0, ymax=10] {\basepathhh2_\component.png};%
		\noexpand\draw[black, line width=0.25mm] (axis cs:-0.0775,-0.0775) rectangle (axis cs:10.0775,10.0775);%
		}\tmp}
		
		% Median performance top row
		\pgfplotsforeachungrouped \component/\mytitle in {1/{}, 2/{26.01 [dB]}, 3/{26.06 [dB]}, 4/{25.99 [dB]}, 5/{26.50 [dB]}} {
		\edef\tmp{%
		\noexpand\nextgroupplot[xlabel={\mytitle}, ylabel={\ifnum\component=1 $x_{\noexpand \mathrm{gt}}$ \else\ifnum\component=2 $\noexpand \hat{x}$ \fi\fi}]%
		\noexpand\addplot graphics [xmin=0, xmax=10, ymin=0, ymax=10] {\basepathhh3_\component.png};%
		\noexpand\draw[black, line width=0.25mm] (axis cs:-0.0775,-0.0775) rectangle (axis cs:10.0775,10.0775);%
		}\tmp}
		
		% Median performance bottom row
		\pgfplotsforeachungrouped \component in {1, ..., 5} {
		\edef\tmp{%
		\noexpand\nextgroupplot[yshift=0.65cm, xlabel={\ifnum\component=1 20.82 [dB]\fi}, ylabel={\ifnum\component=1 $\noexpand \hat{x}_{\noexpand \mathrm{naive}}$ \else\ifnum\component=2 $\noexpand \sigma_{\noexpand \hat{x}}$ \fi\fi}]%
		\noexpand\addplot graphics [xmin=0, xmax=10, ymin=0, ymax=10] {\basepathhh4_\component.png};%
		\noexpand\draw[black, line width=0.25mm] (axis cs:-0.0775,-0.0775) rectangle (axis cs:10.0775,10.0775);%
		}\tmp}
		
		% Mean performance top row
		\pgfplotsforeachungrouped \component/\mytitle in {1/{}, 2/{24.68 [dB]}, 3/{25.04 [dB]}, 4/{24.19 [dB]}, 5/{27.16 [dB]}} {
		\edef\tmp{%
		\noexpand\nextgroupplot[xlabel={\mytitle}, ylabel={\ifnum\component=1 $x_{\noexpand \mathrm{gt}}$ \else\ifnum\component=2 $\noexpand \hat{x}$ \fi\fi}]%
		\noexpand\addplot graphics [xmin=0, xmax=10, ymin=0, ymax=10] {\basepathhh5_\component.png};%
		\noexpand\draw[black, line width=0.25mm] (axis cs:-0.0775,-0.0775) rectangle (axis cs:10.0775,10.0775);%
		}\tmp}
		
		% Mean performance bottom row
		\pgfplotsforeachungrouped \component in {1, ..., 5} {
		\edef\tmp{%
		\noexpand\nextgroupplot[yshift=0.65cm, xlabel={\ifnum\component=1 20.71 [dB]\fi}, ylabel={\ifnum\component=1 $\noexpand \hat{x}_{\noexpand \mathrm{naive}}$ \else\ifnum\component=2 $\noexpand \sigma_{\noexpand \hat{x}}$ \fi\fi}]%
		\noexpand\addplot graphics [xmin=0, xmax=10, ymin=0, ymax=10] {\basepathhh6_\component.png};%
		\noexpand\draw[black, line width=0.25mm] (axis cs:-0.0775,-0.0775) rectangle (axis cs:10.0775,10.0775);%
		}\tmp}
		
		% Max performance top row
		\pgfplotsforeachungrouped \component/\mytitle in {1/{}, 2/{29.16 [dB]}, 3/{31.48 [dB]}, 4/{30.72 [dB]}, 5/{33.84 [dB]}} {
		\edef\tmp{%
		\noexpand\nextgroupplot[xlabel={\mytitle}, ylabel={\ifnum\component=1 $x_{\noexpand \mathrm{gt}}$ \else\ifnum\component=2 $\noexpand \hat{x}$ \fi\fi}]%
		\noexpand\addplot graphics [xmin=0, xmax=10, ymin=0, ymax=10] {\basepathhh7_\component.png};%
		\noexpand\draw[black, line width=0.25mm] (axis cs:-0.0775,-0.0775) rectangle (axis cs:10.0775,10.0775);%
		}\tmp}
		
		% Max performance bottom row
		\pgfplotsforeachungrouped \component in {1, ..., 5} {
		\edef\tmp{%
		\noexpand\nextgroupplot[yshift=0.65cm, xlabel={\ifnum\component=1 21.08 [dB]\fi}, ylabel={\ifnum\component=1 $\noexpand \hat{x}_{\noexpand \mathrm{naive}}$ \else\ifnum\component=2 $\noexpand \sigma_{\noexpand \hat{x}}$ \fi\fi}]%
		\noexpand\addplot graphics [xmin=0, xmax=10, ymin=0, ymax=10] {\basepathhh8_\component.png};%
		\noexpand\draw[black, line width=0.25mm] (axis cs:-0.0775,-0.0775) rectangle (axis cs:10.0775,10.0775);%
		}\tmp}
        \end{groupplot}
        % Row 2
  		\foreach [count=\col from 2] \leftnum/\rightnum in {4.18/6.77, 4.11/7.08, 4.12/7.19, 4.29/6.87} {%
    		\node at ($(mygroup c\col r2.south)+(0,-0.245)$) {\drawcolorbarh{5.503cm}};
    		\node at ($(mygroup c\col r2.south)+(-0.65,-0.5)$) {\tiny\leftnum};
    		\node at ($(mygroup c\col r2.south)+(0.65,-0.5)$) {\tiny\rightnum};
  		}
        
        % Row 4
		\foreach [count=\col from 2] \leftnum/\rightnum in {3.26/10.50, 3.15/6.57, 3.12/8.47, 3.16/6.32} {%
    		\node at ($(mygroup c\col r4.south)+(0,-0.245)$) {\drawcolorbarh{5.503cm}};
    		\node at ($(mygroup c\col r4.south)+(-0.65,-0.5)$) {\tiny\leftnum};
    		\node at ($(mygroup c\col r4.south)+(0.65,-0.5)$) {\tiny\rightnum};
  		}
        
		% Row 6
        \foreach [count=\col from 2] \leftnum/\rightnum in {3.47/11.40, 3.36/5.48, 3.43/7.20, 3.3/5.73} {%
    		\node at ($(mygroup c\col r6.south)+(0,-0.245)$) {\drawcolorbarh{5.503cm}};
    		\node at ($(mygroup c\col r6.south)+(-0.65,-0.5)$) {\tiny\leftnum};
    		\node at ($(mygroup c\col r6.south)+(0.65,-0.5)$) {\tiny\rightnum};
  		}
        
		% Row 8
        \foreach [count=\col from 2] \leftnum/\rightnum in {2.55/18.70, 1.62/16.50, 1.86/15.70, 1.69/16.40} {%
    		\node at ($(mygroup c\col r8.south)+(0,-0.245)$) {\drawcolorbarh{5.503cm}};
    		\node at ($(mygroup c\col r8.south)+(-0.65,-0.5)$) {\tiny\leftnum};
    		\node at ($(mygroup c\col r8.south)+(0.65,-0.5)$) {\tiny\rightnum};
  		}
    \end{tikzpicture}
    \caption{Exemplary DCT inpainting results. Here $x_{\mathrm{gt}}$ refers to a ground truth image, $\hat{x}_{\mathrm{naive}}$ to the zero-fill solution, $\hat{x}$ to the conditional mean of the posterior, and $\sigma_{\hat{x}}$ to the pixel-wise standard deviation of the posterior. From top to bottom, the row pairs roughly correspond to the minimum, median, mean, and maximum performance for each model. The standard deviations are scaled by a factor of $100$, while the other images are clipped to the range $\ci{0}{1}$.}
\label{dct-inpainting-results}
\end{figure}

\begin{figure}
\centering
\captionsetup[subfigure]{justification=centering}
\subfloat[Denoising]{%
\begin{tikzpicture}
    \def\basepathhh{./fig/data/posteriorst/denoising}
    \begin{groupplot}[convplots group plot, yticklabel style={/pgf/number format/fixed,/pgf/number format/precision=3}]
    % Laplace conditional means
    \nextgroupplot[title={$\hat{x}$}, ylabel={Laplace}, xmin=0, xmax=10, ymin=-0.0025, ymax=0.4, xtick={0,10}]
    \addplot [name path=upper, draw=none, each nth point=20] table [col sep=comma, x=x, y=upper] {\basepathhh_laplace_means.csv};
    \addplot [name path=lower, draw=none, each nth point=20] table [col sep=comma, x=x, y=lower] {\basepathhh_laplace_means.csv};
    \addplot [maincolor, fill opacity=0.5] fill between[of=upper and lower];
    \addplot [line width=1pt, maincolor, each nth point=20] table[col sep=comma,x=x,y=y]{\basepathhh_laplace_means.csv};

    % Laplace conditional stds
    \nextgroupplot[title={$\sigma_{\hat{x}}$}, xmin=0, xmax=10, ymin=-0.0000125, ymax=0.003, xtick={0,10}]
    \addplot [name path=upper, draw=none, each nth point=20] table [col sep=comma, x=x, y=upper] {\basepathhh_laplace_stds.csv};
    \addplot [name path=lower, draw=none, each nth point=20] table [col sep=comma, x=x, y=lower] {\basepathhh_laplace_stds.csv};
    \addplot [maincolor, fill opacity=0.5] fill between[of=upper and lower];
    \addplot [line width=1pt, maincolor, each nth point=20] table[col sep=comma,x=x,y=y]{\basepathhh_laplace_stds.csv};

    % Student-t conditional means
    \nextgroupplot[ylabel={Student-t}, xmin=0, xmax=0.15, ymin=-0.0025, ymax=0.4, xtick={0,0.15}]
    \addplot [name path=upper, draw=none, each nth point=20] table [col sep=comma, x=x, y=upper] {\basepathhh_student-t_means.csv};
    \addplot [name path=lower, draw=none, each nth point=20] table [col sep=comma, x=x, y=lower] {\basepathhh_student-t_means.csv};
    \addplot [maincolor, fill opacity=0.5] fill between[of=upper and lower];
    \addplot [line width=1pt, maincolor, each nth point=20] table[col sep=comma,x=x,y=y]{\basepathhh_student-t_means.csv};

    % Student-t conditional stds
    \nextgroupplot[xmin=0, xmax=0.15, ymin=-0.0000125, ymax=0.003, xtick={0,0.15}]
    \addplot [name path=upper, draw=none, each nth point=20] table [col sep=comma, x=x, y=upper] {\basepathhh_student-t_stds.csv};
    \addplot [name path=lower, draw=none, each nth point=20] table [col sep=comma, x=x, y=lower] {\basepathhh_student-t_stds.csv};
    \addplot [maincolor, fill opacity=0.5] fill between[of=upper and lower];
    \addplot [line width=1pt, maincolor, each nth point=20] table[col sep=comma,x=x,y=y]{\basepathhh_student-t_stds.csv};

    % GMM conditional means
    \nextgroupplot[ylabel={\gls{gmm}}, xmin=0, xmax=0.75, ymin=-0.0025, ymax=0.4, xtick={0,0.75}]
    \addplot [name path=upper, draw=none, each nth point=20] table [col sep=comma, x=x, y=upper] {\basepathhh_gmm_means.csv};
    \addplot [name path=lower, draw=none, each nth point=20] table [col sep=comma, x=x, y=lower] {\basepathhh_gmm_means.csv};
    \addplot [maincolor, fill opacity=0.5] fill between[of=upper and lower];
    \addplot [line width=1pt, maincolor, each nth point=20] table[col sep=comma,x=x,y=y]{\basepathhh_gmm_means.csv};

    % GMM conditional stds
    \nextgroupplot[xmin=0, xmax=3, ymin=-0.0000125, ymax=0.003, xtick={0,3}]
    \addplot [name path=upper, draw=none, each nth point=20] table [col sep=comma, x=x, y=upper] {\basepathhh_gmm_stds.csv};
    \addplot [name path=lower, draw=none, each nth point=20] table [col sep=comma, x=x, y=lower] {\basepathhh_gmm_stds.csv};
    \addplot [maincolor, fill opacity=0.5] fill between[of=upper and lower];
    \addplot [line width=1pt, maincolor, each nth point=20] table[col sep=comma,x=x,y=y]{\basepathhh_gmm_stds.csv};
    \end{groupplot}
\end{tikzpicture}
}
\subfloat[DCT inpainting]{%
\begin{tikzpicture}
    \def\basepathhh{./fig/data/posteriorst/dct}
    \begin{groupplot}[convplots group plot, yticklabel style={/pgf/number format/fixed,/pgf/number format/precision=3}]
    % Laplace conditional means
    \nextgroupplot[title={$\hat{x}$}, xmin=0, xmax=22, ymin=-0.0025, ymax=0.4, xtick={0,22}]
    \addplot [name path=upper, draw=none, each nth point=20] table [col sep=comma, x=x, y=upper] {\basepathhh_laplace_means.csv};
    \addplot [name path=lower, draw=none, each nth point=20] table [col sep=comma, x=x, y=lower] {\basepathhh_laplace_means.csv};
    \addplot [maincolor, fill opacity=0.5] fill between[of=upper and lower];
    \addplot [line width=1pt, maincolor, each nth point=20] table[col sep=comma,x=x,y=y]{\basepathhh_laplace_means.csv};

    % Laplace conditional stds
    \nextgroupplot[title={$\sigma_{\hat{x}}$}, xmin=0, xmax=24, ymin=-0.0000125, ymax=0.003, xtick={0,24}]
    \addplot [name path=upper, draw=none, each nth point=20] table [col sep=comma, x=x, y=upper] {\basepathhh_laplace_stds.csv};
    \addplot [name path=lower, draw=none, each nth point=20] table [col sep=comma, x=x, y=lower] {\basepathhh_laplace_stds.csv};
    \addplot [maincolor, fill opacity=0.5] fill between[of=upper and lower];
    \addplot [line width=1pt, maincolor, each nth point=20] table[col sep=comma,x=x,y=y]{\basepathhh_laplace_stds.csv};

    % Student-t conditional means
    \nextgroupplot[xmin=0, xmax=0.3, ymin=-0.0025, ymax=0.4, xtick={0,0.3}]
    \addplot [name path=upper, draw=none, each nth point=20] table [col sep=comma, x=x, y=upper] {\basepathhh_student-t_means.csv};
    \addplot [name path=lower, draw=none, each nth point=20] table [col sep=comma, x=x, y=lower] {\basepathhh_student-t_means.csv};
    \addplot [maincolor, fill opacity=0.5] fill between[of=upper and lower];
    \addplot [line width=1pt, maincolor, each nth point=20] table[col sep=comma,x=x,y=y]{\basepathhh_student-t_means.csv};

    % Student-t conditional stds
    \nextgroupplot[xmin=0, xmax=0.3, ymin=-0.0000125, ymax=0.003, xtick={0,0.3}]
    \addplot [name path=upper, draw=none, each nth point=20] table [col sep=comma, x=x, y=upper] {\basepathhh_student-t_stds.csv};
    \addplot [name path=lower, draw=none, each nth point=20] table [col sep=comma, x=x, y=lower] {\basepathhh_student-t_stds.csv};
    \addplot [maincolor, fill opacity=0.5] fill between[of=upper and lower];
    \addplot [line width=1pt, maincolor, each nth point=20] table[col sep=comma,x=x,y=y]{\basepathhh_student-t_stds.csv};

    % GMM conditional means
    \nextgroupplot[xmin=0, xmax=2, ymin=-0.0025, ymax=0.4, xtick={0,2}]
    \addplot [name path=upper, draw=none, each nth point=20] table [col sep=comma, x=x, y=upper] {\basepathhh_gmm_means.csv};
    \addplot [name path=lower, draw=none, each nth point=20] table [col sep=comma, x=x, y=lower] {\basepathhh_gmm_means.csv};
    \addplot [maincolor, fill opacity=0.5] fill between[of=upper and lower];
    \addplot [line width=1pt, maincolor, each nth point=20] table[col sep=comma,x=x,y=y]{\basepathhh_gmm_means.csv};

    % GMM conditional stds
    \nextgroupplot[xmin=0, xmax=8, ymin=-0.0000125, ymax=0.003, xtick={0,8}]
    \addplot [name path=upper, draw=none, each nth point=20] table [col sep=comma, x=x, y=upper] {\basepathhh_gmm_stds.csv};
    \addplot [name path=lower, draw=none, each nth point=20] table [col sep=comma, x=x, y=lower] {\basepathhh_gmm_stds.csv};
    \addplot [maincolor, fill opacity=0.5] fill between[of=upper and lower];
    \addplot [line width=1pt, maincolor, each nth point=20] table[col sep=comma,x=x,y=y]{\basepathhh_gmm_stds.csv};
    \end{groupplot}
\end{tikzpicture}
}
\caption[]{%
\tikzexternaldisable
Gibbs mean squared error over time in seconds with respect to the conditional mean of the posteriors $\hat{x}$ and the pixel-wise standard deviation $\sigma_{\hat{x}}$ of the posteriors.
The solid line denotes the average runtime across all $256$ problem instances.
The shaded areas around the means denote $\pm$ one standard deviation.
\tikzexternalenable
}%
\label{fig:posterior_times}
\end{figure}

\begin{table}[!h]
    \centering
    \begin{tabular}{lrrrr}
        \toprule
        & Normal & Laplace & Student-t & \gls{gmm} \\
        \midrule
        Denoising & 00:00:21 & 113:33:16 & 5:37:48 & 32:26:19 \\
        DCT inpainting & 00:00:30 & 94:30:45 & 4:31:40 & 30:45:53  \\
        \bottomrule
    \end{tabular}
    \caption{Total runtimes of the posterior sampling experiments in hours:minutes:seconds. These correspond to the time to run $128$ chains in parallel on all $256$ problem instances for $3000$ Gibbs iterations in case of denoising and $1000$ Gibbs iterations in case of DCT inpainting. Gibbs was only run for a single iteration in case of the model with normal factors, since it that case it is a direct sampling algorithm. Note that these do not reflect the time required for Gibbs to
converge.}
    \label{tab:denoising_inpainting}
\end{table}

%% file: sections/conclusion.tex
\section{Conclusions}
\label{sec:conclusions}
We introduced the Gaussian latent machine, a flexible latent variable model that leads to efficient sampling algorithms for a wide class of distributions.  We also showed that many standard models found in Bayesian imaging can be lifted into an an exact or arbitrarily accurate \gls{glm} representation. This allows for efficient two-block Gibbs sampling in the general case,  where the first subproblem simplifies to sampling from an implictly defined multivariate Gaussian distribution, and the second subproblem simplifies to independently sampling from univariate distributions.  Finally,  we have shown the efficacy and scalability of the proposed approach on a wide range of prior and posterior sampling problems. 

\paragraph{Future Work} There are many potential directions for future work. For instance, various extensions of the proposed Gibbs sampling approach to more general settings are possible. Among those, the most tantalizing direction would be to support nonlinear features in the sense that we can perform Gibbs sampling on distributions of the form
\begin{align*}
	\fx{x} \propto \prod_{i = 1}^{m} \phi_i\br*{\br{\psi\br{x}}_i},
\end{align*}
where $\psi: \R^n \to \R^m$ is a general nonlinear mapping. Such an extension should, in principle, be possible by constructing a Markov Chain where, on each iteration, we linearize $\psi$, use the linearized model as proposal distribution, sample from the proposal distribution by using our Gibbs sampling approach, and correct the proposal via Metropolis-Hastings. It remains unclear, however, whether such an approach leads to tractable computations or whether the impressive sampling performance observed in the linear setting carries over to the nonlinear setting.

We showed in the baseline approximation experiments that different types of \gls{glm} parametrizations of the same factor lead to drastically different convergence rates in the Gibbs sampler. Hence, another important direction would be to identify and characterize the exact aspects of a \gls{glm} parametrization that influence this behaviour, and to quantify their impact.

Similarly, we have observed incredibly fast empirical convergence rates in our baseline and prior sampling experiments.  Another interesting research direction would therefore be to derive exact convergence rate bounds in general or at least for specific subclasses of distributions relevant for imaging. 

Finally, another exciting research direction is to learn image priors that are parametrized as \glspl{glm}. This would allow us to obtain principled generative models that, in contrast to some modern deep learning based generative models, can be used directly as priors in Bayesian imaging. The simple preliminary learned model from the posterior sampling experiments already achieved impressive reconstruction performance, but could not generate images that look like natural images. Hence, a critical aspect of this research direction would be to bridge the gap between image analysis and synthesis. 

\FloatBarrier

%% file: sections/appendix.tex
\section{Notation} 
\label{sec:notation}
We rely on many univariate distributions throughout this work.  Some of these distributions are not standard, while others might have multiple competing parametrizations commonly found in the literature.  Therefore,  to eliminate any source of ambiguity,  we provide a complete summary of all univariate distributions used throughout this work in \Cref{tab:distributions}.  The table contains the univariate distributions, their probability density functions, parameters,   support and shorthand notation.
{
\renewcommand{\arraystretch}{2.}
\begin{table}[t!]
\centering
\begin{adjustbox}{width=\columnwidth,center}
\begin{threeparttable}
\begin{tabular}{lllll}
	\toprule
	Name & Distribution & Parameter(s) & Support & Notation \\ 
	\midrule	
	Gaussian & $\frac{1}{\sqrt{2 \pi \sigma^2}} \cdot \exp\br*{-\frac{\br*{x - \mu}^2}{\sigma^2}}$ & $\mu \in \R, \sigma^2 \in \Rpp$ & $\R$ & $\nd$ \\
	Gaussian mixture model & $\sum_{i = 1}^d w_i \cdot \nd{x; \mu_i, \sigma^2_i}$ & $w \in \Delta_d, \mu \in \R^d, \sigma^2 \in \Rpp^d$ & $\R$ & $\GMMDist$ \\
	Exponential & $\lambda \cdot \exp\br*{-\lambda x}$ & $\lambda \in \Rpp$ & $\Rp$ & $\ExpDist$ \\
	Laplace & $\frac{1}{2b} \cdot \exp\br*{-\frac{\abs{x}}{b}}$ & $b \in \Rpp$ & $\R$ & $\LaplaceDist$ \\
	Student-t & $\frac{\Gamma\br*{\frac{\nu + 1}{2}}}{\sqrt{\pi \nu} \cdot \Gamma\br*{\frac{\nu}{2}}} \cdot \br*{1 + \frac{x^2}{\nu}}^{-\frac{\nu + 1}{2}}$ & $\nu \in \Rpp$ & $\R$ & $\tDist$ \\
	Gamma & $\frac{\beta^\alpha}{\Gamma\br*{\alpha}} \cdot x^{\alpha - 1} \cdot \exp\br*{-\beta x}$ & $\alpha, \beta \in \Rpp$ & $\Rpp$ & $\GammaDist$ \\ 
	Symmetrized Gamma & $\frac{\sqrt{2} \cdot \beta^{\alpha}}{\sqrt{\pi} \cdot \Gamma\br{\alpha}} \cdot \br*{\frac{\abs{x}}{\sqrt{2 \beta}}}^{\alpha - \frac{1}{2}} \cdot K_{\alpha - \frac{1}{2}}\br*{\sqrt{2 \beta} \cdot \abs{x}}$ & $\alpha \in \Rpp$ & $\R$ & $\SymGammaDist$ \\
	Generalized inverse Gaussian & $\frac{\br*{\frac{a}{b}}^{\frac{p}{2}}}{2 K_p\br*{\sqrt{ab}}} \cdot x^{p - 1} \cdot \exp\br*{-\frac{ax + \frac{b}{x}}{2}}$ & $a, b \in \Rpp, p \in \R$ & $\Rpp$ & $\GIGDist$ \\
	Categorical & $\sum_{i = 1}^d p_i \cdot \ib{x = i}$ & $p \in \Delta_d$ & $\set{1, \ldots, d}$ & $\CatDist$ \\
    \bottomrule
   \end{tabular}
   \begin{tablenotes}
       \item $\Gamma$ denotes the gamma function defined as $\Gamma\br{x} = \int^{\infty}_{0} t^{x - 1} \exp\br{-t} \,\mathrm{d}t$ for any $x \in \Rpp$.
       \item $K_{\nu}$ denotes the modified Bessel function of the second kind with parameter $\nu$.
       \item $\ib{\cdot}$ denotes Iverson brackets which evaluate to $1$ if the argument proposition is true, and to $0$ otherwise.
   \end{tablenotes}
\end{threeparttable}
\end{adjustbox}
\caption{Summary of univariate distributions used throughout this work.}
\label{tab:distributions}
\end{table}
}

For the sake of simplicity,  we overload our notation to describe a distribution with specific parameters and the probability density function of that distribution evaluated at a particular point.  More precisely,  we use notation of the form $\nd{\mu, \sigma^2}$ to denote a normal distribution with parameters $\mu$ and $\sigma^2$,  while we use the notation $\nd{x; \mu, \sigma^2}$ to denote the probability density function of that distribution evaluated at the point $x$.  

Similarly, we overload out notation to describe both univariate and multivariate normal distributions,  where the meaning can be inferred from the dimensionality of the arguments.  For instance, we use the notation $\nd{\mu,  \Sigma}$ to denote a multivariate normal distribution with mean vector $\mu \in \R^n$ and covariance matrix $\Sigma \in \symmpp^n$ (here $\symmpp^n$ refers to the set of all real-valued symmetric positive matrices of dimension $n \times n$), while we use the notation $\nd{x; \mu,  \Sigma}$ to denote the probability density function of that distribution evaluated at the point $x \in \R^n$.  

The symmetrized Gamma distribution is constructed through its characteristic function (\ie, the Fourier transform of its probability density function).  Therefore, for the sake of completeness, we provide a derivation of its probability density function in Section SM4 of the supplementary materials.

\newpage
\section{Proofs}
\label{sec:proofs}
\subsection{Proof of \texorpdfstring{\cref{prop:nullspace_improper_dist}}{Proposition \ref{prop:nullspace_improper_dist}}}\label{proof:nullspace_improper_dist}
\begin{proof}
	We proof the claims as follows.
	\begin{enumerate}[label=\alph*)]
	\item Assume first that $K$ is injective, that is, $N=\{0\}$. Note that this implies that $m\geq n$. Applying Givens rotations, we can find an orthonormal matrix $Q\in \R^{m\times m}$ such that 
	\begin{align*}
		Q^T K^T = L^T
	\end{align*}
	with $L\in\R^{m\times n}$ a lower triangular matrix (so that $L^T$ is an upper triangular matrix), that is, $L_{i,j}=0$ for $i<j$. 
	Moreover, by injectivity of $K$ we can assume without loss of generality that the first $n$ rows of $K$ are linearly independent, otherwise we may simply permute the columns of $K$ and analogously the functions $(\phi_i)_i$ in \eqref{def_poe}. Thus, $L_{i,i}\neq 0$ for $i=1,\dots,n$.
	Using the transformation theorem for integrals with the transformation $x=Qy$ we find that
	\begin{align*}
		\int_{\R^n} \fx{x}\mathrm{d}x &= \int_{\R^n} \prod_{i=1}^{m}\phi_i((Ly)_i)\mathrm{d}y \leq \prod_{i=n+1}^{m}\|\phi_i\|_\infty \int_{\R^n} \prod_{i=1}^{n}\phi_i((Ly)_i)\mathrm{d}y.
	\end{align*}
	By the structure of $L$ as a lower triangular matrix, we can successively integrate the $\phi_i$ for $i=n,\dots,1$ leading to 
	\begin{align*}
		\begin{aligned}
			\int_{\R^n} \fx{x}\mathrm{d}x \leq \prod_{i=n+1}^{m}\|\phi_i\|_\infty \prod_{i=1}^n\frac{\|\phi_i\|_1}{|L_{i,i}|}<\infty.
		\end{aligned}
	\end{align*}
	\item Assume now that $K$ is not injective and let $N=\ker(K)$. Let
	\begin{align*}
		B=\begin{bmatrix}
			B_N, B_{N^\perp}
		\end{bmatrix}
	\end{align*}
	be an orthonormal matrix where the columns of $B_N$ are a basis of $N$ and the columns of $B_{N^\perp}$ a basis of $N^{\perp}$. Using the transformation $x=B_{N^\perp}w$ it follows 
	\begin{align*}
		\begin{aligned}
			\int_{N^\perp} \fx{x}\mathrm{d}x = \int_{\R^{\ell}} \prod_{i=1}^{m}\phi_i((KB_{N^\perp}w)_i) \mathrm{d}w
		\end{aligned}
	\end{align*}
	where $\ell=\dim(N^\perp)$. Since $KB_{N^\perp}$ is injective finiteness of the integral follows from the injective case above. Lastly, for the integral over $\R^n$ we obtain via the transformation $x = B_Nv + B_{N^\perp}w$ and with $r = \dim(N)$ that
	\begin{align*}
		\begin{aligned}
			\int_{\R^n} \fx{x}\mathrm{d}x = \int_{\R^{r}} \int_{\R^{\ell}} \fx{B_{N^\perp}w}\mathrm{d}w\mathrm{d}v,
		\end{aligned}
	\end{align*}
	which cannot be finite if $r>0$ since $\fx>0$ by assumption on the factors $\phi_i$.
	\end{enumerate}
\end{proof}

\subsection{Proof of \texorpdfstring{\cref{prop:tie_breaking_works}}{Proposition \ref{prop:tie_breaking_works}}}\label{proof:tie_breaking_works}
\begin{proof}
Properness of $f_{\bar{X}}$ can be shown analogously to the proof of \cref{prop:nullspace_improper_dist}.
Now let $A\subset N^\perp$ be an arbitrary measurable set. Using the transformation theorem for integrals with the diffeomorphism $N\times N^\perp\rightarrow \R^d$, $(u,v)\mapsto u+v$ and Fubini's theorem, we can compute
\begin{equation}
\begin{aligned}
	\P[P_{N^\perp}\bar{X}\in A] = &\int\limits_{\R^n} \1_A (P_{N^\perp}x) \fx{x}f_0(x)\,\mathrm{d}x 
	\overset{*}{=} \int_N \int_{N^\perp} \1_A (v) \fx{v}f_0(u)\,\mathrm{d}v\,\mathrm{d}u\\
	= &\int_N \int_A \fx{v}f_0(u)\,\mathrm{d}v\,\mathrm{d}u = \int_A \fx{v}\,\mathrm{d}v \underbrace{\int_N f_0(u)\,\mathrm{d}u}_{=1}
\end{aligned}
\end{equation}
where we used the assumption on $f_0$ in the equality marked with $*$.
\end{proof}

\subsection{Proof of \texorpdfstring{\cref{prop:smart_tie_breaking_works_too}}{Proposition \ref{prop:smart_tie_breaking_works_too}}}\label{proof:smart_tie_breaking_works_too}
\begin{proof}
	Let $S$ denote the block matrix
\begin{align*}
	S \ceq 
	\begin{bmatrix}
		K \\
		\bar{K}
	\end{bmatrix}
\end{align*}
and let $B_{N} \in \R^{n \times r}$ denote a matrix whose columns are an orthonormal basis of $\ker(K)$. Since both $S$ and $B_N$ are injective, it follows that $S B_{N} v = 0$ if and only if $v=0$. 
By definition of $B_N$, on the other hand, $KB_N=0$ and, thus, 
\begin{align*}
	S B_N = \begin{bmatrix}
		0 \\
		\bar{K} B_N
	\end{bmatrix},
\end{align*}
and, consequently, that $\bar{K} B_{N} v = 0$ if and only if $v = 0$. Therefore, the square matrix $\bar{K} B_{N} \in \R^{r \times r}$ has the trivial kernel $\ker(\bar{K} B_{N}) = \set{0}$, which immediately implies that $\bar{K} B_{N}$ is invertible.

Now let in addition $B_{N^\perp} \in \R^{n \times \br{n - r}}$ be such that its columns are an orthonormal basis of $N^\perp$ so that $B \ceq \begin{bmatrix} B_{N^\perp} & B_N \end{bmatrix} \in \R^{n \times n}$ is an orthonormal basis of $\R^n$. Let $U \in \R^n$ denote the vector
\begin{align*}
	U \ceq 
	\begin{bmatrix}
		U_{N^\perp} \\
		U_N
	\end{bmatrix},
\end{align*}
where $U_{N^\perp} \in \R^{n - r}$, and $U_N \in \R^r$. We have by assumption that
\begin{align*}
	f_{\bar{X}} \br{x} &\propto \fx{x} \cdot f_0(x) \\
		&= \br*{\prod_{i = 1}^{m} \phi_i\bigl(\br*{K x}_i\bigr)} \cdot \br*{\prod_{i = 1}^{r} \bar{\phi}_i\bigl((\bar{K} x)_i\bigr)},
\end{align*}
and thus by the change of variables 
\begin{align*}
	B U = \bar{X},
\end{align*}
and the fact that $BU = B_{N^\perp} U_{N^\perp} + B_N U_N$ it follows that
\begin{align*}
	&f_{U_{N^\perp}, U_N}\br{u_{N^\perp}, u_N} \\
	&\qquad\propto \br*{\prod_{i = 1}^{m} \phi_i\br*{\br*{K B_{N^\perp} u_{N^\perp} + K B_N u_N}_i}} \cdot \br*{\prod_{i = 1}^{r} \bar{\phi}_i\br*{\br*{\bar{K} B_{N^\perp} u_{N^\perp} + \bar{K} B_N u_N}_i}} \\
	&\qquad= \br*{\prod_{i = 1}^{m} \phi_i\br*{\br*{K B_{N^\perp} u_{N^\perp}}_i}} \cdot \br*{\prod_{i = 1}^{r} \bar{\phi}_i\br*{\br*{\bar{K} B_{N^\perp} u_{N^\perp} + \bar{K} B_N u_N}_i}}
\end{align*}
since \( K B_N = 0 \).
Marginalizing out over $U_N$ yields
\begin{align*}
	f_{U_{N^\perp}}\br{u_{N^\perp}} &\propto \int_{R^r} f_{U_{N^\perp}, U_N}\br{u_{N^\perp}, u_N} \;\mathrm{d}u_N \\
		&= \br*{\prod_{i = 1}^{m} \phi_i\br*{\br*{K B_{N^\perp} u_{N^\perp}}_i}} \cdot \int_{\R^r} \br*{\prod_{i = 1}^{r} \bar{\phi}_i\br*{\br*{\bar{K} B_{N^\perp} u_{N^\perp} + \bar{K} B_N u_N}_i}} \;\mathrm{d}u_N.
\end{align*}
We have already established that $\bar{K} B_N$ is an invertible matrix and therefore by the change of variables $z = \bar{K} B_N u_N$ it follows that
\begin{align*}
		f_{U_{N^\perp}}\br{u_{N^\perp}} &\propto \br*{\prod_{i = 1}^{m} \phi_i\br*{\br*{K B_{N^\perp} u_{N^\perp}}_i}} \cdot \int_{\R^r} \br*{\prod_{i = 1}^{r} \bar{\phi}_i\br*{\br*{\bar{K} B_{N^\perp} u_{N^\perp}}_i + z_i}} \;\mathrm{d}z. \\
			&= \br*{\prod_{i = 1}^{m} \phi_i\br*{\br*{K B_{N^\perp} u_{N^\perp}}_i}} \cdot \br*{\prod_{i = 1}^{r} \int_{\R} \bar{\phi}_i\br*{\br*{\bar{K} B_{N^\perp} u_{N^\perp}}_i + z_i} \;\mathrm{d}z_i} \\
			&= \prod_{i = 1}^{m} \phi_i\br*{\br*{K B_{N^\perp} u_{N^\perp}}_i}, 
\end{align*}
since
\begin{align*}
\int_{\R} \bar{\phi}_i\br*{\br*{\bar{K} B_{N^\perp} u_{N^\perp}}_i + z_i} \;\mathrm{d}z_i = 1
\end{align*} 
for any $\br*{\bar{K} B_{N^\perp} u_{N^\perp}}_i$ and $i = 1,2, \ldots, r$ due to the assumption that $\bar{\phi}_i$ are univariate densities. The desired result then trivially follows from the fact that $P_{N^\perp} \bar{X} = B_{N^\perp} U_{N^\perp}$.
\end{proof}

\subsection{Proof of \texorpdfstring{\cref{prop:marginalization}}{Proposition \ref{prop:marginalization}}}\label{proof:marginalization}
\begin{proof}
Defining the \gls{glm} $\fxz{x, z} \coloneqq \prod_{i = 1}^m g_i\br*{\br{Kx}_i, z_i} \cdot \f{i}{z_i}$ we can simply compute
\begin{align*}
	\int_{\mathcal{Z}} \fxz{x, z}\,\mathrm{d}z &\propto \int_{\mathcal{Z}} \br*{\prod_{i = 1}^m g_i\br*{\br{Kx}_i, z_i} \cdot \f{i}{z_i}}\,\mathrm{d}z = \prod_{i = 1}^m \underbrace{\br*{\int_{\mathcal{Z}_i} g_i\br*{\br{Kx}_i, z_i} \cdot \f{i}{z_i}\,\mathrm{d}z_i}}_{= \phi_i\br*{\br*{K x}_i} \text{ by \eqref{def_fmp}}} \\
	&= \prod_{i = 1}^m \phi_i\br*{\br*{K x}_i},
\end{align*}
which concludes the proof.
\end{proof}

\subsection{Proof of \texorpdfstring{\cref{prop:xgz_mvg}}{Proposition \ref{prop:xgz_mvg}}}\label{proof:xgz_mvg}
\begin{proof} Observe that $K$ is a $m \times n$ matrix with full rank and $m \geq 0$, and $\Sigma_0\br*{z}$ is a $m \times m$ diagonal matrix with positive entries for any given value of $z$. Thus, $\Sigma_0\br*{z}$ is a symmetric positive definite matrix and, consequently, its inverse $\Sigma^{-1}_0\br*{z}$ exists and is a symmetric positive definite matrix. Similarly, the matrices $K^\top K$ and $K^\top \Sigma^{-1}_0\br*{z} K$ are symmetric positive definite by construction and therefore their inverses exist and are symmetric positive definite matrices. Based on this, it follows from \eqref{def_glm} that
	\begin{align*}
		\fxgz{\xgz} &\propto \nd*{Kx; \mu_0\br*{z}, \Sigma_0\br*{z}} \cdot \prod_{i = 1}^m \f{i}{z_i} \propto \nd*{Kx; \mu_0\br*{z}, \Sigma_0\br*{z}} \\
		&\propto \exp\bigl(-\tfrac{1}{2} \norm*{Kx - \mu_0\br*{z}}^2_{\Sigma^{-1}_0\br*{z}}\bigr) = \exp\bigl(-\tfrac{1}{2} \br*{Kx - \mu_0\br*{z}}^\top \Sigma^{-1}_0\br*{z} \br*{Kx - \mu_0\br*{z}}\bigr) \\
		&\propto \exp\bigl(-\tfrac{1}{2} \bigl(\norm*{Kx}^2_{\Sigma^{-1}_0\br*{z}} - 2 \cdot x^\top K^\top \Sigma^{-1}_0\br*{z} \mu_0\br*{z}\bigr)\bigr) \\
		&= \exp\bigl(-\tfrac{1}{2} \bigl(\norm*{x}^2_{\underbrace{K^\top \Sigma^{-1}_0\br*{z} K}_{\Sigma^{-1}\br*{z}}} - 2 \cdot x^\top K^\top \Sigma^{-1}_0\br*{z} \mu_0\br*{z}\bigr)\bigr) \\
		&= \exp\bigl(-\tfrac{1}{2} \bigl(\norm*{x}^2_{\Sigma^{-1}\br*{z}} - 2 \cdot x^\top K^\top \Sigma^{-1}_0\br*{z} \mu_0\br*{z}\bigr)\bigr) \\
		&= \exp\bigl(-\tfrac{1}{2} \bigl(\norm*{x}^2_{\Sigma^{-1}\br*{z}} - 2 \cdot x^\top \underbrace{\Sigma^{-1}\br*{z} \Sigma\br*{z}}_{I_n} K^\top \Sigma^{-1}_0\br*{z} \mu_0\br*{z}\bigr)\bigr) \\
		&= \exp\bigl(-\tfrac{1}{2} \bigl(\norm*{x}^2_{\Sigma^{-1}\br*{z}} - 2 \cdot x^\top \Sigma^{-1}\br*{z} \underbrace{\Sigma\br*{z} K^\top \Sigma^{-1}_0\br*{z} \mu_0\br*{z}}_{\mu\br*{z}}\bigr)\bigr) \\
		&= \exp\bigl(-\tfrac{1}{2} \bigl(\norm*{x}^2_{\Sigma^{-1}\br*{z}} - 2 \cdot x^\top \Sigma^{-1}\br*{z} \mu\br*{z}\bigr)\bigr) \propto \exp\bigl(-\tfrac{1}{2} \norm*{x - \mu\br*{z}}^2_{\Sigma^{-1}\br*{z}}\bigr) \\
		&\propto \nd*{x; \mu\br*{z}, \Sigma\br*{z}},
	\end{align*}
	where
	\begin{align*}
		\Sigma\br*{z} \ceq \br*{K^\top \Sigma^{-1}_0\br*{z} K}^{-1} \quad \text{and} \quad \mu\br*{z} \ceq \Sigma\br*{z} K^\top \Sigma^{-1}_0\br*{z} \mu_0\br*{z},
	\end{align*}
	as desired.
\end{proof}

\subsection{Proof of \texorpdfstring{\cref{prop:latent_distribution}}{Proposition \ref{prop:latent_distribution}}}\label{proof:latent_distribution}
\begin{proof}
Note that by the same arguments as in the proof of \cref{prop:xgz_mvg} it follows that
\begin{align*}
	\nd*{Kx; \mu_0\br*{z}, \Sigma_0\br*{z}} &\deq \frac{1}{\sqrt{\br*{2 \pi}^m \det \Sigma_0\br*{z}}} \cdot \exp\br*{-\frac{1}{2} \norm*{Kx - \mu_0\br*{z}}^2_{\Sigma^{-1}_0\br*{z}}} \\
	&=\frac{\exp\br*{\frac{1}{2} \norm*{\mu\br*{z}}^2_{\Sigma^{-1}\br*{z}}}}{\sqrt{\br*{2 \pi}^m \det \Sigma_0\br*{z}}} \cdot \exp\br*{-\frac{1}{2} \norm*{Kx - \mu_0\br*{z}}^2_{\Sigma^{-1}_0\br*{z}} - \frac{1}{2} \norm*{\mu\br*{z}}^2_{\Sigma^{-1}\br*{z}}} \\
	&=\frac{\exp\br*{\frac{1}{2} \norm*{\mu\br*{z}}^2_{\Sigma^{-1}\br*{z}} - \frac{1}{2} \norm*{\mu_0\br*{z}}^2_{\Sigma^{-1}_0\br*{z}}}}{\sqrt{\br*{2 \pi}^m \det \Sigma_0\br*{z}}} \cdot \exp\br*{-\frac{1}{2} \norm*{x - \mu\br*{z}}^2_{\Sigma^{-1}\br*{z}}} \\
	% &= \frac{\exp\br*{\frac{1}{2} \norm*{\mu\br*{z}}^2_{\Sigma^{-1}\br*{z}} - \frac{1}{2} \norm*{\mu_0\br*{z}}^2_{\Sigma^{-1}_0\br*{z}}}}{\sqrt{\br*{2 \pi}^m \det \Sigma_0\br*{z}}} \cdot \nd*{x; \mu\br*{z}, \Sigma\br*{z}}.
\end{align*}
From this, \eqref{def_glm} and the definitions of $\mu_0$ and $\Sigma_0$ it follows that 
\begin{align*}
	\fz{z} &= \int_{\R^n} \fxz{x, z}\,\mathrm{d}x
	 \propto \int_{\R^n} \br*{\nd*{Kx; \mu_0\br*{z}, \Sigma_0\br*{z}} \cdot \prod_{i = 1}^m \f{i}{z_i}}\,\mathrm{d}x \\
		   &=  \prod_{i = 1}^m \f{i}{z_i} \cdot \int_{\R^n} \nd*{Kx; \mu_0\br*{z}, \Sigma_0\br*{z}}\,\mathrm{d}x\\
		   &= \frac{\exp\br*{\frac{1}{2} \norm*{\mu\br*{z}}^2_{\Sigma^{-1}\br*{z}} - \frac{1}{2} \norm*{\mu_0\br*{z}}^2_{\Sigma^{-1}_0\br*{z}}}}{\sqrt{\br*{2 \pi}^m \det \Sigma_0\br*{z}}} \cdot \prod_{i = 1}^m \f{i}{z_i} \cdot \underbrace{\int_{\R^n} \exp\br*{-\frac{1}{2} \norm*{x - \mu\br*{z}}^2_{\Sigma^{-1}\br*{z}}}\,\mathrm{d}x}_{= \sqrt{\br*{2 \pi}^n \det \Sigma\br*{z}}} \\
	&\propto \sqrt{\frac{\det \Sigma\br*{z}}{\det \Sigma_0\br*{z}}} \cdot \frac{\exp\br*{\frac{1}{2} \norm*{\mu\br*{z}}^2_{\Sigma^{-1}\br*{z}}}}{\exp\br*{\frac{1}{2} \norm*{\mu_0\br*{z}}^2_{\Sigma^{-1}_0\br*{z}}}} \cdot \prod_{i = 1}^m \f{i}{z_i} \\
	&= \sqrt{\det \Sigma\br*{z}} \cdot \exp\br*{\frac{1}{2} \norm*{\mu\br*{z}}^2_{\Sigma^{-1}\br*{z}}} \cdot \prod_{i = 1}^m \frac{\f{i}{z_i}}{\sigma_i\br*{z_i} \cdot \exp\br*{\frac{1}{2} \frac{\mu^2_i\br*{z_i}}{\sigma^2_i\br*{z_i}}}} \\
	&= \g{z} \cdot \prod_{i = 1}^m \frac{\f{i}{z_i}}{\sigma_i\br*{z_i} \cdot \exp\br*{\frac{1}{2} \frac{\mu^2_i\br*{z_i}}{\sigma^2_i\br*{z_i}}}},
\end{align*}
where
\begin{align*}
	\g{z} \ceq \sqrt{\det \Sigma\br*{z}} \cdot \exp\br*{\frac{1}{2} \norm{\mu\br*{z}}^2_{\Sigma^{-1}\br*{z}}},
\end{align*}
as desired.
\end{proof}

\subsection{Proof of \texorpdfstring{\cref{prop:fz_decomposes}}{Proposition \ref{prop:fz_decomposes}}}\label{proof:fz_decomposes}
\begin{proof}
It trivially follows from \eqref{def_glm} that
	\begin{align*}
		\fzgx{\zgx} &= \frac{\prod_{i = 1}^m g_i\br*{\br{Kx}_i,  z_i} \cdot \f{i}{z_i}}{\int_{\R^m} \br*{\prod_{i = 1}^m g_i\br*{\br{Kx}_i,  z_i} \cdot \f{i}{z_i}}\,\mathrm{d}z} 
		= \frac{\prod_{i = 1}^m g_i\br*{\br{Kx}_i,  z_i} \cdot \f{i}{z_i}}{\prod_{i = 1}^m \br*{\int_{\R} g_i\br*{\br{Kx}_i,  z_i} \cdot \f{i}{z_i}\,\mathrm{d}z_i}}  \\
		&= \prod_{i = 1}^m \frac{g_i\br*{\br{Kx}_i,  z_i} \cdot \f{i}{z_i}}{\int_{\R} g_i\br*{\br{Kx}_i,  z_i} \cdot \f{i}{z_i}\,\mathrm{d}z_i} 
			= \prod_{i = 1}^m \f_{Z_i \mid X}\br*{z_i \mid x},
	\end{align*}
	where
	\begin{align*}
		\f_{Z_i \mid X}\br*{z_i \mid x} \propto g_i\br*{\br{Kx}_i,  z_i} \cdot \f{i}{z_i} \quad \text{for} \quad i = 1, \ldots, m,
	\end{align*}
	as desired.
\end{proof}

\subsection{Proof of \texorpdfstring{\cref{prop:mvg_perturb_map}}{Proposition \ref{prop:mvg_perturb_map}}}\label{proof:mvg_perturb_map}
\begin{proof}
Note that by the same argument as in the proof of \cref{prop:xgz_mvg}, it follows that the matrices $\Sigma_0$ and $K^\top \Sigma^{-1}_0 K$ are symmetric positive definite and that their inverses exist and are also symmetric positive definite matrices. The random variable $X$ is defined as a linear transformation of a multivariate Gaussian distribution, which implies that $X$ is a multivariate Gaussian distribution. Thus, we have that $X \da \nd{\mu, \Sigma}$ with mean
\begin{align*}
	\mu &= \E{X} = \E{\underbrace{\br{K^\top \Sigma^{-1}_0 K}^{-1}}_{\Sigma \ceq} K^\top \Sigma^{-1}_0 Y} = \E{\Sigma K^\top \Sigma^{-1}_0 Y} 
		= \Sigma K^\top \Sigma^{-1}_0 \underbrace{\E{Y}}_{\mu_0} = \Sigma K^\top \Sigma^{-1}_0 \mu_0,
\end{align*}
and covariance matrix
\begin{align*}
	\cov X &= \cov\br{\underbrace{\br{K^\top \Sigma^{-1}_0 K}^{-1}}_{\Sigma} K^\top \Sigma^{-1}_0 Y} = \cov\br{\Sigma K^\top \Sigma^{-1}_0 Y} 
		= \Sigma K^\top \Sigma^{-1}_0 \cov{Y} \br{\Sigma K^\top \Sigma^{-1}_0}^\top \\
		&= \Sigma K^\top \Sigma^{-1}_0 \underbrace{\cov{Y}}_{\Sigma_0} \Sigma^{-1}_0 K \Sigma 
		= \Sigma K^\top \Sigma^{-1}_0 \underbrace{\Sigma_0 \Sigma^{-1}_0}_{I_m} K \Sigma 
		= \Sigma \underbrace{K^\top \Sigma^{-1}_0 K}_{\Sigma^{-1}} \Sigma 
		= \Sigma \underbrace{\Sigma^{-1} \Sigma}_{I_n} \\
		&= \Sigma,
\end{align*}
as desired.
\end{proof}

\subsection{Proof of \texorpdfstring{\cref{prop:precond}}{Proposition \ref{prop:precond}}}\label{proof:precond}
\begin{proof}
	We prove the claims as follows.
	\begin{enumerate}[label=\alph*)]
		\item From the definitions of $K$ and $\Sigma^{-1}$ it follows that $K^\top \Sigma^{-1}_0 K = \sum_{i = 1}^k K^\top_i \Sigma^{-1}_i K_i$. This immediately implies that $\diag\br{K^\top \Sigma^{-1}_0 K} = \sum_{i = 1}^k \diag{K^\top_i \Sigma^{-1}_i K_i}$, as desired.
		\item Let $k^{\br{i}}_j \in \R^{m_i}$ denote the $j$th column of the matrix $K_i$ for $j = 1, \ldots, n$. Since $\Sigma^{-1}_i$ is assumed to be a diagonal matrix, it follows that the $j$th diagonal entry of the matrix $K^\top_i \Sigma^{-1}_i K_i$ is given by the inner product $\ip{k^{\br{i}}_j \odot \diag \Sigma^{-1}_i }{k^{\br{i}}_j}$, where $\odot$ denotes the standard Hadamard (elementwise) product of vectors. This immediately implies that $\diag{K^\top_i \Sigma^{-1}_i K_i} = \br{K^{\circ 2}_i}^\top \diag \Sigma^{-1}_i$, as desired. 
	\end{enumerate}
\end{proof}

\subsection{Proof of \texorpdfstring{\cref{prop:direct_sampling}}{Proposition \ref{prop:direct_sampling}}}\label{proof:direct_sampling}
\begin{proof}
Since $K$ is invertible, it follows that the inverse transformation $X = \inv{K} U$ exists and consequently by the change of variables formula that
	\begin{align*}
		\f{U}{u} =& \frac{1}{\abs{\det K}} \cdot \fx{\underbrace{\inv{K} u}_{= x}} \propto \frac{1}{\abs{\det K}} \cdot \fx{x} \deq \frac{1}{\abs{\det K}} \cdot \prod^n_{i = 1} \phi_i\br{\underbrace{\br*{Kx}_i}_{= u_i}} = \frac{1}{\abs{\det K}} \cdot \prod^n_{i = 1} \phi_i\br*{u_i}.
	\end{align*}
	The desired result trivially follows from the observations that $\phi_1, \ldots, \phi_n$ are normalized univariate densities and that $\abs{\det K} > 0$ (since $K$ is invertible) is a constant that is independent of \( u \).
\end{proof}

\subsection{Proof of \texorpdfstring{\cref{prop:tree_efficient_sampler}}{Proposition \ref{prop:tree_efficient_sampler}}}\label{proof:tree_efficient_sampler}
\begin{proof}
	We begin with the system of equations
	\begin{align*}
		x_j - x_i =& u_{ij} \quad \text{for all} \quad ij \in \edge.
	\end{align*}
	By recursively substituting from the root node toward the leaf nodes,  it follows that
	\begin{align*}
		x_i = x_1 + z_i \quad \text{for all} \quad i \in \node.
	\end{align*}
	This expresses the value $x_i$ at every node $i \in \node$ in terms of $x_1$ (the value at the root node) and its total distance $z_i$ from the root node. Substituting these expressions back into the quality $\sum_{i = 1}^n x_i = u_0$ yields
	\begin{align*}
		x_1 = \frac{1}{n}\br*{u_0 - \sum_{i = 1}^n z_i},
	\end{align*}
	as desired.
\end{proof}

\subsection{Proof of \texorpdfstring{\cref{prop:complete_independent_latents}}{Proposition \ref{prop:complete_independent_latents}}}\label{proof:complete_independent_latents}
\begin{proof}
	Since $K$ is invertible, it follows that the inverse transformation $X = \inv{K} U$ exists and consequently by the change of variables formula that we can reparametrize $f_{X, Z}$ as
	\begin{align*}
		f_{U,Z}\br{u, z} \propto \prod^n_{i = 1} g_i\br{u_i, z_i} \cdot f_i\br{z_i}.
	\end{align*}
	Therefore, it follows that
	\begin{align*}
		\fz{z} \propto& \int_{\R^n} \prod^n_{i = 1} g_i\br{u_i, z_i} \cdot f_i\br{z_i} \,\mathrm{d}u = \br*{\prod^n_{i = 1} \f_i\br{z_i}} \cdot \int_{\R^n} \prod^n_{i = 1} g_i\br{u_i, z_i} \,\mathrm{d}u \\
		=& \br*{\prod^n_{i = 1} \f_i\br{z_i}} \cdot \prod^n_{i = 1} \int_{\R} g_i\br{u_i, z_i} \,\mathrm{d}u_i = \prod^n_{i = 1} \f_i\br{z_i},
	\end{align*}	
	since
	\begin{align*}
		\int_{\R} g_i\br{u_i, z_i} \,\mathrm{d}u_i = 1 \quad \text{for all } z_i \in \mathcal{Z}_i \text{ and } i = 1, \ldots, n.
	\end{align*}
	The desired result trivially follows by marginalizing out over the components of $Z$. 
\end{proof}

%% file: sections/acknowledgements.tex
This project has received funding from the European Union’s EIC Pathfinder Challenges 2022 programme under grant agreement No~101115317 (NEO). Views and opinions expressed are however those of the author(s) only and do not necessarily reflect those of the European Union or European Innovation Council. Neither the European Union nor the
European Innovation Council can be held responsible for them. 
\begin{center}
\includegraphics[width=0.95\textwidth]{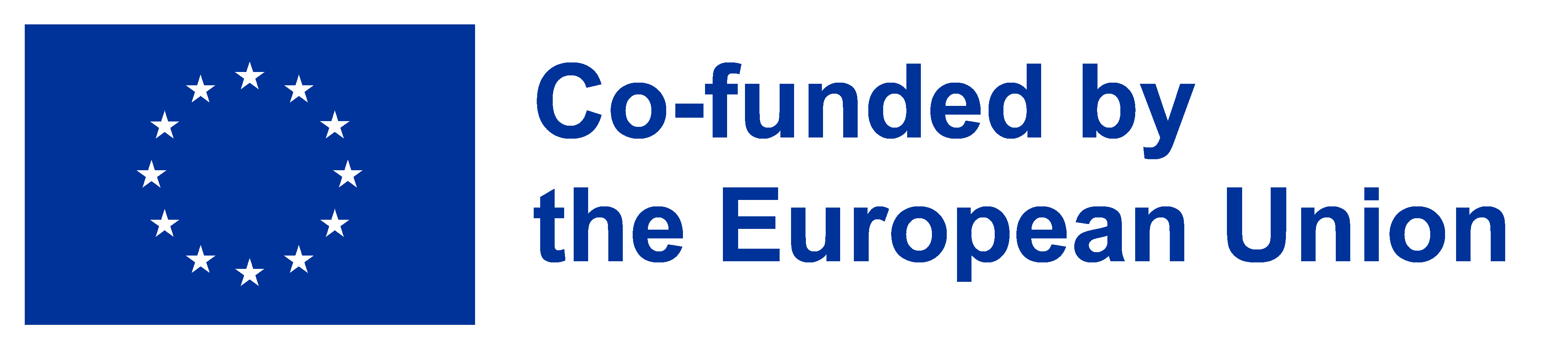}
\end{center}
The authors acknowledge the use of AI tools to assist with editing and polishing the text for grammar, spelling, and style.

%% file: references.bib
@article{Aharon:2006,
  author  = {Aharon, M. and Elad, M. and Bruckstein, A.},
  doi     = {10.1109/TSP.2006.881199},
  title   = {{K-SVD}: {A}n {A}lgorithm for {D}esigning {O}vercomplete {D}ictionaries for {S}parse {R}epresentation},
  journal = {IEEE Transactions on Signal Processing},
  year    = {2006},
  volume  = {54},
  number  = {11},
  pages   = {4311--4322},
}

@article{Andrews:1974,
  author  = {Andrews, D. F. and Mallows, C. L.},
  doi     = {10.1111/j.2517-6161.1974.tb00989.x},
  title   = {Scale {M}ixtures of {N}ormal {D}istributions},
  journal = {Journal of the Royal Statistical Society: Series B (Methodological)},
  year    = {1974},
  volume  = {36},
  number  = {1},
  pages   = {99--102},
}

@article{Arbelaez:2011,
  author  = {Arbeláez, Pablo and Maire, Michael and Fowlkes, Charless and Malik, Jitendra},
  doi     = {10.1109/TPAMI.2010.161},
  title   = {Contour {D}etection and {H}ierarchical {I}mage {S}egmentation},
  journal = {IEEE Transactions on Pattern Analysis and Machine Intelligence},
  year    = {2011},
  volume  = {33},
  number  = {5},
  pages   = {898--916},
}

@article{Bardsley:2012,
  author  = {Bardsley, Johnathan M.},
  doi     = {10.1137/11085760X},
  title   = {{MCMC}-{B}ased {I}mage {R}econstruction with {U}ncertainty {Q}uantification},
  journal = {SIAM Journal on Scientific Computing},
  year    = {2012},
  volume  = {34},
  number  = {3},
  pages   = {A1316--A1332},
}

@article{Bardsley:2014,
  author  = {Bardsley, Johnathan M. and Solonen, Antti and Haario, Heikki and Laine, Marko},
  doi     = {10.1137/140964023},
  title   = {{R}andomize-{T}hen-{O}ptimize: {A} {M}ethod for {S}ampling from {P}osterior {D}istributions in {N}onlinear {I}nverse {P}roblems},
  journal = {SIAM Journal on Scientific Computing},
  year    = {2014},
  volume  = {36},
  number  = {4},
  pages   = {A1895--A1910},
}

@article{Bohra:2023,
  author  = {Bohra, Pakshal and del Aguila Pla, Pol and Giovannelli, Jean-François and Unser, Michael},
  doi     = {10.1109/TSP.2023.3282062},
  title   = {A {S}tatistical {F}ramework to {I}nvestigate the {O}ptimality of {S}ignal-{R}econstruction {M}ethods},
  journal = {IEEE Transactions on Signal Processing},
  year    = {2023},
  volume  = {71},
  number  = {},
  pages   = {2043--2055},
}

@article{Bredies:2010,
  author  = {Bredies, Kristian and Kunisch, Karl and Pock, Thomas},
  doi     = {10.1137/090769521},
  title   = {Total {G}eneralized {V}ariation},
  journal = {SIAM Journal on Imaging Sciences},
  year    = {2010},
  volume  = {3},
  number  = {3},
  pages   = {492--526},
}

@article{Casella:1992,
  author  = {Casella, George and George, Edward I.},
  doi     = {10.1080/00031305.1992.10475878},
  title   = {Explaining the {G}ibbs {S}ampler},
  journal = {The American Statistician},
  year    = {1992},
  volume  = {46},
  number  = {3},
  pages   = {167--174},
}

@article{Chambolle:1997,
  author  = {Chambolle, Antonin and Lions, Pierre-Louis},
  doi     = {10.1007/s002110050258},
  title   = {Image {R}ecovery via {T}otal {V}ariation {M}inimization and {R}elated {P}roblems},
  journal = {Numerische Mathematik},
  year    = {1997},
  volume  = {76},
  number  = {2},
  pages   = {167--188},
}

@article{Dalalyan:2016,
  author  = {Dalalyan, Arnak S.},
  doi     = {10.1111/rssb.12183},
  title   = {Theoretical {G}uarantees for {A}pproximate {S}ampling from {S}mooth and {L}og-{C}oncave {D}ensities},
  journal = {Journal of the Royal Statistical Society Series B: Statistical Methodology},
  year    = {2016},
  volume  = {79},
  number  = {3},
  pages   = {651--676},
}

@article{Devroye:2014,
  author  = {Devroye, Luc},
  doi     = {10.1007/s11222-012-9367-z},
  title   = {Random {V}ariate {G}eneration for the {G}eneralized {I}nverse {G}aussian {D}istribution},
  journal = {Statistics and Computing},
  year    = {2014},
  volume  = {24},
  number  = {2},
  pages   = {239--246},
}

@article{Diamond:2017,
  author  = {Diamond, Steven and Boyd, Stephen},
  doi     = {10.1007/s10957-016-0990-2},
  title   = {Stochastic {M}atrix-{F}ree {E}quilibration},
  journal = {Journal of Optimization Theory and Applications},
  year    = {2017},
  volume  = {172},
  number  = {2},
  pages   = {436--454},
}

@article{Durmus:2017,
  author  = {Alain Durmus and {\'E}ric Moulines},
  doi     = {10.1214/16-AAP1238},
  title   = {Nonasymptotic {C}onvergence {A}nalysis for the {U}nadjusted {L}angevin {A}lgorithm},
  journal = {The Annals of Applied Probability},
  year    = {2017},
  volume  = {27},
  number  = {3},
  pages   = {1551--1587},
}

@article{Durmus:2018,
  author  = {Durmus, Alain and Moulines, \'{E}ric and Pereyra, Marcelo},
  doi     = {10.1137/16M1108340},
  title   = {Efficient {B}ayesian {C}omputation by {P}roximal {M}arkov {C}hain {M}onte {C}arlo: {W}hen {L}angevin {M}eets {M}oreau},
  journal = {SIAM Journal on Imaging Sciences},
  year    = {2018},
  volume  = {11},
  number  = {1},
  pages   = {473--506},
}

@article{Durmus:2019,
  author  = {Alain Durmus and Szymon Majewski and B{{\l}}a{\.{z}}ej Miasojedow},
  doi     = {},
  title   = {Analysis of {L}angevin {M}onte {C}arlo via {C}onvex {O}ptimization},
  journal = {Journal of Machine Learning Research},
  year    = {2019},
  volume  = {20},
  number  = {73},
  pages   = {1--46},
}

@article{Durmus:2022,
  author  = {Durmus, Alain and Moulines, \'{E}ric and Pereyra, Marcelo},
  doi     = {10.1137/22M1522917},
  title   = {A {P}roximal {M}arkov {C}hain {M}onte {C}arlo {M}ethod for {B}ayesian {I}nference in {I}maging {I}nverse {P}roblems: {W}hen {L}angevin {M}eets {M}oreau},
  journal = {SIAM Review},
  year    = {2022},
  volume  = {64},
  number  = {4},
  pages   = {991--1028},
}

@article{Habring:2024,
  author  = {Habring, Andreas and Holler, Martin and Pock, Thomas},
  doi     = {10.1137/23M1591451},
  title   = {Subgradient {L}angevin {M}ethods for {S}ampling from {N}onsmooth {P}otentials},
  journal = {SIAM Journal on Mathematics of Data Science},
  year    = {2024},
  volume  = {6},
  number  = {4},
  pages   = {897--925},
}

@article{Gilavert:2015,
  author  = {Gilavert, Clément and Moussaoui, Saïd and Idier, Jérôme},
  doi     = {10.1109/TSP.2014.2367457},
  title   = {Efficient {G}aussian {S}ampling for {S}olving {L}arge-{S}cale {I}nverse {P}roblems {U}sing {MCMC}},
  journal = {IEEE Transactions on Signal Processing},
  year    = {2015},
  volume  = {63},
  number  = {1},
  pages   = {70--80},
}

@article{Green:2015,
  author  = {Green, Peter J. and Latuszynski, Krzysztof and Pereyra, Marcelo and Robert, Christian P.},
  doi     = {10.1007/s11222-015-9574-5},
  title   = {Bayesian {C}omputation: {A} {S}ummary of the {C}urrent {S}tate, and {S}amples {B}ackwards and {F}orwards},
  journal = {Statistics and Computing},
  year    = {2015},
  volume  = {25},
  number  = {4},
  pages   = {835--862},
}

@article{Goujon:2023,
  author  = {Goujon, Alexis and Neumayer, Sebastian and Bohra, Pakshal and Ducotterd, Stanislas and Unser, Michael},
  doi     = {10.1109/TCI.2023.3306100},
  title   = {A {N}eural-{N}etwork-{B}ased {C}onvex {R}egularizer for {I}nverse {P}roblems},
  journal = {IEEE Transactions on Computational Imaging},
  year    = {2023},
  volume  = {9},
  number  = {?},
  pages   = {781--795},
}

@article{Hinton:2002,
  author  = {Hinton, Geoffrey E.},
  doi     = {10.1162/089976602760128018},
  title   = {Training {P}roducts of {E}xperts by {M}inimizing {C}ontrastive {D}ivergence},
  journal = {Neural Computation},
  year    = {2002},
  volume  = {14},
  number  = {8},
  pages   = {1771--1800},
}

@article{Hoermann:2014,
  author  = {Hörmann, Wolfgang and Leydold, Josef},
  doi     = {10.1007/s11222-013-9387-3},
  title   = {Generating {G}eneralized {I}nverse {G}aussian {R}andom {V}ariates},
  journal = {Statistics and Computing},
  year    = {2014},
  volume  = {24},
  number  = {4},
  pages   = {547--557},
}

@article{Kolmogorov:2016,
  author  = {Kolmogorov, Vladimir and Pock, Thomas and Rolinek, Michal},
  doi     = {10.1137/15M1010257},
  title   = {Total {V}ariation on a {T}ree},
  journal = {SIAM Journal on Imaging Sciences},
  year    = {2016},
  volume  = {9},
  number  = {2},
  pages   = {605--636},
}

@article{Kschischang:2001,
  author  = {Kschischang, F. R. and Frey, B. J. and Loeliger, H.-A.},
  doi     = {10.1109/18.910572},
  title   = {Factor {G}raphs and the {S}um-{P}roduct {A}lgorithm},
  journal = {IEEE Transactions on Information Theory},
  year    = {2001},
  volume  = {47},
  number  = {2},
  pages   = {498--519},
}

@article{Kuric:2024,
  author  = {Kuric, Muhamed and Ahmetspahic, Jan and Pock, Thomas},
  doi     = {10.1137/23M1556915},
  title   = {Total {G}eneralized {V}ariation on a {T}ree},
  journal = {SIAM Journal on Imaging Sciences},
  year    = {2024},
  volume  = {17},
  number  = {2},
  pages   = {1040--1077},
}

@article{Lewicki:2000,
  author  = {Lewicki, Michael S. and Sejnowski, Terrence J.},
  doi     = {10.1162/089976600300015826},
  title   = {Learning {O}vercomplete {R}epresentations},
  journal = {Neural Computation},
  year    = {2000},
  volume  = {12},
  number  = {2},
  pages   = {337--365},
}

@article{Nikolova:2005,
  author  = {Nikolova, Mila and Ng, Michael K.},
  doi     = {10.1137/030600862},
  title   = {Analysis of {H}alf-{Q}uadratic {M}inimization {M}ethods for {S}ignal and {I}mage {R}ecovery},
  journal = {SIAM Journal on Scientific Computing},
  year    = {2005},
  volume  = {27},
  number  = {3},
  pages   = {937--966},
}

@article{Nickolls:2008,
  author  = {Nickolls, John and Buck, Ian and Garland, Michael and Skadron, Kevin},
  doi     = {10.1145/1365490.1365500},
  title   = {Scalable {P}arallel {P}rogramming with {CUDA}: {I}s {CUDA} the {P}arallel {P}rogramming {M}odel that {A}pplication {D}evelopers have been waiting for?},
  journal = {Queue},
  year    = {2008},
  volume  = {6},
  number  = {2},
  pages   = {40--53},
}

@article{Orieux:2012,
  author  = {Orieux, F. and Feron, O. and Giovannelli, J.-F.},
  doi     = {10.1109/LSP.2012.2189104},
  title   = {Sampling {H}igh-{D}imensional {G}aussian {D}istributions for {G}eneral {L}inear {I}nverse {P}roblems},
  journal = {IEEE Signal Processing Letters},
  year    = {2012},
  volume  = {19},
  number  = {5},
  pages   = {251--254},
}

@article{Pereyra:2016,
  author  = {Pereyra, Marcelo and Schniter, Philip and Chouzenoux, Emilie and Pesquet, Jean-Christophe and Tourneret, Jean-Yves and Hero, Alfred O. and McLaughlin, Steve},
  doi     = {10.1109/JSTSP.2015.2496908},
  title   = {A {S}urvey of {S}tochastic {S}imulation and {O}ptimization {M}ethods in {S}ignal {P}rocessing},
  journal = {IEEE Journal of Selected Topics in Signal Processing},
  year    = {2016},
  volume  = {10},
  number  = {2},
  pages   = {224--241},
}

@article{Pereyra:2023,
  author  = {Pereyra, Marcelo and Vargas-Mieles, Luis A. and Zygalakis, Konstantinos C.},
  doi     = {10.1137/22M1506122},
  title   = {The {S}plit {G}ibbs {S}ampler {R}evisited: {I}mprovements to {I}ts {A}lgorithmic {S}tructure and {A}ugmented {T}arget {D}istribution},
  journal = {SIAM Journal on Imaging Sciences},
  year    = {2023},
  volume  = {16},
  number  = {4},
  pages   = {2040--2071},
}

@article{Roberts:1996,
  author  = {Roberts, Gareth O. and Tweedie, Richard L.},
  doi     = {10.2307/3318418},
  title   = {Exponential {C}onvergence of {L}angevin {D}istributions and {T}heir {D}iscrete {A}pproximations},
  journal = {Bernoulli},
  year    = {1996},
  volume  = {2},
  number  = {4},
  pages   = {341--363},
}

@article{Roth:2009,
  author  = {Roth, Stefan and Black, Michael J.},
  doi     = {10.1007/s11263-008-0197-6},
  title   = {Fields of {E}xperts},
  journal = {International Journal of Computer Vision},
  year    = {2009},
  volume  = {82},
  number  = {2},
  pages   = {205--229},
}

@article{Rubinstein:2013,
  author  = {Rubinstein, Ron and Peleg, Tomer and Elad, Michael},
  doi     = {10.1109/TSP.2012.2226445},
  title   = {Analysis {K-SVD}: {A} {D}ictionary-{L}earning {A}lgorithm for the {A}nalysis {S}parse {M}odel},
  journal = {IEEE Transactions on Signal Processing},
  year    = {2013},
  volume  = {61},
  number  = {3},
  pages   = {661--677},
}

@article{Rudin:1992,
  author  = {Rudin, Leonid I. and Osher, Stanley and Fatemi, Emad},
  doi     = {10.1016/0167-2789(92)90242-F},
  title   = {Nonlinear {T}otal {V}ariation based {N}oise {R}emoval {A}lgorithms},
  journal = {Physica D: Nonlinear Phenomena},
  year    = {1992},
  volume  = {60},
  number  = {1},
  pages   = {259--268},
}

@article{Rue:2001,
  author  = {Rue, Havard},
  doi     = {10.1111/1467-9868.00288},
  title   = {Fast {S}ampling of {G}aussian {M}arkov {R}andom {F}ields},
  journal = {Journal of the Royal Statistical Society: Series B (Statistical Methodology)},
  year    = {2001},
  volume  = {63},
  number  = {2},
  pages   = {325--338},
}

@article{Vono:2022,
  author  = {Vono, Maxime and Dobigeon, Nicolas and Chainais, Pierre},
  doi     = {10.1137/20M1371026},
  title   = {High-{D}imensional {G}aussian {S}ampling: {A} {R}eview and a {U}nifying {A}pproach {B}ased on a {S}tochastic {P}roximal {P}oint {A}lgorithm},
  journal = {SIAM Review},
  year    = {2022},
  volume  = {64},
  number  = {1},
  pages   = {3--56},
}

@article{Wang:2017,
  author  = {Wang, Zheng and Bardsley, Johnathan M. and Solonen, Antti and Cui, Tiangang and Marzouk, Youssef M.},
  doi     = {10.1137/16M1080938},
  title   = {Bayesian {I}nverse {P}roblems with l1 {P}riors: {A} {R}andomize-{T}hen-{O}ptimize {A}pproach},
  journal = {SIAM Journal on Scientific Computing},
  year    = {2017},
  volume  = {39},
  number  = {5},
  pages   = {S140--S166},
}

@article{West:1987,
  author  = {West, Mike},
  doi     = {10.1093/biomet/74.3.646},
  title   = {On {S}cale {M}ixtures of {N}ormal {D}istributions},
  journal = {Biometrika},
  year    = {1987},
  volume  = {74},
  number  = {3},
  pages   = {646--648},
}

@article{Zach:2023,
  author  = {Zach, Martin and Knoll, Florian and Pock, Thomas},
  doi     = {10.1109/TMI.2023.3311345},
  title   = {Stable {D}eep {MRI} {R}econstruction {U}sing {G}enerative {P}riors},
  journal = {IEEE Transactions on Medical Imaging},
  year    = {2023},
  volume  = {42},
  number  = {12},
  pages   = {3817--3832},
}

@article{Zhu:1997a,
  author  = {Zhu, Song Chun and Wu, Ying Nian and Mumford, David},
  doi     = {10.1162/neco.1997.9.8.1627},
  title   = {Minimax {E}ntropy {P}rinciple and {I}ts {A}pplication to {T}exture {M}odeling},
  journal = {Neural Computation},
  year    = {1997},
  volume  = {9},
  number  = {8},
  pages   = {1627--1660},
}

@article{Zhu:1997b,
  author  = {Zhu, Song Chun and Mumford, D.},
  doi     = {10.1109/34.632983},
  title   = {Prior {L}earning and {G}ibbs {R}eaction-{D}iffusion},
  journal = {IEEE Transactions on Pattern Analysis and Machine Intelligence},
  year    = {1997},
  volume  = {19},
  number  = {11},
  pages   = {1236--1250},
}

@article{Zhu:1998,
  author  = {Zhu, Song Chun and Wu, Yingnian and Mumford, David},
  doi     = {10.1023/A:1007925832420},
  title   = {Filters, {R}andom {F}ields and {M}aximum {E}ntropy ({FRAME}): {T}owards a {U}nified {T}heory for {T}exture {M}odeling},
  journal = {International Journal of Computer Vision},
  year    = {1998},
  volume  = {27},
  number  = {2},
  pages   = {107--126},
}

@inproceedings{Cheng:2018,
  booktitle = {Proc. Conf. On Learning Theory (COLT)},
  title     = {Underdamped {L}angevin {MCMC}: {A} {N}on-asymptotic {A}nalysis},
  author    = {Cheng, Xiang and Chatterji, Niladri S. and Bartlett, Peter L. and Jordan, Michael I.},
  doi       = {},
  number    = {},
  volume    = {75},
  year      = {2018},
  pages     = {300--323}
}

@inproceedings{Du:2019,
  booktitle = {Advances in Neural Information Processing Systems (NeurIPS)},
  title     = {Implicit {G}eneration and {M}odeling with {E}nergy-{B}ased {M}odels},
  author    = {Du, Yilun and Mordatch, Igor},
  doi       = {},
  number    = {},
  volume    = {},
  year      = {2019},
  pages     = {}
}

@inproceedings{Hinton:1999,
  booktitle = {Proc. Int. Conf. on Artificial Neural Networks (ICANN)},
  title     = {Products of {E}xperts},
  author    = {Hinton, Geoffrey E.},
  doi       = {10.1049/cp:19991075},
  number    = {},
  volume    = {1},
  year      = {1999},
  pages     = {1--6}
}

@inproceedings{Papandreou:2010,
  booktitle = {Advances in Neural Information Processing Systems (NeurIPS)},
  title     = {Gaussian {S}ampling by {L}ocal {P}erturbations},
  author    = {Papandreou, George and Yuille, Alan L.},
  doi       = {},
  number    = {},
  year      = {2010},
  volume    = {23},
  pages     = {},
}

@inproceedings{Paszke:2019,
  booktitle = {Advances in Neural Information Processing Systems (NeurIPS)},
  title     = {{PyTorch}: {A}n {I}mperative {S}tyle, {H}igh-{P}erformance {D}eep {L}earning {L}ibrary},
  author    = {Paszke, Adam and Gross, Sam and Massa, Francisco and Lerer, Adam and Bradbury, James and Chanan, Gregory and Killeen, Trevor and Lin, Zeming and Gimelshein, Natalia and Antiga, Luca and Desmaison, Alban and Kopf, Andreas and Yang, Edward and DeVito, Zachary and Raison, Martin and Tejani, Alykhan and Chilamkurthy, Sasank and Steiner, Benoit and Fang, Lu and Bai, Junjie and Chintala, Soumith},
  doi       = {},
  number    = {},
  volume    = {},
  year      = {2019},
  pages     = {}
}

@inproceedings{Roth:2005,
  booktitle = {Proc. IEEE Conf. on Computer Vision and Pattern Recognition (CVPR)},
  title     = {Fields of {E}xperts: {A} {F}ramework for {L}earning {I}mage {P}riors},
  author    = {Roth, S. and Black, M. J.},
  doi       = {10.1109/CVPR.2005.160},
  number    = {},
  volume    = {2},
  year      = {2005},
  pages     = {860--867}
}

@inproceedings{Schmidt:2010,
  booktitle = {Proc. IEEE Conf. on Computer Vision and Pattern Recognition (CVPR)},
  title     = {A {G}enerative {P}erspective on {MRF}s in {L}ow-{L}evel {V}ision},
  author    = {Schmidt, Uwe and Gao, Qi and Roth, Stefan},
  doi       = {10.1109/CVPR.2010.5539844},
  number    = {},
  volume    = {},
  year      = {2010},
  pages     = {1751--1758}
}

@inproceedings{Song:2019,
  booktitle = {Advances in Neural Information Processing Systems (NeurIPS)},
  title     = {Generative {M}odeling by {E}stimating {G}radients of the {D}ata {D}istribution},
  author    = {Song, Yang and Ermon, Stefano},
  doi       = {},
  number    = {},
  volume    = {},
  year      = {2019},
  pages     = {}
}

@inproceedings{Wainwright:1999,
  booktitle = {Advances in Neural Information Processing Systems (NeurIPS)},
  title     = {Scale {M}ixtures of {G}aussians and the {S}tatistics of {N}atural {I}mages},
  author    = {Wainwright, Martin J. and Simoncelli, Eero},
  doi       = {},
  number    = {},
  volume    = {},
  year      = {1999},
  pages     = {}
}

@inproceedings{Weiss:2007,
  booktitle = {Proc. IEEE Conf. on Computer Vision and Pattern Recognition (CVPR)},
  title     = {What {M}akes a {G}ood {M}odel of {N}atural {I}mages?},
  author    = {Weiss, Yair and Freeman, William T.},
  doi       = {10.1109/CVPR.2007.383092},
  number    = {},
  volume    = {},
  year      = {2007},
  pages     = {1--8}
}

@inproceedings{Welling:2002,
  booktitle = {Advances in Neural Information Processing Systems (NeurIPS)},
  title     = {Learning {S}parse {T}opographic {R}epresentations with {P}roducts of {S}tudent-t {D}istributions},
  author    = {Welling, Max and Osindero, Simon and Hinton, Geoffrey E.},
  doi       = {},
  number    = {},
  volume    = {},
  year      = {2002},
  pages     = {}
}

@book{Barber:2012,
  title     = {Bayesian {R}easoning and {M}achine {L}earning},
  author    = {Barber, David},
  doi       = {10.1017/CBO9780511804779},
  publisher = {Cambridge University Press},
  address   = {Cambridge},
  year      = {2012}
}

@book{Box:1992,
  title     = {Bayesian {I}nference in {S}tatistical {A}nalysis},
  author    = {Box, George E. P. and Tiao, George C.},
  doi       = {10.1002/9781118033197},
  publisher = {John Wiley \& Sons},
  address   = {},
  year      = {1992}
}

@book{Gelman:2013,
  title     = {Bayesian Data Analysis},
  author    = {Gelman, Andrew and Carlin, John B. and Stern, Hal S. and Dunson, David B. and Vehtari, Aki and Rubin, Donald B.},
  doi       = {10.1201/b16018},
  publisher = {CRC Press},
  address   = {},
  year      = {2013},
  edition   = {3rd}
}

@book{Gradshteyn:2007,
  title     = {Table of Integrals, Series, and Products},
  author    = {Gradshteyn, I. S. and Ryzhik, I. M.},
  doi       = {},
  publisher = {Academic Press},
  address   = {},
  year      = {2007},
  edition   = {7th}
}

@book{Koller:2009,
  title     = {Probabilistic Graphical Models: Principles and Techniques},
  author    = {Koller, Daphne and Friedman, Nir},
  publisher = {The MIT Press},
  address   = {Cambridge, MA},
  year      = {2009},
}

@book{Nocedal:2006,
  title     = {Numerical {O}ptimization},
  edition   = {2nd},
  author    = {Nocedal, Jorge and Wright, Stephen J.},
  doi       = {10.1007/978-0-387-40065-5},
  publisher = {Springer},
  address   = {New York, NY},
  year      = {2006}
}

@book{Robert:2004,
  title     = {Monte {C}arlo {S}tatistical {M}ethods},
  edition   = {2nd},
  author    = {Robert, Christian P. and Casella, George},
  doi       = {10.1007/978-1-4757-4145-2},
  publisher = {Springer},
  address   = {New York, NY},
  year      = {2004}
}

@book{Rue:2005,
  title     = {Gaussian {M}arkov {R}andom {F}ields: {T}heory and {A}pplications},
  author    = {Rue, Havard and Held, Leonhard},
  doi       = {10.1201/9780203492024},
  publisher = {Chapman and Hall/CRC},
  address   = {New York},
  year      = {2005}
}

@book{Villani:2009,
  title     = {Optimal {T}ransport},
  subtitle  = {Old and New},
  author    = {Villani, C{'e}dric},
  doi       = {10.1007/978-3-540-71050-9},
  publisher = {Springer},
  address   = {Heidelberg},
  year      = {2009}
}

@article{Fruehwirth:2024,
  title={{E}rgodicity of {L}angevin {D}ynamics and its {D}iscretizations for {N}on-smooth {P}otentials},
  author={Fruehwirth, Lorenz and Habring, Andreas},
  doi={10.48550/arXiv.2411.12051},
  journal={arXiv preprint},
  year={2024}
}

@article{Habring:2025,
  title={Diffusion at {A}bsolute {Z}ero: {L}angevin {S}ampling {U}sing {S}uccessive {M}oreau {E}nvelopes},
  author={Habring, Andreas and Falk, Alexander and Zach, Martin and Pock, Thomas},
  doi={10.48550/arXiv.2503.22258},
  journal={arXiv preprint},
  year={2025}
}

@article{Kingma:2013,
  title={Auto-{E}ncoding {V}ariational {B}ayes},
  author={Kingma, Diederik P and Welling, Max},
  doi = {10.48550/arXiv.1312.6114},  
  journal={arXiv preprint},
  year={2013}
}

@incollection{Geyer:2011,
  title = {{I}mportance {S}ampling, {S}imulated {T}empering, and {U}mbrella {S}ampling},
  author = {Charles J. Geyer},
  booktitle = {Handbook of Markov Chain Monte Carlo},
  editor = {Steve Brooks and Andrew Gelman and Galin Jones and Xiao-Li Meng},
  publisher = {Chapman and Hall/CRC},
  year = {2011},
  chapter = {11},
  pages = {295--311},
  doi = {10.1201/b10905},
  isbn = {9780429138508}
}

@incollection{Papandreou:2016,
  title = {Perturb-and-{MAP} {R}andom {F}ields},
  author = {George Papandreou and Alan L. Yuille},
  booktitle = {Perturbations, Optimization, and Statistics},
  editor = {Tamir Hazan and George Papandreou and Daniel Tarlow},
  publisher = {The MIT Press},
  year = {2016},
  chapter = {2},
  pages = {21--48},
  doi = {10.7551/mitpress/10761.003.0003},
  isbn = {9780262337939}
}

@incollection{Plataniotis:2001,
  title = {Gaussian {M}ixtures and {T}heir {A}pplications to {S}ignal {P}rocessing},
  author = {Kostantinos N. Plataniotis and Dimitris Hatzinakos},
  booktitle = {Advanced Signal Processing Handbook},
  editor = {Stergios Stergiopoulos},
  publisher = {CRC Press},
  year = {2001},
  chapter = {3},
  pages = {},
  doi = {10.1201/9781315149790},
  isbn = {9781315149790}
}

@incollection{Robert:2011,
  title = {A {S}hort {H}istory of {MCMC}: {S}ubjective {R}ecollections from {I}ncomplete {D}ata},
  author = {Christian Robert and George Casella},
  booktitle = {Handbook of Markov Chain Monte Carlo},
  editor = {Steve Brooks and Andrew Gelman and Galin Jones and Xiao-Li Meng},
  publisher = {Chapman and Hall/CRC},
  year = {2011},
  chapter = {2},
  pages = {49--66},
  doi = {10.1201/b10905},
  isbn = {9780429138508}
}

@unpublished{Bromiley:2014,
  type = {Technical Report},
  title = {Products and {C}onvolutions of {G}aussian {P}robability {D}ensity {F}unctions},
  author = {Bromiley, Paul A.},
  address = {},
  publisher = {},
  year = {2014},
  url = {}
}

@misc{Chierchia,
  author       = {Chierchia, G. and Chouzenoux, E. and Combettes, P. L. and Pesquet, J.-C},
  title        = {The {P}roximity {O}perator {R}epository},
  howpublished = {\url{https://proximity-operator.net/}},
  note         = {Accessed: 2025-03-06},
  year         = {}
}

@article{Zach:2024,
  title = {{Product} of {Gaussian} {Mixture} {Diffusion} {Models}},
  volume = {66},
  ISSN = {1573-7683},
  DOI = {10.1007/s10851-024-01180-3},
  number = {4},
  journal = {Journal of Mathematical Imaging and Vision},
  publisher = {Springer Science and Business Media LLC},
  author = {Zach, Martin and Kobler, Erich and Chambolle, Antonin and Pock, Thomas},
  year = {2024},
  month = mar,
  pages = {504–528}
}
